%% file: main.tex
\documentclass[acmsmall]{acmart}

\AtBeginDocument{%
  \providecommand\BibTeX{{%
    \normalfont B\kern-0.5em{\scshape i\kern-0.25em b}\kern-0.8em\TeX}}}

\input{publication-metadata}

\input{layout}
\usepackage{listings}

\def\oursolution{DYLAN}
\usepackage{color}
\definecolor{dkgreen}{rgb}{0,0.6,0}
\definecolor{gray}{rgb}{0.5,0.5,0.5}
\definecolor{mauve}{rgb}{0.58,0,0.82}

\newcommand{\update}[1]{#1}

\usepackage{subcaption}
\usepackage[ruled,linesnumbered]{algorithm2e}
\usepackage{enumitem}
\usepackage{multirow}
\usepackage{amsmath}

\begin{document}
\title{Dynamic Graph Indexing for Label-Filtered Approximate Nearest Neighbor Search}

\input{authors}

\input{section/0-abstract}

\input{classification}
\input{article-history}

 \maketitle

\input{command}

\input{section/1-introduction}

\input{section/2-preliminaries}

\input{section/3-solution}

\input{section/5-experiments}

\input{section/4-relatedwork}

\input{section/6-conclusion}

\section*{Acknowledgments}
This work is supported by the “1+1+1” CUHK-CUHK(SZ)-GDST Joint Collaboration Fund (No. 2025A0505000045) and a CUHK direct grant.

\newpage
\bibliographystyle{ACM-Reference-Format}
\bibliography{reference}

\newpage
\input{section/appendix}

\end{document}

%% file: publication-metadata.tex
\setcopyright{cc}
\setcctype{by}
\copyrightyear{2026}
\acmJournal{PACMMOD}
\acmYear{2026}
\acmVolume{4}
\acmNumber{6 (SIGMOD)}
\acmArticle{335}
\acmMonth{12}
\acmPrice{}
\acmDOI{10.1145/3856330}

%% file: layout.tex
\newsavebox{\CRbox}
\newenvironment{CRfit}{\begin{lrbox}{\CRbox}}{%
\end{lrbox}%
\ifdim\wd\CRbox>\linewidth
  \resizebox{\linewidth}{!}{\usebox{\CRbox}}%
\else
  \usebox{\CRbox}%
\fi}

\AtBeginDocument{\SetAlFnt{\footnotesize}}

%% file: authors.tex
\author{Mengxu Jiang}
\email{mxjiang@se.cuhk.edu.hk}
\orcid{0009-0007-5525-682X}
\affiliation{%
  \institution{The Chinese University of Hong Kong}
  \country{Hong Kong}
}

\author{Zhi Yang}
\email{zyang@se.cuhk.edu.hk}
\orcid{0009-0005-5931-4412}
\affiliation{%
  \institution{The Chinese University of Hong Kong}
  \country{Hong Kong}
}

\author{Fangyuan Zhang}
\orcid{0009-0004-3253-4661}
\affiliation{
\institution{Harbin Institute of Technology}\city{Shenzhen}\country{China}}
\email{zzzzzfy@link.cuhk.edu.hk}

\author{Yin Yang}
\orcid{0000-0002-0549-3882}
\affiliation{\institution{Hamad Bin Khalifa University}\city{Doha}\country{Qatar}}
\email{yyang@hbku.edu.qa}

\author{Sibo Wang}
\email{swang@se.cuhk.edu.hk}
\orcid{0000-0003-1892-6971}
\affiliation{
  \institution{The Chinese University of Hong Kong}
  \country{Hong Kong}
}
\authornote{Sibo Wang is the corresponding author.}

%% file: section/0-abstract.tex
\begin{abstract}

Approximate nearest neighbor search (ANNS) over high-dimensional vectors is a core primitive in modern retrieval systems. In practice, objects are often also associated with labels (e.g., attributes or keywords), and retrieval must combine vector similarity with explicit label-based filtering, giving rise to label-filtered ANNS. This task is challenging because label filtering and ANNS
favor very different index structures, selective filters fragment the admissible
search space, and updates complicate index maintenance.

To address these challenges, we propose DYLAN, a dynamic composite index for label-filtered ANNS supporting
containment, overlap, and their conjunctions and disjunctions. DYLAN organizes
label sets by shared prefixes in a frequency-ordered trie and selectively shares
HNSW-style graph indices across trie-induced subsets, so that each trie region
has a constant-factor covering index while each object participates in only
logarithmically many indices. To mitigate fragmentation, DYLAN performs a
cross-region multi-graph search that exploits ancestor-level indices as bridges
between otherwise isolated trie regions. To support insertions, it adopts a
threshold-based maintenance scheme that rebuilds shared indices only when
necessary, yielding low amortized update cost.

Extensive experiments with real data show that DYLAN consistently improves query throughput at comparable recall, by up to two orders of magnitude on challenging label-rich datasets, while retaining competitive memory, construction, and update costs.

\end{abstract}

%% file: classification.tex
\ccsdesc[500]{Information systems~Retrieval models and ranking}

%% file: article-history.tex
\received{17 April 2026}
\received[revised]{19 August 2026}
\received[accepted]{12 September 2026}

%% file: command.tex
\def\header{\vspace{1mm} \noindent}
\def\done{\hspace*{\fill} $\framebox[2mm]{}$ \medskip}

\def\diffsym{{{}_\Delta}}
\newcommand{\tuple}[1]{\left\langle #1 \right\rangle}
\newcommand{\vect}[1]{\boldsymbol{#1}}
\newcommand{\mceil}[1]{\left\lceil #1 \right\rceil}
\newcommand{\mfloor}[1]{\left\lfloor #1 \right\rfloor}

\newcommand{\bigO}[1]{O\left(#1\right)}
\newcommand{\bigTheta}[1]{\Theta\left(#1\right)}
\newcommand{\bigOmega}[1]{\Omega\left(#1\right)}
\newcommand{\x}{\boldsymbol{x}}

\renewcommand{\O}{\mathcal{O}}

\makeatletter
\newcommand*{\setunitlength}{\setlength{\unitlength}{\f@size pt}}
\makeatother
\newcommand*{\butterfly}{%
    \setunitlength%
    \begin{picture}(.6,.6)
        \roundcap
        \put( 0, 0){\line( 2,2){.6}}
        \put(.6, 0){\line(-2,2){.6}}
        \put( 0, 0){\line( 1.5,0){.6}}
        \put( 0,.6){\line( 1.5,0){.6}}
    \end{picture}%
}

\newcommand*{\butterflyhorizon}{%
    \setunitlength%
    \begin{picture}(.6,.6)
        \roundcap
        \put( 0, 0){\line( 2,2){.6}}
        \put(.6, 0){\line(-2,2){.6}}
        \put( 0, 0){\line( 0,1.5){.6}}
        \put( 0.6,0){\line( 0,1.5){.6}}
    \end{picture}%
}

\newcommand*{\bitriangle}{%
    \setunitlength%
    \begin{picture}(.6,.6)
        \roundcap
        \put( 0, 0){\line( 2,1){.6}}
        \put(.6, 0){\line(-2,1){.6}}
        \put( 0, 0.3){\line( 2,1){.6}}
        \put(.6, 0.3){\line(-2,1){.6}}
        \put( 0, 0){\line( 1.5,0){.6}}
        \put( 0,.6){\line( 1.5,0){.6}}
    \end{picture}%
}

\newcommand*{\bitriangletall}{%
    \setunitlength%
    \begin{picture}(.8,.8)
        \roundcap
        \put( 0, 0){\line( 2,1){.8}}
        \put(.8, 0){\line(-2,1){.8}}
        \put( 0, 0.4){\line( 2,1){.8}}
        \put(.8, 0.4){\line(-2,1){.8}}
        \put( 0, 0){\line( 1.5,0){.8}}
        \put( 0,.8){\line( 1.5,0){.8}}
    \end{picture}%
}

\newcommand*{\bitrianglehorizon}{%
    \setunitlength%
    \begin{picture}(.8,.6)
        \roundcap
        \put( 0, 0){\line( 1,1.5){.4}}
        \put(.8, 0){\line(-1,1.5){.4}}
        \put( 0, 0.6){\line( 1,-1.5){.4}}
        \put(.8, 0.6){\line(-1,-1.5){.4}}
        \put( 0, 0){\line( 0,1.5){.6}}
        \put( .8,0){\line( 0,1.5){.6}}
    \end{picture}%
}

\newcommand*{\caterpillar}{%
    \setunitlength%
    \begin{picture}(.6,.6)
        \roundcap
        \put( 0, 0){\line( 2,2){.6}}
        \put(.6, 0){\line(-2,2){.6}}
        \put( 0,.6){\line( 1.5,0){.6}}
    \end{picture}%
}

\newcommand*{\fourpath}{%
    \setunitlength%
    \begin{picture}(.8,.6)
        \roundcap
        \put( 0, 0.6){\line( 1,-1.5){.4}}
        \put(.8, 0.6){\line(-1,-1.5){.4}}
        \put( 0, 0){\line( 0,1.5){.6}}
        \put( .8,0){\line( 0,1.5){.6}}
    \end{picture}%
}

\def\Cater{\caterpillar}
\def\BF{\butterfly }
\def\BT{\bitrianglehorizon}
\def\fourp{\fourpath}

\def\Var{\mathbb{V}\mathrm{ar}}

\def\Cov{\mathbb{C}\mathrm{ov}}
\def\E{\mathbb{E}}

\def\C{\mathcal{C}}
\def\Pr{\mathrm{Pr}}
\def\pspace{\mathcal{S}}
\def\evspace{\mathcal{F}}
\newcommand{\1}[1]{\mathds{1}_{\left[#1\right]}}

\def\Ngh{\mathcal{N}}
\def\Edge{\mathcal{E}}

\def\taume{{\tau\text{-}1}}
\newcommand{\past}[1]{#1_\taume}
\newcommand{\cur}[1]{#1_\tau}

\def\emalpha{{1\mathrm{-}\alpha}}
\def\rmax{r_{max}}
\def\pih{\hat{\pi}}
\def\pitil{\tilde{\pi}}
\newcommand{\funcname}[1]{\operatorname{\bf #1}}

\def\srev{}

\def\algoem{\texttt}
\def\rev{\color{blue}}
\def\revh{\color{blue}}
\newcommand{\delText}[1]{}

\newenvironment{proofsketch}{%
  \renewcommand{\proofname}{Proof Sketch}\proof}{\endproof}

\def\agendastar{$\textrm{Agenda}^\#$}
\newcommand{\tr}[1]{}

%% file: section/1-introduction.tex
\section{Introduction}
\label{sec:intro}

Recent advances in machine learning, especially large language models (LLMs), have
significantly increased the demand for managing high-dimensional vector data. Through a
machine learning model, an unstructured object, such as an image or a full-text article,
can be mapped to a high-dimensional vector called its \textit{embedding}, in such a way
that objects close in the vector space are often semantically related. For example, under
a vision-language model such as CLIP~\cite{radford2021learningtransferablevisualmodels},
the image of a pill and the textual instructions of the corresponding medicine map to
nearby embeddings. This enables semantic retrieval over unstructured data, typically
through \textit{approximate nearest neighbor search} 
(\textit{ANNS})~\cite{IndykM98}, which is supported effectively in modern vector database
systems~\cite{chen2026nova, Chen2024SingleStoreV, JiangWLTDWZWLYRLC25}.

In practice, however, retrieval is rarely based on vector similarity alone. Each object is often also associated with a set of \textit{labels}, such as categories, tags, or keywords. For example, medical articles may be tagged as ``papers in reputed medical journals'', ``manufacturer instructions'', or ``social posts''. Accordingly, ANNS is often augmented with explicit label-based filtering, e.g., retrieving only papers in reputed medical journals and manufacturer instructions, so as to provide stronger control over the source of information. Such a query is referred to as \textit{label-filtered ANNS}~\cite{GollapudiKSKBRL23}.

\header
\textbf{Main challenges.}
Although both ANNS and label-based filtering have been studied extensively, their combination %
remains challenging because the two tasks favor fundamentally different index structures. Label filtering is typically supported by set-oriented structures such as inverted lists~\cite{ZobelMoffat1998}, whereas ANNS is usually supported by graph-based indices such as HNSW~\cite{MalkovY20}. A straightforward approach is to %
choose one of these structures, which leads to either \textit{pre-filtering} (i.e., employing a label index to filter objects but performing ANNS index-free) or \textit{post-filtering} (index-based ANNS with post-processing to filter the results with labels). 
The former may incur a costly scan over a large candidate set, whereas the latter often requires substantial over-retrieval to maintain recall.

\input{table/comparison}

A more integrated direction, called \textit{in-filtering}~\cite{patel2024acorn}, performs filtering during graph traversal. Specifically, when processing a label-filtered ANNS query on an HNSW-style graph~\cite{MalkovY20}, the search expands only neighbors that satisfy the filter. The main difficulty is that a selective filter may eliminate a large fraction of graph edges, effectively sparsifying the admissible subgraph and making many qualified objects unreachable, which leads to false dismissals and low recall. Existing techniques alleviate this issue by exploring larger local neighborhoods, e.g., via two-hop expansion~\cite{patel2024acorn, DBLP:journals/pvldb/SehgalS25}, but this increases traversal cost and still does not exploit the structure of label-set relationships. In particular, such methods are largely \textit{label-agnostic}: they treat the filter as a generic predicate rather than explicitly indexing how label sets relate to one another.

\noindent
\textbf{Existing label-aware solutions.}
To our knowledge, only a limited number of \textit{label-aware} ANNS indices have been proposed, and existing solutions still address only part of the problem. Table~\ref{tab:label_aware} summarizes representative methods in terms of label awareness, query efficiency across predicates, update support, and label scalability. Label filter types (i.e., containment/overlap/complex) are formally defined in Section~\ref{sec:pre:problem_definition}. 
Update support and label scalability are assessed from each method's reported behavior.

A first line of work augments a \emph{single graph index} with label
information. Filtered-DiskANN~\cite{GollapudiKSKBRL23} incorporates labels into both index
construction and search and supports incremental updates, but focuses on overlap
filters, performs poorly on containment filters, and does not extend to complex
predicates. RWalks~\cite{aitAomar2025rwalks} instead guides retrieval by a hybrid score
combining vector proximity with label similarity, avoiding a hard admissibility
check during expansion; however, it encodes labels into vector-like
representations, so its index expands substantially with many distinct label
types, and updates are not explicitly addressed. As reflected in Table~\ref{tab:label_aware},
both fail to preserve enough edges that guide the search from one qualified
object to another, and hence struggle with highly selective label filters, as
shown in our experiments.

Another line of work adopts a \emph{partitioned multi-graph} design. Notably,
UNG~\cite{UNG2024} groups objects by distinct label sets and builds a separate
graph index for each group, which fragments the vector space and weakens global
navigability. UNG alleviates this with a label navigating graph (LNG) connecting
related subgraphs, but cross-region connectivity may still be insufficient for
queries spanning multiple regions, hurting recall. Inserting a new label set may
also trigger widespread changes in the LNG, slowing down updates.

A third category materializes filter states, e.g., UniFilter~\cite{Unifilter}
combines an HNSW graph with an AND/OR-style graph for label relationships.
Targeting containment queries, it caps the size of its auxiliary label index, so
its memory stays bounded as the label universe grows; the price, as reflected in
Table~\ref{tab:label_aware}, is that new labels are only partially absorbed and
query performance degrades substantially on label-rich datasets.

Finally, Curator~\cite{Curator} applies an \emph{IVF-based} index for filtered ANNS under low selectivity, whose search effectiveness is weaker than that of graph-based ANNS indices as acknowledged in \cite{Curator}. 

\noindent
{\bf Our solution.} We propose \textit{DYnamic Label-filtered Approximate Nearest neighbor search} (\textit{{\oursolution}}), an index for label-filtered ANNS with efficient updates. {\oursolution} supports containment, overlap, and complex predicates composed via conjunction and disjunction. \update{In this paper, we assume the index fits in main memory, and our implementation is single-threaded.}

Our starting point is simple: organizing data objects in a shared-prefix label trie
(efficient for label-based filtering), and then materializing graph indices (effective
for ANNS) over trie-induced subsets of vectors. The challenge, however, is that
materializing a graph index for every trie node incurs prohibitive space and update costs
for a large trie, whereas maintaining graph indices only at lower-level (i.e.,
fine-grained) trie nodes fragments the search space and restricts the global navigability
required by graph-based ANNS; moreover, updates may substantially change the trie
structure and invalidate pre-built graph indices. {\oursolution} therefore builds on three
key ideas: {\em selective graph materialization}, {\em cross-region graph navigation}, and
{\em threshold-based maintenance of shared graph indices}.

Concretely, {\oursolution} is built in two phases. The first constructs a label trie $T$
capturing the relationships among label sets in the data: each non-root node $u \in T$
carries a label and a prefix formed by the labels on the root-to-$u$ path, each object is
assigned to the leaf whose prefix matches its label set, and the objects assigned to
leaves in the subtree rooted at $u$ form the \emph{trie region} of $u$. The second phase
{\em selectively materializes multiple HNSW-style graph indices} embedded in $T$, rather
than building one for every trie node. Each object participates in only logarithmically
many graph indices, keeping the total index size near-linear (Theorem~\ref{thm:build}),
while the shared index covering a trie region contains at most a constant factor more
objects than the region itself, thereby avoiding substantial degradation in search
efficiency. Intuitively, graph indices near the root capture more global vector
relationships across broader label regions, whereas those near the leaves preserve denser
local connectivity among objects with highly similar label sets.

Given a query vector $\mathbf{q}$ with a label predicate $\Phi$, the filtered search space
often spans multiple trie regions. {\oursolution} first identifies a set $\mathcal{N}$ of
trie nodes whose subtrees jointly cover the filtered object set: the regions of
containment and overlap predicates are located directly on the trie, whereas a complex
predicate is recursively decomposed into sub-predicates combined by conjunction or
disjunction, whose regions are computed separately and then combined. Searching the
corresponding graph indices in isolation is ineffective, because the search may become
trapped in local optima and miss promising candidates in other regions. {\oursolution}
therefore performs a {\em cross-region multi-graph search}, further exploiting the graph
indices at common ancestors of the nodes in $\mathcal{N}$; these ancestor-level indices
may contain cross-subtree connections that serve as effective bridges across otherwise
isolated trie regions, restoring the global navigability needed by graph-based ANNS.

{\oursolution} also supports dynamic updates. Following common practice in dynamic
graph-based ANNS indices~\cite{jiang2025digra,zhang2025efficient,Unifilter}, deletions are
handled lazily by marking deleted objects as invalid and skipping them during search.
Insertions are more challenging: a single insertion may affect multiple shared graph
indices along the corresponding root-to-leaf path, and an eager strategy would trigger the
construction of new graph indices or the rebuilding of existing ones. {\oursolution}
therefore adopts a {\em threshold-based maintenance} scheme, which rebuilds a shared graph
index only when its size mismatch with the target region exceeds a prescribed threshold,
yielding a provably bounded amortized insertion cost while preserving the search
effectiveness of the maintained indices.

Extensive experiments on multiple real datasets validate the effectiveness of {\oursolution}. Our main contributions are as follows.

\begin{itemize}[topsep=0mm, partopsep=0pt, itemsep=0pt, leftmargin=10pt]

    \item We propose the {\oursolution} index, which introduces a {\em selective graph materialization} scheme on a label trie for effective label-filtered ANNS, while keeping the index size near-linear.
    
    \item We design a {\em cross-region multi-graph search} algorithm that exploits ancestor graph indices to bridge fragmented trie regions, balancing query efficiency and accuracy.
    
    \item We develop a {\em threshold-based maintenance} scheme for dynamic updates,  with a
provably bounded amortized insertion cost. %

    \item Extensive experiments show that {\oursolution} achieves superior query efficiency at similar recall levels, with favorable update and indexing costs across multiple datasets for label-filtered ANNS.
\end{itemize}

%% file: table/comparison.tex
\begin{table*}[t]
\centering
\caption[Comparison of label-filtered ANNS methods.]{Comparison of label-filtered ANNS methods. \ $\checkmark$: full support; $\times$: no support; ``Partial'': limited support. \textcolor{black}{Query-efficiency ratings summarize the empirical trends in Sec.\ \ref{sec:exp}. ``High'': consistently reaches the target recall and is among the fastest across reported datasets; ``Medium'': reaches the recall but with clearly lower throughput or only on some datasets; ``Low'': often falls short or is substantially slower. Label scalability is rated by how index space grows with the label universe. ``High'': near-constant or slowly growing; ``Low'': a growth rate that becomes prohibitive.}}\vspace{-2mm}
\label{tab:label_aware}

\scalebox{0.90}{
\begin{CRfit}\begin{tabular}{@{} l c *{4}{c} *{2}{c} c c @{}}
\toprule
\multirow{2}{*}{\textbf{Method}} & \multirow{2}{*}{\textbf{Label Aware}} & \multicolumn{3}{c}{\textbf{Query Efficiency}} & \multicolumn{2}{c}{\textbf{Update Support}} & \multirow{2}{*}{\textbf{Label Scalability}} \\
\cmidrule(lr){3-5} \cmidrule(lr){6-7}
& & \textbf{Containment} & \textbf{Overlap} & \textbf{Complex} & \textbf{Vector} & \textbf{Label set} & \\
\midrule
ACORN~\cite{patel2024acorn} & $\times$  & Low & Low & Low & $\checkmark$  & $\checkmark$  & High \\
NaviX~\cite{DBLP:journals/pvldb/SehgalS25} & $\times$  & Low & Low & Low & $\checkmark$  & $\checkmark$  & High \\
Filtered-DiskANN~\cite{GollapudiKSKBRL23} & $\checkmark$ & Low & Medium & $\times$ & $\checkmark$  & $\checkmark$  & High \\
RWalks~\cite{aitAomar2025rwalks} & $\checkmark$ & Medium & Medium & $\times$ & $\times$ & $\times$ & Low \\
UNG~\cite{UNG2024} & $\checkmark$ &  Medium & Medium & $\times$ & $\checkmark$ & $\times$ & High \\
UniFilter~\cite{Unifilter} & $\checkmark$ & Low & $\times$ & $\times$ & $\checkmark$ & Partial & High \\

Curator~\cite{Curator} & $\checkmark$ & Medium &  Medium &  Medium & $\checkmark$ & $\checkmark$ & High\\

{\oursolution} (Ours) & $\checkmark$ & High & High& High & $\checkmark$ & $\checkmark$ & High \\
\bottomrule
\end{tabular}\end{CRfit}
}
\vspace{-1mm}
\end{table*}

%% file: section/2-preliminaries.tex
\section{Preliminaries}
\subsection{Problem Definition}\label{sec:pre:problem_definition}
Let $O$ be a dataset containing $n$ objects, where each object is represented as a vector in $\mathbb{R}^d$. Given two vectors $\mathbf{p}, \mathbf{q} \in \mathbb{R}^d$, let $\delta(\mathbf{p}, \mathbf{q})$ represent their distance, e.g., Euclidean distance in $\mathbb{R}^d$. We define nearest neighbor search as follows.

\begin{definition}[Nearest Neighbor Search (NNS)]\label{def:nns}
Given an object set $O$ consisting of $n$ objects, a query vector $\mathbf{q} \in \mathbb{R}^d$, and a positive integer $k \le n$, the nearest neighbor search returns a set of vectors $R^* = \{\mathbf{o}_1^*, \mathbf{o}_2^*, \dots, \mathbf{o}_k^*\} \subseteq O$ 
with the top-$k$ smallest distances to $\mathbf{q}$, such that $\forall \mathbf{o}^* \in R^*, \forall \mathbf{o} \notin R^*$, we have  $\delta(\mathbf{o}^*, \mathbf{q}) \leq \delta(\mathbf{o}, \mathbf{q})$.
\end{definition}

\begin{figure}[!tbp]
\centering
\includegraphics[height=35mm]{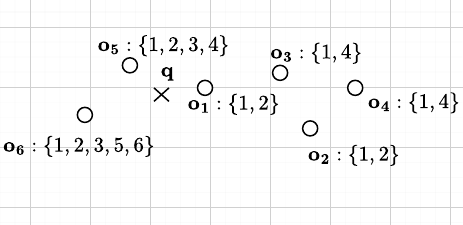}%
\caption{Example of a label-filtered nearest neighbor search}
\label{Fig:Vector}
\end{figure}

Next, we focus on the situation in which each object in $O$ is further associated with a label set. For instance, Figure~\ref{Fig:Vector} shows a set of objects \( O=\{\mathbf{o}_1, \mathbf{o}_2, \dots, \mathbf{o}_6\} \) where each object is associated with a set of integer-valued labels, e.g., \( \mathbf{o}_1 \) has a label set $\{1,2\}$.

We define label-filtered NNS as follows.

\begin{definition}[Label-Filtered NNS]\label{def:lfnns}
Given an object set $O$ where each object $\mathbf{o} \in O$ is associated with
a label set $f_{\mathbf{o}}$, a query is defined as $Q=(\mathbf{q},\Phi, k)$,
where $\mathbf{q}\in\mathbb{R}^d$ is a query vector and $\Phi$ is a label
filtering predicate. The label-filtered NNS returns the $\min(k, |O_\Phi|)$-nearest neighbors
to $\mathbf{q}$ within the subset
\(
O_{\Phi}=\{\mathbf{o}\in O \mid f_{\mathbf{o}} \models \Phi\},
\)
i.e., the objects satisfying the label filter constraint $\Phi$.
\end{definition}

In the above definition, a common notion for the filter predicate $\Phi$
(e.g., \cite{UNG2024}) is
$\Phi ::= \mathsf{cont}(F)\allowbreak | \mathsf{over}(F)\allowbreak | (\Phi \land \Phi)\allowbreak | (\Phi \lor \Phi)$, which
include
(i) \emph{containment}: given a query label set $f_q$, an object
$\mathbf{o}$ with label set $f_{\mathbf{o}}$ satisfies
$f_{\mathbf{o}} \models \mathsf{cont}(f_q)$ iff $f_q \subseteq f_{\mathbf{o}}$,
and
(ii) \emph{overlap}:
$f_{\mathbf{o}} \models \mathsf{over}(f_q)$ iff
$f_q \cap f_{\mathbf{o}} \neq \emptyset$.
In addition, we can combine predicates using logical connectors to form
complex filters:
\emph{Conjunction}:
$f_{\mathbf{o}} \models (\Phi_1 \land \Phi_2)$ iff
$f_{\mathbf{o}} \models \Phi_1$ and $f_{\mathbf{o}} \models \Phi_2$;
\emph{Disjunction}:
$f_{\mathbf{o}} \models (\Phi_1 \lor \Phi_2)$ iff
$f_{\mathbf{o}} \models \Phi_1$ or $f_{\mathbf{o}} \models \Phi_2$.
Here $\Phi_1$ and $\Phi_2$ can themselves be complex label filters.

In the example of Figure~\ref{Fig:Vector}, given a query vector $\mathbf{q}$
(represented by a cross) and $k=2$, an NNS on $O$ returns $\mathbf{o}_1$ and $\mathbf{o}_5$, which are
closest to $\mathbf{q}$. Meanwhile, given a query label set
$f_q=\{1,3\}$ and the predicate $\Phi=\mathsf{cont}(f_q)$, the
label-filtered NNS returns $\mathbf{o}_5$ and $\mathbf{o}_6$, which are the
closest objects to $\mathbf{q}$ whose label sets contain both query labels $1$
and $3$. Now consider two query label sets $f_{q,1}=\{3,6\}$ and $f_{q,2}=\{1,2\}$, and
let $\Phi=\mathsf{cont}(f_{q,1})\lor\mathsf{cont}(f_{q,2})$. Then
$\mathbf{o}_6$ satisfies $\mathsf{cont}(f_{q,1})$, while
$\mathbf{o}_1,\mathbf{o}_2,\mathbf{o}_5,\mathbf{o}_6$ satisfy
$\mathsf{cont}(f_{q,2})$. Hence, the filtered object set is
\(
\{\mathbf{o}_1,\mathbf{o}_2,\mathbf{o}_5,\mathbf{o}_6\},
\)
and the label-filtered NNS returns the top-$k$ objects from this set, namely
$\mathbf{o}_1$ and $\mathbf{o}_5$.

Exact NNS in a high-dimensional space is computationally expensive due to the curse of dimensionality~\cite{IndykM98}. Hence, approximate nearest neighbor search (ANNS) is widely used~\cite{aitAomar2025rwalks,UNG2024}, which offers a tradeoff between query efficiency and result quality (commonly measured by recall). 
In particular, let $R^*$ be the exact label-filtered NNS result and $R$ is the ANNS result for the same query. We define $\text{Recall}(R) = \frac{|R \cap R^*|}{|R^*|}$, which is $\frac{|R \cap R^*|}{k}$ when $|O_\Phi| \ge k$. 

\subsection{HNSW}\label{sec:pre:HNSW}

Among existing ANNS algorithms, graph-based methods~\cite{FuXWC19, MalkovY20, SubramanyaDSKK19}, notably hierarchical navigable small world (HNSW)~\cite{MalkovY20}, have demonstrated superior performance in terms of the trade-off between query time and accuracy. Given a set $O$ of $n$ vectors in $\mathbb{R}^d$, it builds a \textit{proximity graph} $G$, in which each vertex corresponds to a vector in $O$ and each directed edge connects a pair of vertices according to a pre-defined proximity rule under the distance metric $\delta$ of Section~\ref{sec:pre:problem_definition}. In the particular case of HNSW, $G$ consists of $L$ layers, each of which contains a proximity graph known as a navigable small world (NSW); let $H[i]$ be the graph at layer $i$. The number of vertices decreases exponentially from the bottom layer $H[0]$ to the top layer $H[L-1]$, and the maximum out-degree is $2M$ at $H[0]$ and $M$ at all higher layers, where $M$ is a system parameter.

\noindent\textbf{Search.} To process an ANNS query $\mathbf{q}$, HNSW starts from an entry vertex in the highest layer $H[L-1]$ and iteratively descends the layers, using the vertex closest to $\mathbf{q}$ at the current layer as the entry point for the next. At the bottom layer $H[0]$, it performs a greedy search maintaining a candidate queue $C$ of discovered vertices to be expanded and a result queue $W$ of up to $ef$ closest vertices to $\mathbf{q}$ seen so far. At each step, the vertex closest to $\mathbf{q}$ is extracted from $C$; the search terminates once it is farther from $\mathbf{q}$ than the farthest vertex in $W$, and otherwise its out-neighbors are examined and eligible vertices are inserted into both queues, with $W$ truncated to size $ef$. Finally, the top-$k$ closest vertices in $W$ are returned. A larger value of the system parameter $ef$ increases the number of vertices visited before termination, leading to higher recall with increased query cost, and vice versa.

\noindent\textbf{Construction and updates.} The proximity graph $G$ is built incrementally, by sequentially inserting vertices into an initially empty $G$. To insert a vertex $u$, the algorithm first determines its highest level $L_u$ using a truncated geometric distribution, and then inserts $u$ from level $H[L_u]$ down to $H[0]$. At each level, a greedy layer search with construction parameter $ef_c$ collects a bounded candidate set of vertices close to $u$, which is then pruned to form the out-neighbor set $N(u)$ within the out-degree limit: candidates are sorted by their distance to $u$, and a candidate $v$ is discarded if it is \emph{dominated} by an already selected $w$, i.e., $\delta(u,w) < \delta(u,v)$ and $\delta(v,w) < \delta(u,v)$. Each selected neighbor $v$ adds $u$ to $N(v)$, and $N(v)$ is pruned in the same way if it exceeds the limit. Since $L_u$ has a constant expected value, building an HNSW graph of $n$ vectors takes $O(n \cdot c_{\text{upd}})$ time and $O(nM)$ space, where $c_{\text{upd}}$ denotes the insertion cost of a single vector. Insertions are thus already covered by the construction procedure; for deletions, HNSW commonly employs \emph{soft deletion}, marking a vertex as deleted without removing it from the graph structure and skipping it in future searches.

%% file: section/3-solution.tex
\section{Our Solution: {\oursolution} }\label{sec:solution}

Next, we present the proposed {\oursolution} indexing scheme. Sec.~\ref{subsec:index} introduces the label trie and the selective graph materialization scheme. Secs.~\ref{subsec:query} and~\ref{subsec:update} describe the cross-region query processing algorithm and the update mechanism based on threshold-based maintenance, respectively. These three components correspond to the three core challenges outlined in the introduction: index materialization, cross-region navigability, and dynamic maintenance.
Table~\ref{tab:notation_summary} summarizes the common notations used in the paper.

\subsection{Index Structure}\label{subsec:index}

Recall from Sec.~\ref{sec:intro} that {\oursolution} organizes objects in a frequency-ordered label trie and selectively materializes HNSW-style graph indices over trie-induced subsets: the trie captures how label sets relate to one another, while the graph indices preserve vector-space navigability within and across the resulting subsets. Materializing a graph index for every trie node, however, would lead to prohibitive space and update costs. {\oursolution} therefore selectively shares graph indices among trie nodes, while bounding the additional search space introduced by such sharing. We next describe the label trie and the selective graph materialization scheme.

\header
\textbf{Label trie.} 
The {\oursolution} index employs a label trie $T$ to organize the labels of the given object set $O$.

To avoid confusion, in the following we use different terms to distinguish elements of the trie and the HNSW graph: in particular, a \textit{node} always resides in the trie, whereas a \textit{vertex} appears in the graph. 

Each graph vertex represents an object. In the following, $o_i$ denotes the ID of the object represented
by a vertex, whereas $\mathbf{o}_i\in\mathbb{R}^d$ denotes its
corresponding vector. As detailed later, graph
adjacency lists store only object IDs, while the vectors are stored
separately in a global vector array.
Let $\mathcal{L}=\bigcup_{o\in O}f_o$ be the label universe, and let $\mathrm{freq}(l)=|\{o\in O\mid l\in f_o\}|$ denote the frequency of label $l$. We sort the labels in descending order of frequency and assign them integer IDs $1,2,\ldots,|\mathcal{L}|$ accordingly. 
 We further introduce a special terminal label $\#$ with frequency zero and treat it as larger than every ordinary label ID.
 
During trie construction, the labels in each $f_o$ are traversed in increasing ID order, followed by the terminal label $\#$. 

This ordering is purely frequency-based and assumes no semantic hierarchy among labels.
Then, the label trie $T$ is defined as follows.
\begin{definition}[Label trie]\label{def:label-trie}
Given the label universe $\mathcal{L}$ of $O$, the label trie $T$ is a rooted tree whose root is unlabeled and carries no ordinary label on its path, whose other internal nodes represent ordinary labels in $\mathcal{L}$, and whose leaves represent the terminal label $\#$; ordinary-label IDs strictly increase along every root-to-leaf path.

\end{definition} 

\begin{figure}[!tbp]
\centering
\includegraphics[width=.80\linewidth]{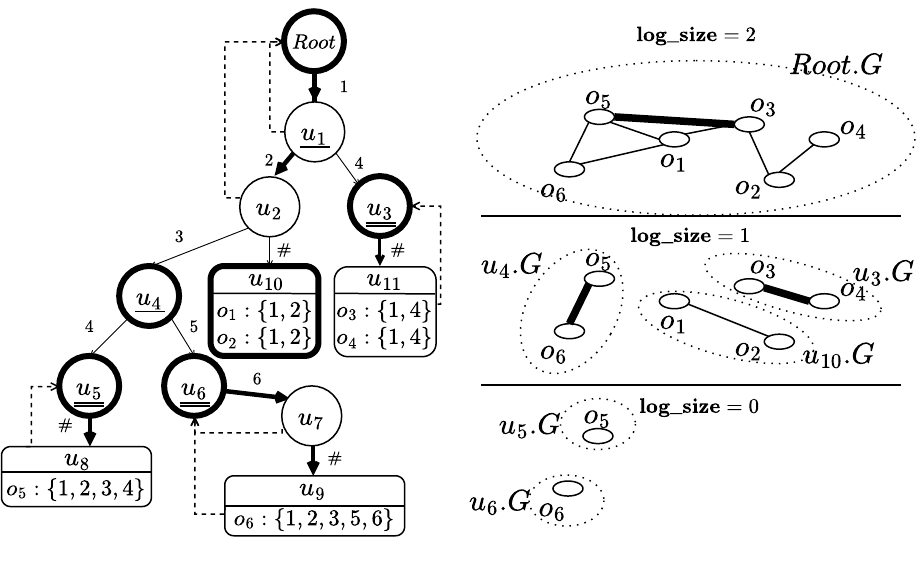}\par
\makebox[0.35\linewidth][c]{(a)}\makebox[0.35\linewidth][c]{(b)}\par
\caption{The {\oursolution} index for object set $O$ in Fig.~\ref{Fig:Vector}.}
\label{fig:graph_exm}
\end{figure}

For a node $u\in T$, let $\mathrm{prefix}(u)\subseteq\mathcal{L}$ denote the set of ordinary labels on the path from the root to $u$, and let $O_u$ denote the objects assigned to leaves in the subtree rooted at $u$. Thus, each node $u$ represents a \emph{trie region} $O_u$ consisting of objects that share $\mathrm{prefix}(u)$; this region serves as the basic unit for graph materialization described next. We use $u.p$ to denote the parent of $u$, and $u.ch=\{(l,c)\}$ to denote its children, where each pair $(l,c)$ indicates that child $c$ corresponds to label $l$.
\begin{example}
Consider again the object set $O$ in Fig.~\ref{Fig:Vector}. Its label universe contains six original labels, and we append the special terminal label $\#$ to every object's label set. Fig.~\ref{fig:graph_exm}a shows the label trie \( T \) in the {\oursolution} index. Node $u_2$, for instance, corresponds to label 2 and $prefix(u_2) = \{1, 2\}$, following the path root-$u_1$-$u_2$. Its parent is $u_2.p = u_1$, and its two children $u_2.ch$ are $u_4$ (with label 3) and $u_{10}$ (with terminal label $\#$). Accordingly, $O_{u_2}=\{\mathbf{o}_1, \mathbf{o}_2, \mathbf{o}_5, \mathbf{o}_6\}$, since the ordered label sequences of all four objects begin with labels
$1$ and $2$.

In particular, $\mathbf{o}_1$ and $\mathbf{o}_2$ terminate at $u_{10}$, while $\mathbf{o}_5$ and $\mathbf{o}_6$ are assigned to $u_8$ and $u_9$, respectively.
\end{example}

\input{table/notations}

\noindent
\textbf{Graph indices integrated in $\boldsymbol{T}$.}
As will be clarified later in Sec.~\ref{subsec:query}, given a query label filter $\Phi$, we decompose the filtered object set $O_{\Phi}$ into the union of subsets, each corresponding to a subtree rooted at a node in \(T\). Formally, we can find a node set $\mathcal{N}$ in $T$, where $\bigcup_{u \in \mathcal{N}} O_u = O_{\Phi}$.
We then aim to assign each node $u \in \mathcal{N}$ a graph index through which all objects in $O_u$ can be searched efficiently, while keeping the total index space bounded. Rather than materializing a separate graph index for every trie node, {\oursolution} materializes graph indices only at selected nodes and allows other nodes to share them. We bound the additional search space incurred by such sharing using the notion of an $\alpha$-approximate index.

\begin{definition}[$\alpha$-Approximate Index]\label{def:alpha-approximate-index}
    Given a graph index $G$ built on an object set $O_G$ and a target object set $O$, we say $G$ is an $\alpha$-approximate index for $O$ iff. $O \subseteq O_G$ and $|O_G| / |O| \le \alpha$.
\end{definition} 

Here, $\alpha$ bounds the size of the indexed object set relative to the target object set. Specifically, if $G$ is an $\alpha$-approximate index for $O_u$, then $G$ contains all objects in $O_u$ and at most $\alpha|O_u|$ vertices in total, including at most $(\alpha-1)|O_u|$ vertices outside $O_u$. To bound the number of such additional vertices under graph sharing, we group trie nodes by the sizes of their corresponding trie regions.

For each trie node $u$ and materialized graph $G$, let $\mathrm{log\_size}(u)=\lfloor\log_2|O_u|\rfloor$ and $\mathrm{log\_size}(G)=\lfloor\log_2|O_G|\rfloor$, respectively.
If $u$ is the root or $\mathrm{log\_size}(u)< \mathrm{log\_size}(u.p)$,
we materialize an HNSW index $u.G$ on $O_u$ and set $u.index=u$. Otherwise, we set $u.index=u.p.index$, so that $u$ shares the graph index assigned to its parent. Here, $u.index$ points to the trie node whose materialized graph is assigned to $u$; hence, $u.index.G$ denotes the graph index used to
search $O_u$. This design yields the following guarantee\footnote{All proofs can be found in Appendix~\ref{app:proofs}.}.

\begin{lemma}\label{lemma:index_quality}
     For any node $u$ in $T$, the initial graph index $u.index.G$ built by Alg.~\ref{algo:index_build} is a $2$-approximate index for $O_u$.
\end{lemma}

By Lem.\ \ref{lemma:index_quality} and Def.\ \ref{def:alpha-approximate-index}, the initially constructed index $u.index.G$ contains all objects in $O_u$ and at most $2|O_u|$ vertices in total, i.e., at most $|O_u|$ additional vertices outside $O_u$. After dynamic insertions, this local bound is relaxed to $4|O_u|$, as will be established in Thm.~\ref{thm:insert}. To record graph sharing in the trie, if a child $c$ of node $u$ satisfies $\mathrm{log\_size}(c)=\mathrm{log\_size}(u)$, then $c.index=u.index$, and we set $u.ch_{\mathrm{heavy}}=c$. Such a child is unique, as stated below.

\begin{lemma}\label{lemma:unique-heavy}
For any node \(u\) in the trie \(T\), there exists at most one child \(c\) of \(u\) such that \( \mathsf{log\_size}(c) = \mathsf{log\_size}(u)\).
\end{lemma}

\begin{figure}[!tbp]
\centering
\includegraphics[width=0.9\linewidth]{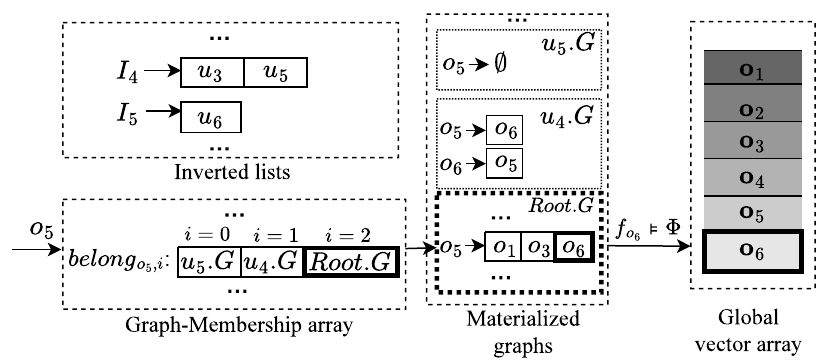}%
\vspace{2mm}
\caption{Memory Layout of {\oursolution}.}
\label{fig:graph_chasing}
\end{figure}

Although an object may appear as a vertex in multiple materialized graph indices, its vector is stored only once. As illustrated in Fig.~\ref{fig:graph_chasing}, {\oursolution} uses a separated memory layout, in which graph adjacency lists store only neighbor object IDs, while all vectors are stored in a global vector array. This additional level of indirection may incur extra memory accesses during graph traversal. However, our experiments on HNSW show that it has negligible impact (less than $1\%$) on query performance. See Appendix~\ref{app:memory-layout} for details.

\begin{example}
Continuing Fig.~\ref{fig:graph_exm}, we have $|O_{u_1}|=6$ and $|O_{u_3}|=2$, giving $\mathrm{log\_size}(u_1)=2$ and $\mathrm{log\_size}(u_3)=1$. Since $u_1=u_3.p$ and their logarithmic sizes differ, we materialize $u_3.G$ on $O_{u_3}$. In contrast, $u_{11}$ has the same logarithmic size as its parent $u_3$ and therefore shares $u_3.G$, i.e., $u_{11}.index=u_3.index=u_3$ and $u_3.ch_{\mathrm{heavy}}=u_{11}$. As illustrated in Fig.~\ref{fig:graph_chasing}, although $o_6$ appears in multiple
materialized graphs, its vector $\mathbf{o}_6$ is stored only once in
the global vector array and retrieved through its object ID. Note that in each example we only show the layer 0 of each graph index.

\end{example}

\noindent
\textbf{Auxiliary structures.}
To efficiently access the selectively materialized graphs during query processing, {\oursolution} maintains two auxiliary structures. 
For each object ID $o$, an array $belong_{o,i}$ points to the materialized graph $u.G$ with $\log\_size(u)=i$ that contains $o$ as a vertex, or is \textsc{null} if no such graph exists. 

We also maintain an inverted list $I_l$ for each label $l$: if the edge from a node $u$ to its child $c$ is labeled $l$, then $c$ is inserted into $I_l$.
During query processing, $belong$ locates the graphs containing a given object and $I_l$ locates the trie nodes reachable by label $l$ (Sec.~\ref{subsec:query}).

\begin{example}
Continuing Figs.~\ref{fig:graph_exm} and~\ref{fig:graph_chasing}, one can check that $I_4=\{u_3,u_5\}$ and $I_5=\{u_6\}$, while $belong$ records the materialized graphs containing $o_5$: $u_5.G$, $u_4.G$, and $Root.G$.

\end{example}

\noindent
{\bf{Base index construction.}}
Alg.~\ref{algo:index_build} inserts all objects into the label trie and updates the inverted lists (Lines 1--3 and 10--18), then recursively materializes or shares graph indices according to $\log\_size$, updating $belong$ and the heavy-child pointers (Lines 19--29).

\noindent
\textbf{Connectivity enhancement.} Finally, we apply \textsf{supplementEdges} as an additional connectivity enhancement (Lines~5--8 and 30--35). For an object $o$ with fewer than $2M$ bottom-layer neighbors in $u.index.G$, \textsc{supplementEdges} fills the missing slots using nearby objects sharing label $l$ (Lines~5--8 and~30--35). For $k'$ missing edges, it uses search parameter $\lfloor ef_c k'/(2M)\rfloor$ and is applied only to the first $l_{\mathrm{th}}=O(\log n)$ labels. Supplemented neighbors are stored as object-ID references rather than vertices of $u.index.G$; hence, they do not change $O_G$ or its approximation bound, and $belong$ locates their resident graphs when expanded.

Lem.~\ref{lemma:appear} and Thm.~\ref{thm:build}  give the time and space bounds of Alg.~\ref{algo:index_build}.

\begin{lemma}\label{lemma:appear}
For an object $o$, there are $\O(\min\{|f_o|, \log n\})$ distinct nodes $u$ in the label trie $T$ such that $u.index=u$ and $o \in O_u$.

\end{lemma}

\begin{theorem}\label{thm:build}
Alg.~\ref{algo:index_build} takes $\O(n\log n+\sum_{\mathbf{o}\in O}  (M\cdot \min\{|f_o|,\log n\}+|f_o| ))$ space and $\O(n\log n+\sum_{\mathbf{o}\in O} |f_o|+\sum_{\mathbf{o}\in O} \min\{|f_o|,\log n\}(c_{upd} + c_q))$ time to build the {\oursolution} index, where \( c_{upd} \) is the cost of inserting an object into the HNSW graph and \( c_q \) is the cost of a query in \oursolution.
    
\end{theorem}

\subsection{Query Processing}\label{subsec:query}

\input{algorithm/build_index}

Query processing consists of three steps.
First, given a filter $\Phi$, we identify a set $\mathcal{N}$ of trie nodes whose corresponding trie regions $O_u$ form an exact disjoint cover of $O_\Phi$. Second, since the graph indices associated with these trie nodes may be disconnected, we augment them with graph indices from selected common ancestors, obtaining a graph set $\mathcal{G}$ that improves cross-region connectivity.

Finally, we jointly traverse the graphs in $\mathcal{G}$ and apply $\Phi$ to filter the visited objects. We describe these three steps below.

\header
{\bf Disjoint cover computation.} Given a label filter $\Phi$, we first identify a set $\mathcal{N}$ of trie nodes whose corresponding trie regions are pairwise disjoint and collectively cover all qualifying objects in $O_\Phi$.

\begin{definition}[Disjoint Cover Set]
A node set
$\mathcal{N}$ ($\mathcal{N} \subseteq T$) is a \textit{disjoint cover set} of a label filter \( \Phi \) if  its corresponding trie regions form a partition of $O_\Phi$, i.e.,
\(
\bigcup_{u\in\mathcal{N}}O_u=O_\Phi
\text{ and }
O_{u_i}\cap O_{u_j}=\emptyset
\text{ for all }i\neq j \text{ and } u_i, u_j \in \mathcal{N}.
\)

\end{definition}

\input{algorithm/get_nodes}

The disjoint cover $\mathcal{N}$ identifies the exact trie regions for $\Phi$. For each $u\in\mathcal{N}$, {\oursolution} searches $O_u$ through $u.index.G$, which remains a
$2$-approximate index for $O_u$ initially by Lem.~\ref{lemma:index_quality}, and a $4$-approximate one after dynamic insertions by Thm.~\ref{thm:insert}.

Alg.~\ref{algo:index_getnodes} computes $\mathcal{N}$
(Lines~8--36). For containment, it scans $I_l$ for
$l=\max f_q$, the query label with the largest ID and hence the least frequent one under the build-time ordering, and retains $u$ iff
$f_q\subseteq\mathrm{prefix}(u)$ (Lines~9--13). For overlap, it
scans $I_l$ for each $l\in f_q$ and retains $u$ iff
$\mathrm{prefix}(u)\cap(f_q\setminus\{l\})=\emptyset$, which removes
nested duplicates and yields an exact disjoint cover
(Lines~15--20). 
For a complex predicate, the algorithm recursively computes the two
covers. Since trie regions are either nested or disjoint, conjunction
keeps the smaller region in each overlapping pair, whereas
disjunction keeps the larger one (Lines~24--32). The recursion ends
at containment or overlap predicates (Lines~35--36).

\input{algorithm/multigraph}

\header
{\bf Key-ancestor graph selection.} Once the disjoint cover set $\mathcal{N}$ is computed, it is still challenging to process the query.
For each node $u\in\mathcal{N}$, the $\alpha$-approximate guarantee ensures that $u.index.G$ contains all objects in $O_u$ while introducing only a constant-factor amount of additional search space. Thus, the selected graph indices primarily serve to maintain search quality. However, these graphs may be disconnected. 
A naive approach is to perform a separate ANN search on each graph and subsequently merge the results, which is inefficient as the cost grows with the number of nodes in $\mathcal{N}$. Another approach is to perform a single HNSW search over all graphs with a shared priority queue initialized with the entry point of each graph. However, in the absence of connecting edges between different graphs, the search may become trapped in one graph and overlook closer candidates in other regions.

The composite structure of {\oursolution} provides a natural solution to this problem. 
In particular, if nodes in $\mathcal{N}$ occur in the subtrees rooted
at two or more distinct children of a trie node $a$, the HNSW graph
associated with $a$ may contain edges spanning these child subtrees.

Such edges can connect otherwise isolated graph indices and guide the search across different coverage regions. We refer to selected common ancestors that provide such cross-region connectivity as \emph{key ancestor nodes}, defined below.

\begin{definition}[Key Ancestor Node]
Given a disjoint cover $\mathcal{N}$, a trie node $a$ is a
\emph{key ancestor} if nodes in $\mathcal{N}$ occur in the subtrees
rooted at two or more distinct children of $a$.
\end{definition}

By Lem.\ \ref{lemma:index_quality}, $a.index.G$ contains all objects in $O_a$. It therefore includes objects from the distinct child subtrees containing these cover nodes and may contain edges that bridge their search spaces. 
Note that this constant-factor bound applies separately to the graph
assigned to each selected trie region to ensure search quality, rather than to the union of all
graphs in $\mathcal{G}$. Therefore, it does not bound the total number
of vertices visited during multi-graph search.

Alg.~\ref{algo:index_getnodes} first computes the disjoint cover set $\mathcal{N}$ and initializes the graph index set $\mathcal{G}$ (Line~1). It then adds the graph assigned to each cover node (Lines~2--3) and the graphs assigned to key ancestors encountered during the upward traversals (Lines~4--6). The resulting graph index set $\mathcal{G}$, together with $\mathcal{N}$, is returned in Line~7 for the multi-graph search to be explained shortly. During the upward traversals, we record for each visited ancestor the child subtree from which it is first reached; the ancestor is marked as a key ancestor once it is reached from a second child subtree.

\begin{theorem}\label{thm:find_nodes}
\emergencystretch=3em\relax
Alg.~\ref{algo:index_getnodes} runs in
$\O\bigl(\sum_{u \in I_{l_{\max}}} |\mathsf{prefix}(u)|\bigr)$
time for a containment query,
$\O\bigl(\sum_{l \in f_q} \sum_{u \in I_l} |\mathsf{prefix}(u)|\bigr)$
for an overlap query, and
$\O\bigl(\sum_{l \in \mathcal{L}_\Phi} \sum_{u \in I_l}
|\mathsf{prefix}(u)| \allowbreak\cdot |\Phi|\bigr)$
for a complex query, where $l_{\max}$ is the query label with the
largest ID, $\mathcal{L}_\Phi$ the set of labels appearing in $\Phi$,
and $|\Phi|$ the number of containment and overlap predicates in $\Phi$.
\end{theorem}

\begin{example}
Revisiting the example in Fig.~\ref{fig:graph_exm}, suppose the predicate $\Phi$ is \textit{overlap}, with label set $f_{\mathbf{q}} = \{4,5\}$, meaning that the result objects must contain either label 4 or 5. Using inverted lists, we locate nodes $u_3$, $u_5$ (both corresponding to label 4), $u_6$ (label 5). Since these nodes have disjoint subtrees of trie $T$, they form a disjoint cover set $\mathcal{N}=\{u_3, u_5, u_6\}$. Then we identify $u_4$ and $u_1$ as key ancestor nodes for $\mathcal{N}=\{u_3, u_5, u_6\}$. We add $u_4.G$ and $\mathsf{Root}.G$ (since $u_1.index = Root$) to the index set $\mathcal{G}$. Observe that graph indices of $u_5$, $u_6$, and $u_3$ in $\mathcal{N}$ are isolated; meanwhile, the edge from $o_5$ to $o_3$ in $\mathsf{Root}.G$ and the edge from $o_5$ to $o_6$ in $u_4.G$ connect the three isolated indices, improving search effectiveness, explained next.
\end{example}

\input{algorithm/index_query}

\noindent
{\bf Multi-graph search.}
Given the disjoint cover $\mathcal{N}$ and graph index set $\mathcal{G}$, we jointly search selected graphs with Alg.~\ref{algo:multi_search}. 
We first use upper layers of $u.index.G$ for each $u\in\mathcal{N}$ to obtain entry points of layer 0 (Lines~1--3). The entry points initialize the candidate queue $C$, while their qualifying subset initializes the result queue $W$ (Lines~4--6). As in standard HNSW search, we repeatedly expand the candidate $v$ closest to the query. Once $W$ contains $ef$ vertices, the search ends when $v$ is farther than the current farthest vertex $W_{\max}$ in $W$ (Lines~7--9).

For each expanded vertex $v$, we traverse its graphs in $\mathcal{G}$, identified by $belong$, in descending order of logarithmic size (Lines~11--23).

For each unvisited neighbor $w$, we first check if $f_w\models\Phi$. Only qualifying neighbors trigger vector access and distance computation, and are inserted into $C$ and $W$ when appropriate (Lines~14--21). As $\mathcal{G}$ also contains key-ancestor graphs, the traversal can follow bridge edges across different trie regions. To limit excessive expansion, we stop expanding $v$ once $M_{\mathrm{cnt}}\geq 2M$ (Lines~22--23), where $M_{\mathrm{cnt}}$ counts examined neighbors satisfying $\Phi$.

\begin{example}
In Fig.~\ref{fig:graph_chasing}, suppose that object ID $o_5$ is being expanded. The $belong$ array identifies the selected graphs containing $o_5$, including $Root.G$. When the neighbor ID $o_6$ is encountered in $Root.G$, we first test whether $f_{o_6}\models\Phi$. If so, $o_6$ is used to retrieve its vector $\mathbf{o}_6$ from the global vector array for distance computation.

\end{example}

The complete query procedure (Alg. \ref{algo:index_query}) first invokes Alg.~\ref{algo:index_getnodes} to obtain the disjoint cover set $\mathcal{N}$ and graph set $\mathcal{G}$. It then runs
Alg.~\ref{algo:multi_search} and returns the top-$k$ objects in the resulting queue.

\begin{figure}[!tbp]
\centering
\includegraphics[width=1\linewidth]{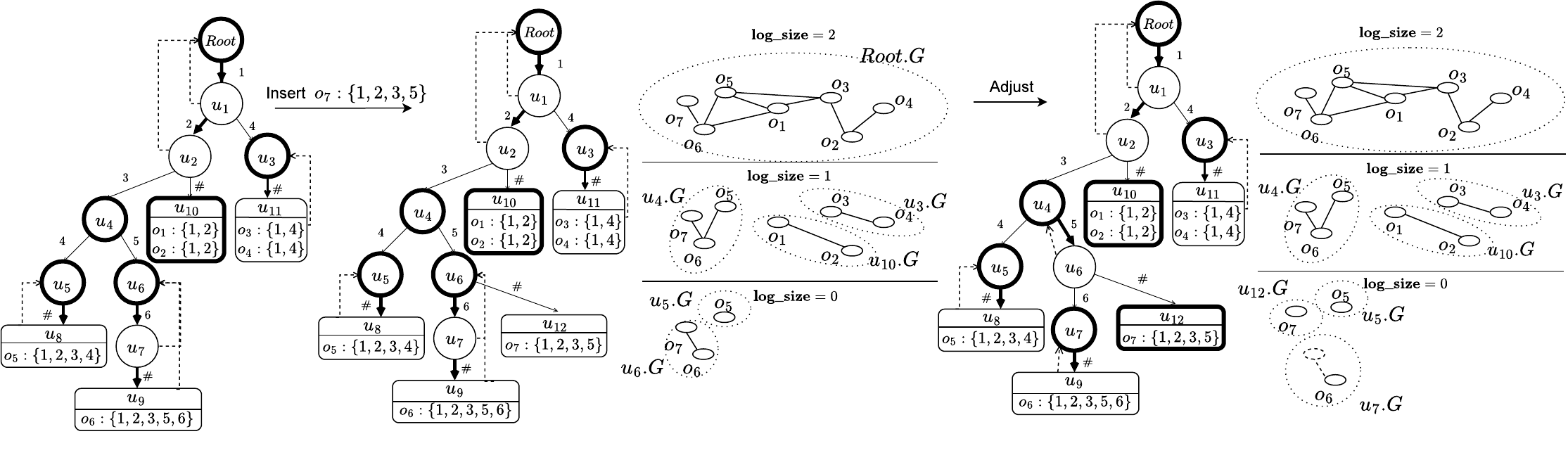}\par
\makebox[0.3333333333333333\linewidth][c]{(a)}\makebox[0.3333333333333333\linewidth][c]{(b)}\makebox[0.3333333333333333\linewidth][c]{(c)}\par
\vspace{2mm}
\caption{An example of the object insertion in the {\oursolution} index from Fig.~\ref{fig:graph_exm}.}
\label{fig:insertion_example}
\end{figure}

\noindent
{\bf Multi-threading extension.}
Although our current implementation is single-threaded, query processing extends naturally
to parallel execution: different queries are independent and can be processed
concurrently without modifying the index, and within a query, the entry-point searches over different cover nodes' graph indices can also run in parallel. Parallelizing the joint multi-graph traversal would require synchronizing shared search states
such as the candidate and result queues, which we leave as future work.

\input{algorithm/insertion}

\subsection{Index Update}\label{subsec:update}
Updating the {\oursolution} index requires keeping the graph index at each label trie node up-to-date.

For deletions, as in most ANNS indices such as HNSW described in Sec.~\ref{sec:pre:HNSW}, we apply soft deletion and simply mark the corresponding vertices as invalid, which are skipped during search. We therefore focus on insertions, which trigger non-trivial index maintenance.
The insertion of an object $\mathbf{o}$ includes three main steps: (i) insert $\mathbf{o}$ into the label trie $T$, 
(ii) update the affected graph indices along the insertion path in $T$, and (iii) adjust the graph index used by each node, and rebuild an index if it becomes insufficiently effective for its associated node.

The first two steps update the existing trie and graph structures locally.
For (i), all operations are lightweight, using only $\O(|f_o|)$ time, where $|f_o|$ is the number of labels in $o$. For step (ii), since an object belongs to at most $O(\min\{|f_o|,\log n\})$ graphs, the cost is bounded by $O(\min\{|f_o|,\log n\}\cdot(c_{\mathrm{upd}}+c_q))$, where $n$ is the number of objects in the dataset $O$, $c_{\mathrm{upd}}$ is the cost for inserting an object into an HNSW index, and $c_q$ is the cost of a query in {\oursolution}.

The main challenge lies in the third step, where the size change of an affected subtree may alter its $\mathrm{log\_size}$ and hence its graph-sharing relationship. Adjusting this relationship may involve expensive index rebuilding. To avoid rebuilding an index immediately after every such change, we adopt a {\em threshold-based maintenance} strategy: we allow the effective index for each node to be a $4$-approximate index (see Def.~\ref{def:alpha-approximate-index}). By tolerating a slightly weaker index guarantee, we postpone expensive rebuilds until the accumulated changes exceed a threshold. The cost of rebuilding is then amortized over later updates, which bounds the amortized time complexity per insertion to  \(\O\bigl(|f_o| M + \min\{|f_o|, \log n\} \cdot (c_{\text{upd}}+c_q)\bigr) \).

Alg.~\ref{algo:index_insert} appends $\#$ and inserts $o$ into the
trie (Lines~1--2), then traverses its root-to-leaf path
(Lines~4--23). We omit the straightforward auxiliary routines:
\textsc{CopyGraph}, \textsc{Rebuild}, and \textsc{DestroyGraph}
copy, rebuild, and release a graph, while \textsc{changeIndexNode}
redirects affected $index$ pointers; $belong$ is updated in Line~19.

New nodes follow the materialization rule (Line~5), and $o$ is
inserted into each graph it materializes (Lines~6--7). If
$\mathrm{log\_size}(u)$ increases, $u.G$ is copied to its heavy child
$c$ and the affected pointers are redirected. Because the copy may
contain objects outside $O_c$, it is rebuilt on $O_c$ only when
$\mathrm{log\_size}(c.G)>\mathrm{log\_size}(c)+1$
(Lines~8--15). If
$\mathrm{log\_size}(u)=\mathrm{log\_size}(u.p)$, $u$ instead shares
its parent's graph; we therefore destroy $u.G$ and update the
pointers and $belong$ entries (Lines~16--19). Finally, we supplement
edges (Lines~20--21).

The following theorem shows the efficiency of Alg.~\ref{algo:index_insert} and the validity of the index structure.

\begin{theorem} \label{thm:insert}
    Alg.~\ref{algo:index_insert} inserts each new object in amortized \(\O\bigl(|f_o| M + \min\{|f_o|, \log n\}  (c_{\text{upd}}+c_q)\bigr) \) time while ensuring that for each node \(u\), the associated index  remains a \(4\)-approximate index for \(O_u\).
    
\end{theorem}

\input{table/dataset}

\begin{example}
Consider again the index in Fig.~\ref{fig:graph_exm} and insert $o_7$ with
label set $\{1,2,3,5\}$. In Fig.~\ref{fig:insertion_example}(b), $o_7$ enters
the trie, the graph indices along its path ($Root.G$, $u_4.G$, $u_6.G$) are
updated, and the new node $u_{12}$ gets its graph assignment. This raises
$|O_{u_6}|$ to $2$, changing $\mathrm{log\_size}(u_6)$, so we adjust the
assignments as in Fig.~\ref{fig:insertion_example}(c): $u_6.G$ is copied to
its heavy child $u_7$, with $u_7.index$ and $u_9.index$ redirected to $u_7$;
meanwhile $u_6$ now matches its parent $u_4$ in $\mathrm{log\_size}$, so $u_6.G$
is removed, $u_6.index$ points to $u_4$, and $u_6$ becomes the heavy child of
$u_4$. The copied $u_7.G$ may temporarily hold vertices outside $O_{u_7}$;
these are retained until the logarithmic-size gap reaches the threshold.
\end{example}

\noindent
\textbf{New labels and frequency drift.}
A label unseen at build time does not restructure the trie: it is assigned the next available ID upon its first insertion, and since this ID is the largest so
far, it is traversed last in $f_\mathbf{o}$, so the insertion path follows existing nodes over the previously seen labels and branches off only near the leaf. Existing IDs are never reassigned, and the terminal symbol~$\#$ ranks after every ordinary label, so the ordering invariant of Def.~\ref{def:label-trie} is preserved. However, the frequency order drifts from the true frequencies as updates accumulate, which affects efficiency only, not correctness: the label trie and the disjoint cover computation of Sec.~\ref{subsec:query} are exact under any ordering, which serves solely to place frequent labels near the root, where larger trie regions admit more graph sharing.
Drift has two effects: \textsc{contSearch} selects $\max\{f_q\}$ as the least frequent query label (Alg.~\ref{algo:index_getnodes}, Line~9), so a frequent new label makes this choice suboptimal and widens the scan of $I_l$; and frequent labels placed deep in the trie make it bushier, increasing trie space. Both effects remain mild in our experiments (Figs.~\ref{fig:shift_yfcc}--\ref{fig:shift_audio}). If desired, rebuilding whenever $n$ doubles refreshes the ordering and amortizes to the same order as Thm.~\ref{thm:insert}.

%% file: table/notations.tex
\begin{table}[t]
    \centering
    {
    \caption{Frequently used notations}
    \vspace{-2mm}
    \scalebox{1}{
    \begin{CRfit}\begin{tabular}{p{35mm}p{95mm}}
        \toprule
        {\bf Symbol} & {\bf Meaning} \\ \hline
        \midrule
         $O, n$ &  Given set of objects and its cardinality \\  \hline
         $f_{\mathbf{o}}$ & Label set of object $\mathbf{o}$ \\  \hline
         $\delta(u,v)$ & Distance between $u$ and $v$ \\  \hline
         $\Phi$, $\mathbf{q}$ & Query label filter and the query vector\\  \hline
        $O_{\Phi}$ & Set of objects whose label sets satisfy $\Phi$\\ \hline
         $M$ & Maximum out-degree of HNSW (except vertices in the bottom layer whose out-degree is $2M$) \\  \hline
         $ef, ef_{c}$ & Size of the candidate set during graph search and graph construction, respectively\\ \hline
         $T$ & Label trie in \oursolution \\   \hline
         $O_u$ & Set of objects falling into the subtree of $u \in T$\\ \hline
         $N_G(v)$ & Out-neighbor set of vertex $v$ in graph index $G$ \\
         \hline
         $\mathsf{log\_size}(u)$ & $\lfloor \log_2|O_u|\rfloor$\\ \hline
         $u.p$, $u.ch$ &  Parent and children of node $u \in T$ \\ \hline
        $u.index$ &  Trie node whose materialized graph used by $u$  \\ \hline
        \(u.\mathsf{ch_{heavy}}\) & Child with the same $\mathsf{log\_size}$ as node $u \in T$ \\ \hline
        $belong_{o,i}$ & Materialized graph of log size \(i\) that contains \(o\)\\ 
         \bottomrule
    \end{tabular}\end{CRfit}
    }
    \label{tab:notation_summary}
    }
\end{table}

%% file: algorithm/build_index.tex
\begin{algorithm}[t]
	\caption{{\oursolution} index construction}
	\label{algo:index_build}
 \footnotesize
	
	\KwIn{Object set $O$ with label set $\mathcal{L}$, HNSW parameters $M$ and $ef_{c}$}
    \KwOut{Root node of the {\oursolution} index} 
\SetAlgoNoEnd

    Create an empty trie $T$ with $Root$; initialize $I$ and $belong$ as empty

    \For{$\mathbf{o} \in O$}{
        $f_{\mathbf{o}} \leftarrow f_{\mathbf{o}} \cup\{\#\}$,     
        \textsc{TrieInsert}$(Root,\mathbf{o})$
    }
    
    \textsc{buildGraph}$(Root, O, M,ef_{c})$

    \For{$l \leftarrow 1$ to $l_{th}$}{
        \For{$u \in I_l$}{
            \For{$o \in O_u$}{
                \textsc{supplementEdges}$(T,I, l,o, u, M, ef_{c})$
            }
        }
    }

    \Return{Root}

    \SetKwFunction{TrieInsert}{TrieInsert}
	\SetKwProg{Fn}{procedure}{:}{}
	\Fn{\text{TrieInsert}{($Root$,$\mathbf{o}$)}} {
       Initialize node $u \leftarrow Root$
       
       \For{$idx \leftarrow 1$ to  $|f_{\mathbf{o}}|$}{
            \If{$\nexists (l,c) \in u.ch,l=f_{\mathbf{o}}[idx]$}{
                Create new child $c$, and set its parent $c.p\leftarrow u$

                $I_{f_{\mathbf{o}}[idx]} \leftarrow I_{f_{\mathbf{o}}[idx]} \cup \{c\}$
                
                $u.ch \leftarrow u.ch \cup \{(f_{\mathbf{o}}[idx],c)\}$
            }
            $u \leftarrow$ the child of $u$ corresponding to $f_{\mathbf{o}}[idx]$
        }
        assign $\mathbf{o}$ to $u$
        
    }

    \SetKwFunction{buildGraph}{buildGraph}
	\SetKwProg{Fn}{procedure}{:}{}
	\Fn{\text{buildGraph}{($u$, $O_u$, $M$, $ef_{c}$)}} {
       \If{$u=Root$ or $\mathsf{log\_size}(u)<\mathsf{log\_size}(u.p)$}{
            \For{$\mathbf{o} \in O_u$}{
                \textsc{HNSWInsert}$(u.G,\mathbf{o},M, ef_{c})$

                $belong_{\mathbf{o},\mathsf{log\_size}(u)}\leftarrow u.G$
            }
            $u.index \leftarrow u$
       }
       \Else{
            $u.index \leftarrow u.p.index$
            
            $u.p.ch_{\mathsf{heavy}} \leftarrow u$
       }

        \For{$(l,c)\in u.ch$}{
            \textsc{buildGraph}$(c,O_c, M, ef_{c})$
        }
        
    }
    
	\SetKwProg{Fn}{procedure}{:}{}
	\Fn{\text{supplementEdges}{($T$, $I$, $l$, $o$, $u$, $M$, $ef_{c}$)}} {
       $k' \leftarrow 2M-|N_{u.index.G}(o)|$ // only the bottom layer
                
                \If{$k'> 0$ }{
                    $C \leftarrow \textsc{Alg.~\ref{algo:index_query}}(T,I,(\mathbf{o},cont(\{l\})), M, \lfloor\frac{ef_c*k'}{2M}\rfloor,\lfloor\frac{ef_c*k'}{2M}\rfloor) $
                    
                    $N'\leftarrow$ \textsc{Prune}$(o,k',C \setminus(  N_{u.index.G}(o)\cup \{o\} ))$
                    
                    $N_{u.index.G}(o) \leftarrow N_{u.index.G}(o) \cup N'$
                }
        
    }

\end{algorithm}

%% file: algorithm/get_nodes.tex
\begin{algorithm}[t]
	\caption{Disjoint cover set \& graph index set retrieval}
	\label{algo:index_getnodes}
	
\SetAlgoNoEnd
\DontPrintSemicolon 
  \footnotesize
	\KwIn{Label trie $T$, inverted list $I$, query label filter $\Phi$}
    $\mathcal{N} \leftarrow$ \textsc{complexSearch}$(T,I,\Phi)$;   $\mathcal{G}\leftarrow \emptyset$

    \For{$u \in \mathcal{N}$}{
            $\mathcal{G}\leftarrow \mathcal{G} \cup \{u.index.G\}$
            
            \For{$a \in$ ancestors of $u$}{
                \If{$a$ is a key ancestor of $\mathcal{N}$}{
                    $\mathcal{G}\leftarrow \mathcal{G}\cup\{a.index.G\}$\;
                }
    }
    }

    \Return{$\mathcal{N},\mathcal{G}$}

    \SetKwProg{Fn}{function}{:}{}
	\Fn{\text{contSearch}{($T$, $I$, $f_q$)}}    {
               $\mathcal{N}\leftarrow\emptyset$; $l \leftarrow max\{f_\mathbf{q}\}$ 
                
                \For{$u \in I_l$}{
                    \If{$f_q\subseteq prefix(u)$}{
                        $\mathcal{N} \leftarrow \mathcal{N} \cup \{u\}$
                    }
                }
            
            \Return{$\mathcal{N}$}

    }
	\Fn{\text{overSearch}{($T$, $I$, $f_q$)}}    {

            $\mathcal{N}\leftarrow\emptyset$\;
        
                \For{$l \in f_{q}$}{
                    \For{$u \in I_l$}{
                        \If{$prefix(u)\cap(f_q\setminus\{l\})=\emptyset$}{
                            $\mathcal{N} \leftarrow \mathcal{N} \cup \{u\}$
                        }
                    }
                }
                \Return{$\mathcal{N}$}
    }
	\Fn{\text{complexSearch}{($T$, $I$, $\Phi$)}}    {
        $\mathcal{N}\leftarrow\emptyset$\;
    \If{$\Phi$ is $(\Phi_0 \land \Phi_1)$ or $(\Phi_0 \lor \Phi_1)$ }{
            $\mathcal{N}_0 \leftarrow$ \textsc{complexSearch}$(T,I,\Phi_0)$
            
            $\mathcal{N}_1 \leftarrow$ \textsc{complexSearch}$(T,I,\Phi_1)$

            \For{$i\in\{0,1\}$}{

            \For{$u\in \mathcal{N}_i$}{
                $A_u \leftarrow$ ancestors of $u$
                
                \If{$\Phi=\Phi_0\wedge\Phi_1$ and $(\{u\}\cup A_u)\cap\mathcal{N}_{1-i}\neq\emptyset$}{
                    $\mathcal{N} \leftarrow \mathcal{N} \cup \{u\}$
                }

                \If{$\Phi=\Phi_0 \lor \Phi_1$ and $A_u \cap \mathcal{N}_{1-i} = \emptyset $}{
                    $\mathcal{N} \leftarrow \mathcal{N} \cup \{u\}$
                }
            }
            
            }

            \Return{$\mathcal{N}$}

            }
             \Else{ 
                \lIf{$\Phi$ is $cont(f_q)$}{\Return{\textsc{contSearch}$(T,I,f_q)$}}
                \lIf{$\Phi$ is $over(f_q)$}{\Return{\textsc{overSearch}$(T,I,f_q)$}}
                }

}

\end{algorithm}

%% file: algorithm/multigraph.tex
\begin{algorithm}[t]
	\caption{Multi-Graph Search}
	\label{algo:multi_search}
	
\SetAlgoNoEnd
\DontPrintSemicolon 
  \footnotesize
	\KwIn{Label trie $T$, disjoint cover set $\mathcal{N}$, graph index set $\mathcal{G}$, query ($\mathbf{q}$, $\Phi$), HNSW parameters $M$ and $ef$}

    $entries \leftarrow \emptyset$

    \For{$u \in \mathcal{N}$}{
        $entries\leftarrow entries \cup \{\text{\textsc{HNSW\_entry}}(u.index.G, \mathbf{q})\}$
        
    }
    
    Set $C \xleftarrow{}{} entries$ and mark all vertices in $entries$ as visited 
    
    $W \leftarrow \text{up to the } ef \text{ closest qualifying vertices in } entries$ to $\mathrm{q}$
    
    Denote $W_{max}$ as the current farthest vertex in $W$ to $\mathrm{q}$. If $W=\emptyset$, define $\delta (W_{max}, \mathbf{q})=+\infty$
    
    \While{$|C| > 0$}{
    
        $v \xleftarrow{}{}$ extract nearest vertex to $\mathbf{q}$ in $C$; 

        \lIf{$|W|=ef$ and $\delta(\mathbf{v},\mathbf{q}) > \delta(W_{max},\mathbf{q})$}{break}

        $M_{\text{cnt}} \leftarrow 0$

        \For{$i \leftarrow \mathsf{log\_size}(Root)$ downto $0$}{
            \If{$belong_{v,i} \neq NULL$ and $belong_{v,i} \in \mathcal{G}$}{
                $G\leftarrow belong_{v,i}[0]$ // only use the bottom layer
                
                 \For{$w \in N_G(v)$}{
                    \If{$w$ is not visited}{
                        Mark $w$  as visited
                    
                    \If{$f_w \models \Phi$}{

                        $M_{\text{cnt}} \leftarrow M_{\text{cnt}} +1$
                        
                        \If {$|W| < ef$ or $\delta(\mathbf{w},\mathbf{q}) < \delta(W_{max}, \mathbf{q})$}{
                            $C \xleftarrow{}{} C \cup \{w\}$, $W \xleftarrow{}{} W \cup \{w\}$
                        }
                        \lIf {$|W| > ef$}{
                            $W\leftarrow W\setminus \{W_{max}\}$
                        }

                    }
                    \lIf {$M_{\text{cnt}} \geq 2M$}{
                        break
                    }
                }
                }

                \lIf {$M_{\text{cnt}} \geq 2M$}{
                        break
                    }
                
            }
        }

    }

    \Return{W}

\end{algorithm}

%% file: algorithm/index_query.tex
\begin{algorithm}[t]
	\caption{{\oursolution}-Query}
	\label{algo:index_query}
	
\SetAlgoNoEnd
  \footnotesize
	\KwIn{Label trie $T$, inverted index $I$, query ($\mathbf{q}$, $\Phi$), HNSW  parameters  $M$, search parameters $ef$ and $k$}

    $\mathcal{N},\mathcal{G} \leftarrow \textsc{Algorithm~\ref{algo:index_getnodes}}(T,I,\Phi)$\;

    $results \leftarrow \textsc{Algorithm~\ref{algo:multi_search}} (T,\mathcal{N},\mathcal{G},(\mathbf{q},\Phi),M,ef)$\;
    
    \Return{the top-$k$ objects in $results$}

\end{algorithm} 

%% file: algorithm/insertion.tex
\begin{algorithm}[t]
	\caption{{\oursolution}-Insert}
	\label{algo:index_insert}
	
\SetAlgoNoEnd
  \footnotesize
	\KwIn{object $\mathbf{o}$, label set $f_{\mathbf{o}}$, HNSW construction parameters $M$, $ef_{c}$}

    $f_{\mathbf{o}} \leftarrow f_{\mathbf{o}} \cup \{\#\}$

    \textsc{TrieInsert}$(Root,\mathbf{o})$
    
        $u \leftarrow Root$
       
       \For{$idx \leftarrow 1$ to  $|f_\mathbf{o}|+1$}{
            \lIf{ $u$ is newly created}{
                initialize graph assignment of $u$
                }
            
            \If{$u.index=u$}{
                \lIf{$\mathbf{o} \notin u.G$}{\textsc{HNSWInsert}$(u.G, \mathbf{o}, M, ef_{c})$}

                \If{$|O_{u}|>1$ and $\mathsf{log\_size}(u) \neq \lfloor \log_2{(|O_{u}|-1)} 
                \rfloor$}{

                    \If{$u.ch_{\mathsf{heavy}} \neq NULL$}{
                        $u.ch_{\mathsf{heavy}}.G \leftarrow$ \textsc{CopyGraph}$(u.G)$
                        
                        $c \leftarrow u.ch_{\mathsf{heavy}}$

                        \textsc{changeIndexNode}$(c)$

                        \If{$\mathsf{log\_size}(c.G)>\mathsf{log\_size}(c)+1$}{
                            \textsc{Rebuild}$(c.G,O_c)$
                        }

                        $u.ch_{\mathsf{heavy}} \leftarrow NULL$
                        
                    }
                    \If{$u \neq Root$ and $\mathsf{log\_size}(u)=\mathsf{log\_size}(u.p)$}{
                        \textsc{DestroyGraph}$(u.G)$

                        $u.p.ch_{\mathsf{heavy}} \leftarrow u$;
                        $u.index \leftarrow u.p.index$
                    }

                    update $belong$
                    
                }

            }
            \If{$idx>1$ and $f_{\mathbf{o}}[idx-1] \leq l_{th}$}{
                \textsc{supplementEdges}$(T,I,f_{\mathbf{o}}[idx-1],o,u,M,ef_c)$
                }
            \If{$idx\leq|f_o|$}{
            $u \leftarrow$ the child node of $u$ corresponding to $f_{\mathbf{o}}[idx]$;
            }
        }

\end{algorithm}

%% file: table/dataset.tex
\begin{table*}[t]
    \centering
    {
    \caption{Summary of datasets in the experiments.}
    \vspace{-4mm}
    \scalebox{1}{
    \begin{CRfit}\begin{tabular}{ccccccccccccc}
        \toprule
   Dataset & Vector Type & Dim. & \# Vector & \# Query & Attribute Type & $|\mathcal{L}|$ & \multicolumn{3}{c}{$|f_o|$} & \multicolumn{2}{c}{Avg. Selectivity} & $|r|$\\   \cmidrule(lr){8-10} \cmidrule(lr){11-12} & & & & & & & Med. &P90 & Max & $\mathsf{containment}$ & $\mathsf{overlap}$ \\  %
    \midrule
    TripClick & text & 768 &  1,055,976 & 1000 & clinical area & 29 & 1& 2&8 & 0.07 &0.19 & 0.0055\\
    LAION1M &  image & 512 & 1,000,448 & 1000 & entities, locations, etc & 30 &2 &2 &3 &0.09 & 0.19& 0.0428\\
    YFCC & image & 192 &  1,000,000 & 1000 & classes & 181,931 &6 &16 & 1332& 0.038 & 0.19& 0.0217\\
    YTB-Audio & audio & 128 &  5,000,000 & 1000 & classes & 3862 &3 & 5&23 & 0.09&0.19& 0.0009 \\
    YTB-Video & video & 1024 &  1,000,000 & 200 & classes & 3862 &3 &5 & 21& 0.09&0.19&0.0206 \\
    Yelp1M & text chunk  & 384 &  1,000,000 & 1000 & business categories & 1,227 &5 &9 & 36 & 0.09&0.19& 0.4791 \\
    \bottomrule
    \end{tabular}\end{CRfit}
    }
        \label{tab:Dataset}
    }
\end{table*}

%% file: section/5-experiments.tex
\section{Experiments}
\label{sec:exp}

We compare our {\oursolution} with state-of-the-art label-filtered ANNS methods. {\oursolution} is implemented in C++, and all optimizations are enabled for competitor code. All indices reside in memory, and all experiments are conducted single-threaded on a Linux machine with an Intel Xeon Gold 5320 CPU @ 2.20GHz and 1TB memory.

\noindent
{\bf Datasets.}
The experiments involve six real datasets. Five of them, together with their
query settings, are taken from the filtered-ANNS
benchmark~\cite{filteredbenchmark}, which is widely tested in related research for filtered ANNS~\cite{aitAomar2025rwalks, UNG2024, DBLP:journals/pvldb/SehgalS25,
patel2024acorn}: (i) TripClick~\cite{rekabsaz2021TripClick}, a large-scale
health information retrieval dataset comprising click logs from the Trip
Database health web search engine; (ii) LAION1M~\cite{schuhmann2021laion}, a
publicly available dataset of CLIP-filtered~\cite{radford2021learningtransferablevisualmodels}
image-text pairs; (iii) YFCC~\cite{thomee2016YFCC100m}, a 1M-vector subset of
YFCC100M containing image/video embeddings and associated labels; and (iv)
YTB-Audio and (v) YTB-Video~\cite{abu2016youtube}, a benchmark of millions of
videos annotated with thousands of machine-generated topical entities.
We additionally build (vi) Yelp1M for RAG-style workloads: each
review in the Yelp Open Dataset~\cite{Yelp} is treated as one text chunk,
embedded into a 384-dimensional vector using all-MiniLM-L6-v2~\cite{MiniLLM},
and labeled with the multi-valued categories of its associated business.
Table~\ref{tab:Dataset} reports the statistics of these 6 datasets. The last
column reports $|r|$, the absolute Pearson correlation between the label-set
Jaccard similarity and the $L_2$ distance of 1M randomly sampled object pairs
per dataset (using distance instead of similarity only flips the sign). A small $|r|$ thus indicates
that objects with similar label sets are not systematically closer in the
vector space.

\subsection{Experimental Settings}

\noindent
{\bf Evaluation metrics.} Following prior work~\cite{aitAomar2025rwalks, UNG2024, GollapudiKSKBRL23}, we use queries per second (QPS) to measure throughput and recall (defined in Sec.~\ref{sec:pre:problem_definition}) to measure result quality, with $k=10$ in all experiments. We use $selectivity(O,\Phi)=|O_\Phi|/|O|$ to denote the proportion of objects satisfying the filter.

\noindent
{\bf Query generation.} We generate two types of queries for each dataset,
with \emph{containment} and \emph{overlap} label constraints. We first randomly generate candidate query label sets according to label frequencies and compute their actual filter selectivities. We bucket the candidates into selectivity ranges defined by powers of two and sample queries across these buckets to obtain workloads with diverse selectivities while keeping the average selectivity around 0.09 for containment and 0.19 for overlap. Due to their skewed label distributions, the resulting averages on YFCC and TripClick may deviate from these targets.

\noindent
{\bf Competitors.} We include the following methods in our experimental evaluation: 
(i) \textit{ACORN-1} and \textit{ACORN-$\gamma$}~\cite{patel2024acorn},
(ii) \textit{NaviX}~\cite{DBLP:journals/pvldb/SehgalS25},
(iii) \textit{Filtered-DiskANN}~\cite{GollapudiKSKBRL23},
(iv) \textit{RWalks}~\cite{aitAomar2025rwalks},
(v) \textit{UNG}~\cite{UNG2024},
(vi) \textit{UniFilter}~\cite{Unifilter},
(vii) \textit{Curator}~\cite{Curator},
(viii) \textit{Prefilter} (described in Sec.~\ref{sec:intro}), 
(ix) our \textit{\oursolution}.

To ensure fair comparisons, we tune the build-time parameters of every method, {\oursolution} included, on every dataset, selecting the configuration with the best QPS at recall 0.9 across both query types. The tuning details are given in Appendix~\ref{app:tune-parameters}. For each method, we then vary its query-time search parameter to obtain the recall-QPS tradeoff. Our code is available at~\cite{codetr}.

\begin{figure}[!tbp]
\centering
\includegraphics[width=.85\linewidth]{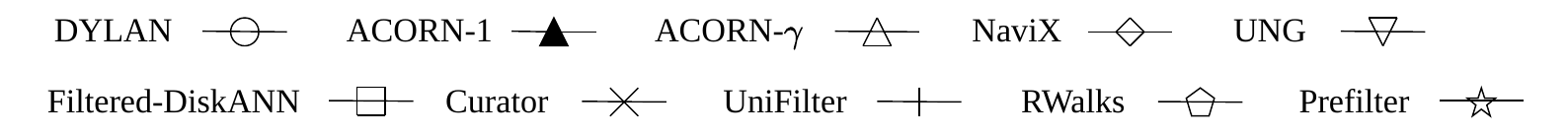}\par
\begin{minipage}[t]{0.50\linewidth}
\vspace{0pt}
\centering
\begin{minipage}[t]{.50\linewidth}
\vspace{0pt}
\centering
\includegraphics[width=\linewidth]{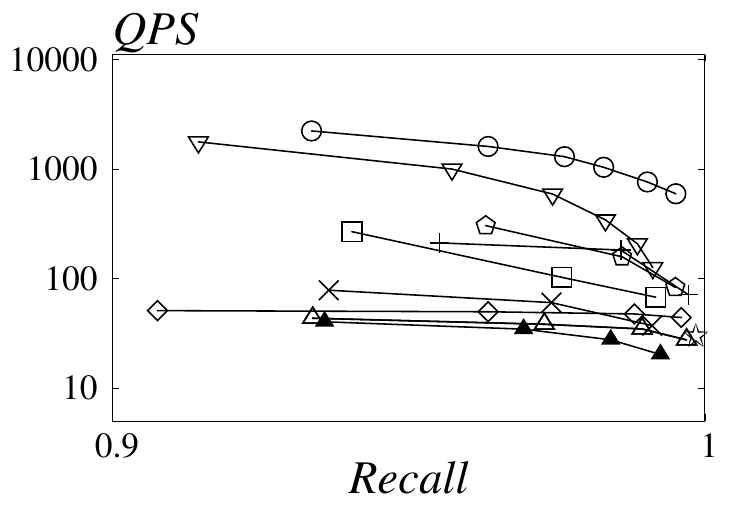}\par
{\small (a) Containment\par}
\end{minipage}\hfill\begin{minipage}[t]{.50\linewidth}
\vspace{0pt}
\centering
\includegraphics[width=\linewidth]{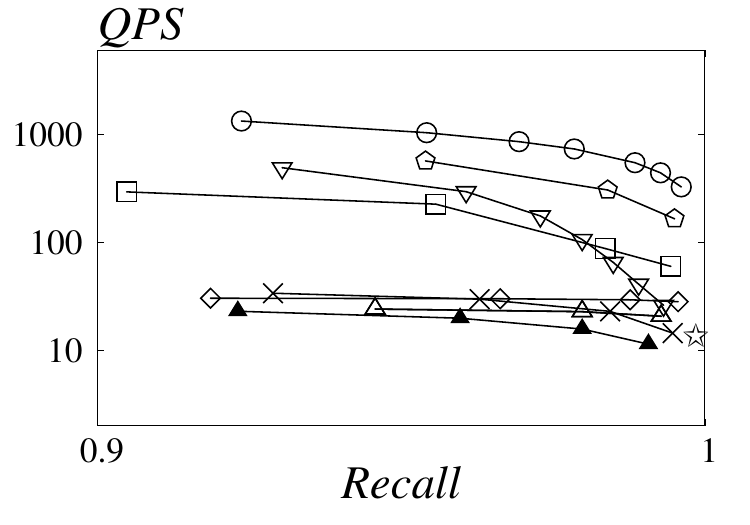}\par
{\small (b) Overlap\par}
\end{minipage}\par
\caption{QPS vs. recall on TripClick.}
\label{fig:TripClick}
\end{minipage}\hfill\begin{minipage}[t]{0.50\linewidth}
\vspace{0pt}
\centering
\begin{minipage}[t]{.50\linewidth}
\vspace{0pt}
\centering
\includegraphics[width=\linewidth]{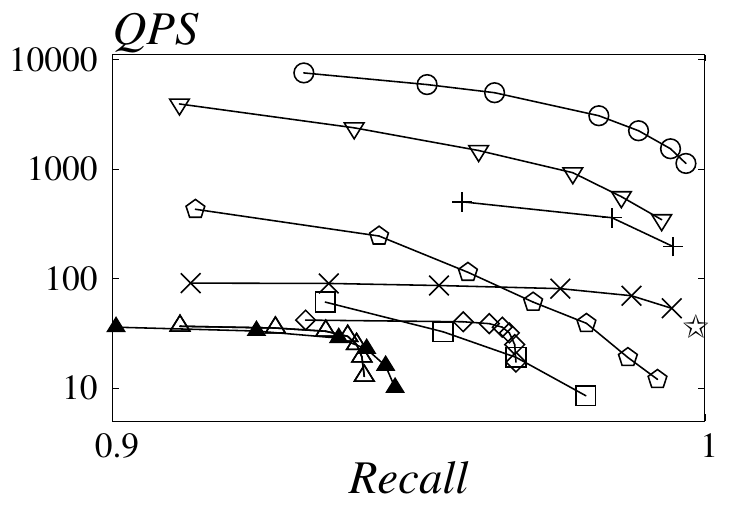}\par
{\small (a) Containment\par}
\end{minipage}\hfill\begin{minipage}[t]{.50\linewidth}
\vspace{0pt}
\centering
\includegraphics[width=\linewidth]{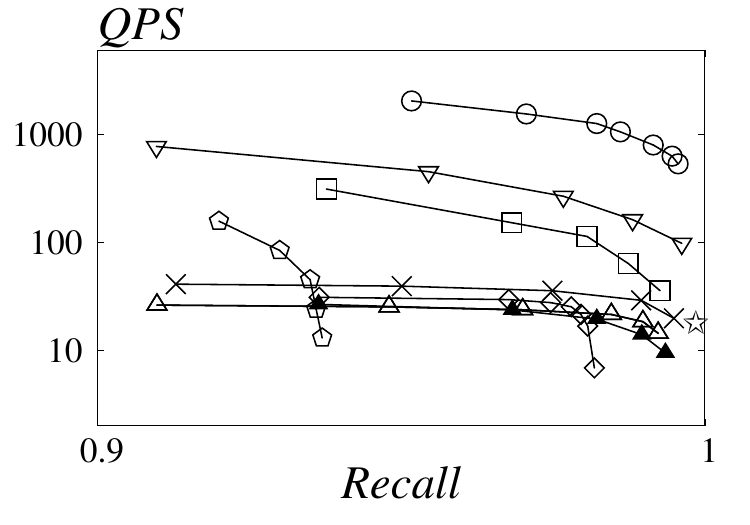}\par
{\small (b) Overlap\par}
\end{minipage}\par
\caption{QPS vs. recall on LAION1M.}
\label{fig:LAION1M}
\end{minipage}
\end{figure}

\begin{figure}[!tbp]
\centering
\begin{minipage}[t]{0.50\linewidth}
\vspace{0pt}
\centering
\begin{minipage}[t]{.50\linewidth}
\vspace{0pt}
\centering
\includegraphics[width=\linewidth]{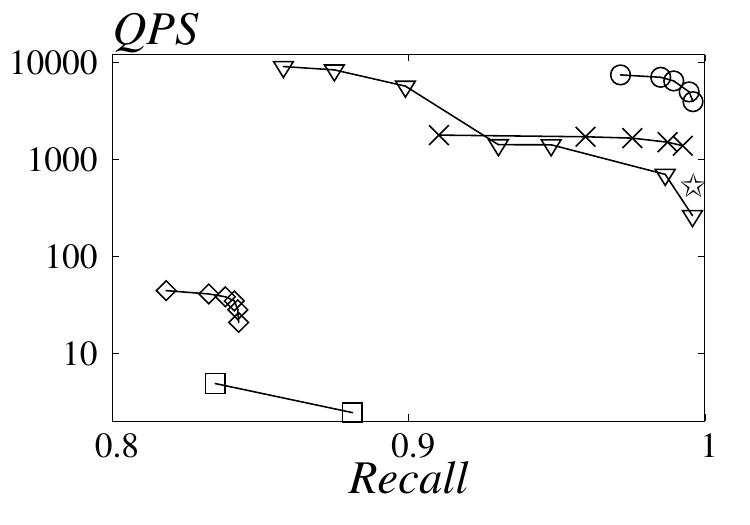}\par
{\small (a) Containment\par}
\end{minipage}\hfill\begin{minipage}[t]{.50\linewidth}
\vspace{0pt}
\centering
\includegraphics[width=\linewidth]{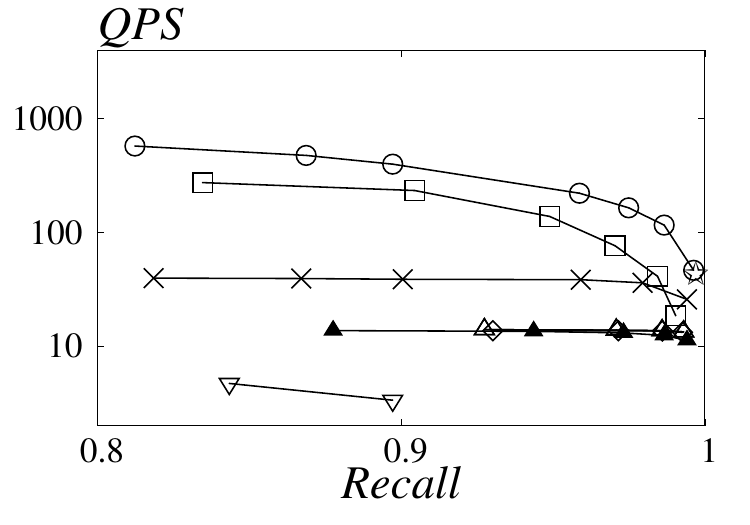}\par
{\small (b) Overlap\par}
\end{minipage}\par
\caption{QPS vs. recall on YFCC.}
\label{fig:YFCC}
\end{minipage}\hfill\begin{minipage}[t]{0.50\linewidth}
\vspace{0pt}
\centering
\begin{minipage}[t]{.50\linewidth}
\vspace{0pt}
\centering
\includegraphics[width=\linewidth]{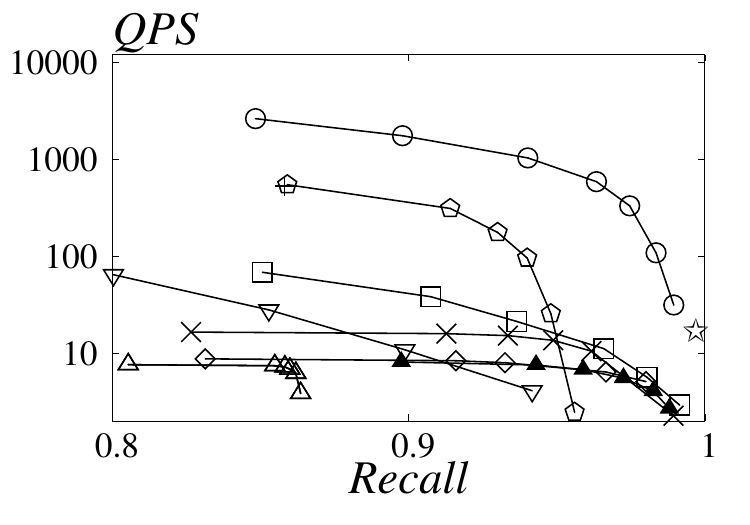}\par
{\small (a) Containment\par}
\end{minipage}\hfill\begin{minipage}[t]{.50\linewidth}
\vspace{0pt}
\centering
\includegraphics[width=\linewidth]{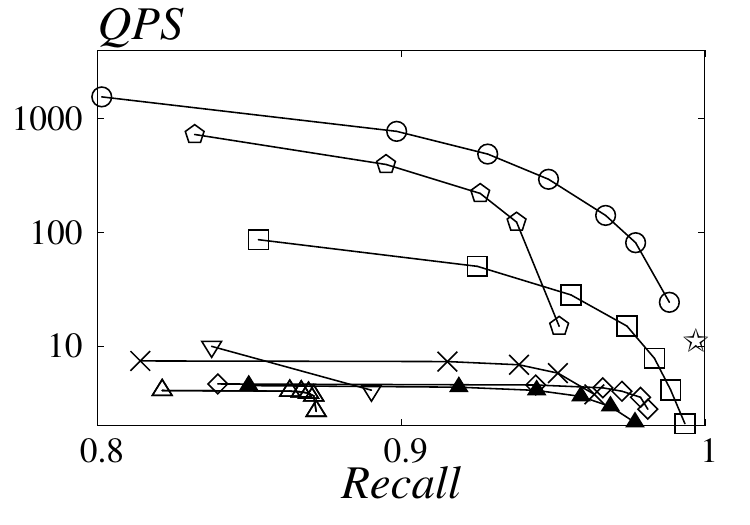}\par
{\small (b) Overlap\par}
\end{minipage}\par
\caption{QPS vs. recall on YTB-Audio.}
\label{fig:YTB-Audio}
\end{minipage}
\end{figure}

\begin{figure}[!tbp]
\centering
\begin{minipage}[t]{0.50\linewidth}
\vspace{0pt}
\centering
\begin{minipage}[t]{.50\linewidth}
\vspace{0pt}
\centering
\includegraphics[width=\linewidth]{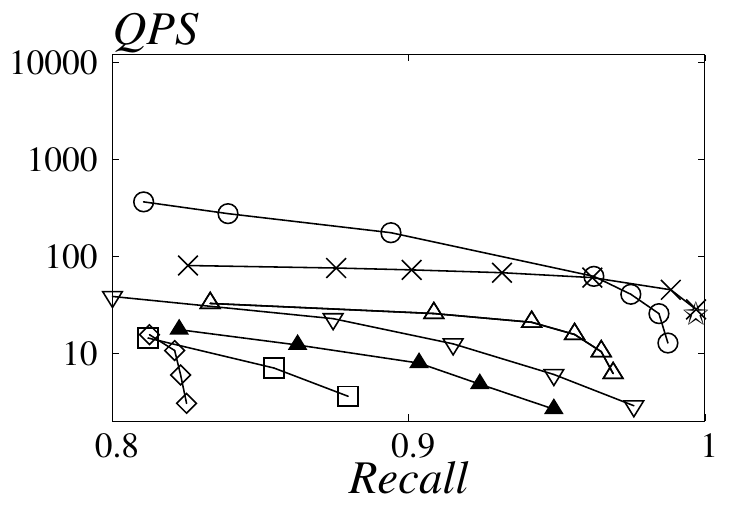}\par
{\small (a) Containment\par}
\end{minipage}\hfill\begin{minipage}[t]{.50\linewidth}
\vspace{0pt}
\centering
\includegraphics[width=\linewidth]{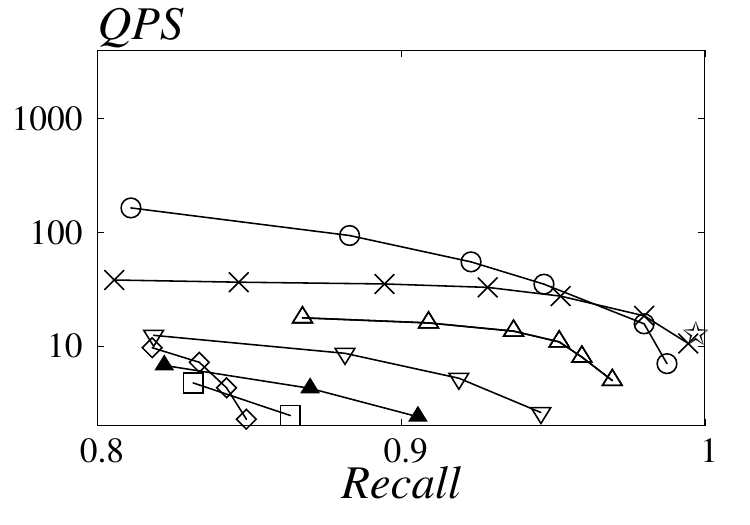}\par
{\small (b) Overlap\par}
\end{minipage}\par
\caption{QPS vs. recall on YTB-Video.}
\label{fig:YTB-Video}
\end{minipage}\hfill\begin{minipage}[t]{0.50\linewidth}
\vspace{0pt}
\centering
\begin{minipage}[t]{.50\linewidth}
\vspace{0pt}
\centering
\includegraphics[width=\linewidth]{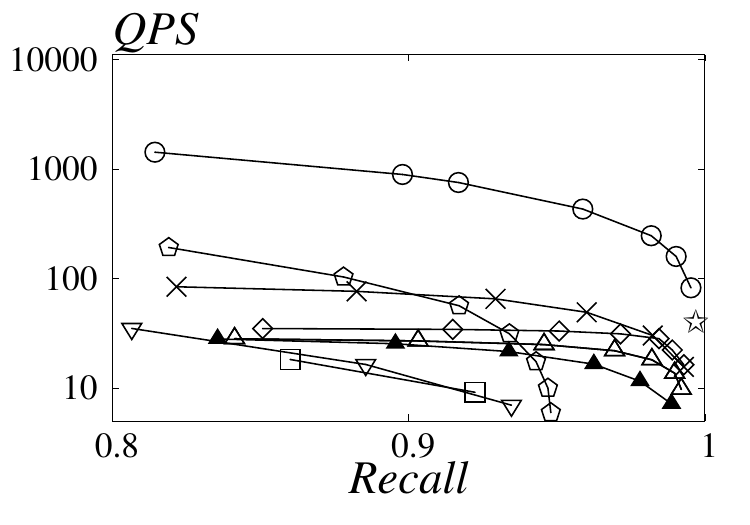}\par
{\small (a) Containment\par}
\end{minipage}\hfill\begin{minipage}[t]{.50\linewidth}
\vspace{0pt}
\centering
\includegraphics[width=\linewidth]{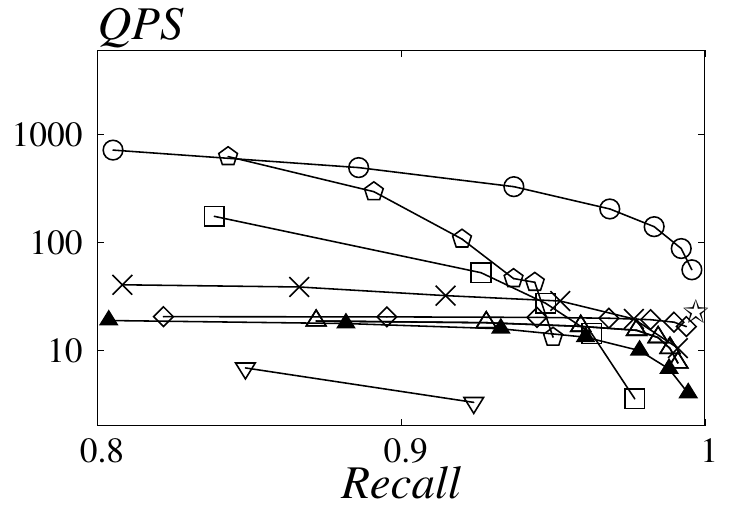}\par
{\small (b) Overlap\par}
\end{minipage}\par
\caption{QPS vs. recall on Yelp1M.}
\label{fig:Yelp1M}
\end{minipage}
\end{figure}
\subsection{Comparison Results}\label{subsec:comparison-results}

{\bf Query performance.} Figs.~\ref{fig:TripClick}-\ref{fig:Yelp1M} plot the QPS-recall curves for the 6 datasets in our experiments. We report methods that achieve at least $0.9$ recall for LAION1M and TripClick. For YFCC, YTB-Audio, YTB-Video, and Yelp1M, which are large datasets with many label categories and are therefore more challenging, most methods fail to achieve strong performance. Hence, we only report methods that achieve at least $0.8$ recall. RWalks is excluded from the experiments on YFCC because its memory consumption exceeds 1 TB. Prefilter is exact and appears as a single point in the figures. The proposed {\oursolution} consistently yields the best overall performance across all datasets. On datasets with relatively few label categories and simple label sets, such as TripClick and LAION1M, {\oursolution} remains highly competitive. At $0.99$ recall, it is $5\times$ faster than the best baseline on LAION1M and TripClick for containment queries. 

On datasets with more label categories and complex label sets, {\oursolution} shows a substantial advantage over its competitors. YFCC contains a large number of labels with an uneven distribution. Other methods perform well only on either containment or overlap queries, but not both. For example, UNG performs well on containment queries but fails to reach high recall on overlap queries. {\oursolution} handles both query types effectively.
On YTB-Audio, {\oursolution} is consistently faster than the fastest baseline, with speedups of $5.7\times$ for containment and $2.3\times$ for overlap at recall at least $0.8$; in the higher-recall region, the speedups are $43.5\times$ and $8.7\times$ at recall at least $0.95$, respectively.
On the YTB-Video dataset, Curator achieves high QPS at high recall, at a level comparable to {\oursolution}. This is because the YTB-Video dataset is inherently difficult to search; regardless of the index type, a large number of vectors must be accessed to achieve high recall. Curator is based on IVF, and compared to graph-based methods, it does not incur excessive random access when scanning many vectors, resulting in faster performance. Yet, on most other datasets, {\oursolution} outperforms Curator by one to two orders of magnitude; the exception is YFCC, where the speedup over Curator is about $5\times$.
On Yelp1M, a text/RAG-style workload, {\oursolution} also consistently outperforms the baselines for both containment and overlap queries, demonstrating its effectiveness on text chunks with real-world multi-valued category labels.

\begin{figure}[!tbp]
\centering
\includegraphics[width=.85\linewidth]{figures/legend/recall_qps_pole_legend_2.pdf}\par
\begin{minipage}[t]{0.50\linewidth}
\vspace{0pt}
\centering
\begin{minipage}[t]{.50\linewidth}
\vspace{0pt}
\centering
\includegraphics[width=\linewidth]{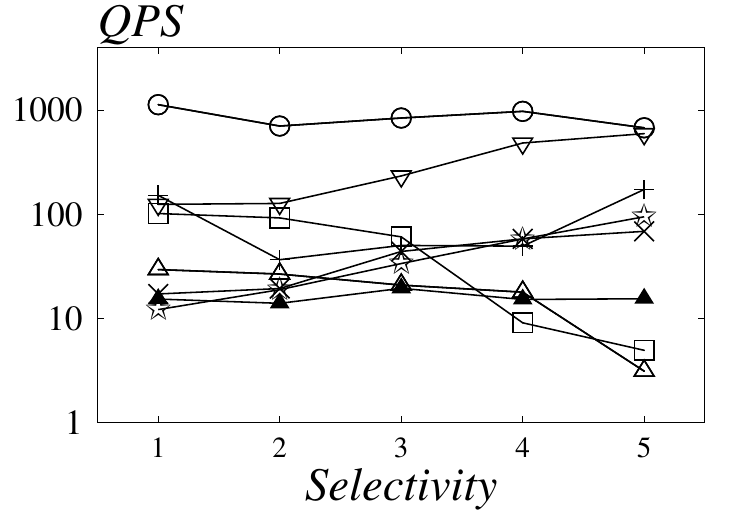}\par
{\small (a) Containment\par}
\end{minipage}\hfill\begin{minipage}[t]{.50\linewidth}
\vspace{0pt}
\centering
\includegraphics[width=\linewidth]{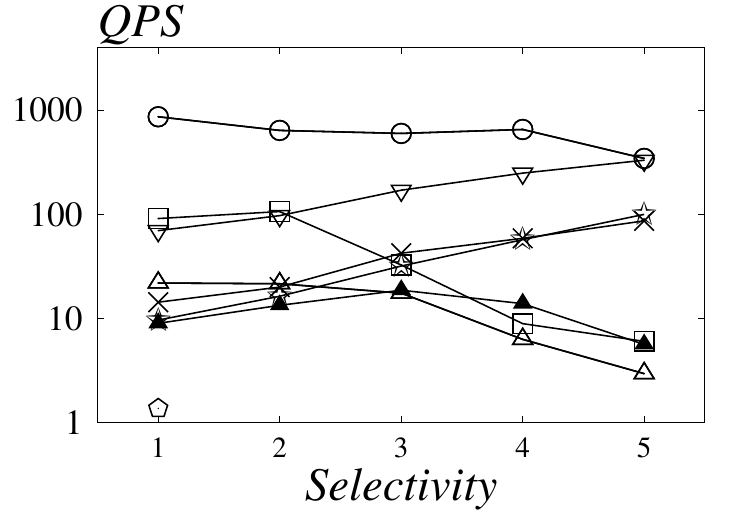}\par
{\small (b) Overlap\par}
\end{minipage}\par
\caption{Impact of selectivity on the LAION1M.}
\label{exp:LAION1M_sel}
\end{minipage}\hfill\begin{minipage}[t]{0.50\linewidth}
\vspace{0pt}
\centering
\begin{minipage}[t]{.50\linewidth}
\vspace{0pt}
\centering
\includegraphics[width=\linewidth]{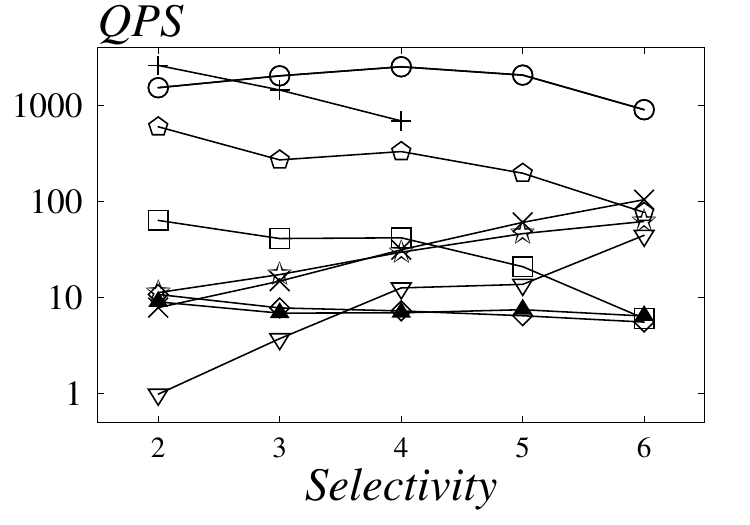}\par
{\small (a) Containment\par}
\end{minipage}\hfill\begin{minipage}[t]{.50\linewidth}
\vspace{0pt}
\centering
\includegraphics[width=\linewidth]{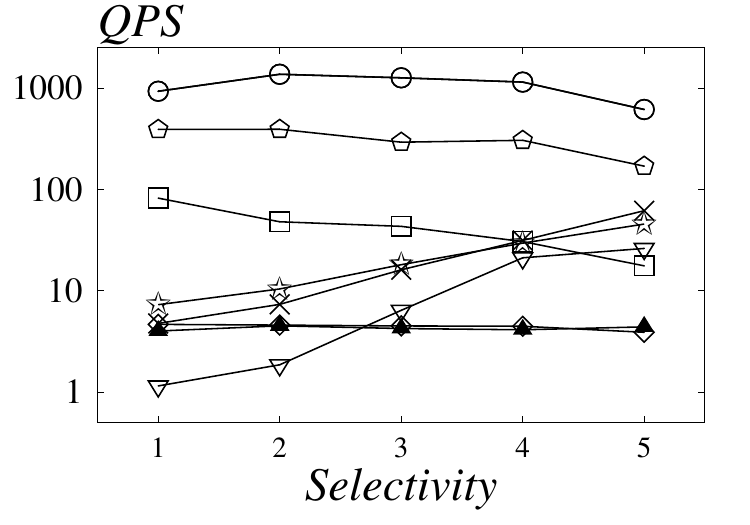}\par
{\small (b) Overlap\par}
\end{minipage}\par
\caption{Impact of selectivity on the YTB-Audio.}
\label{exp:YTB-Audio_sel}
\end{minipage}
\end{figure}

\begin{figure}[!tbp]
\centering
\begin{minipage}[t]{0.50\linewidth}
\vspace{0pt}
\centering
\begin{minipage}[t]{.50\linewidth}
\vspace{0pt}
\centering
\includegraphics[width=\linewidth]{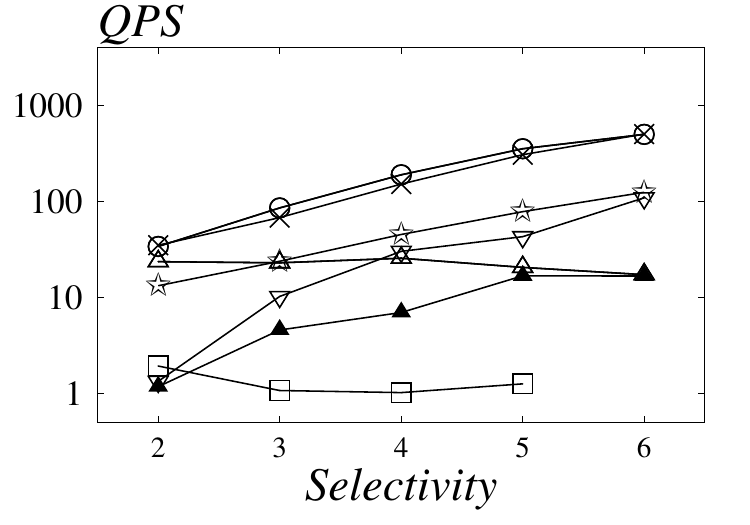}\par
{\small (a) Containment\par}
\end{minipage}\hfill\begin{minipage}[t]{.50\linewidth}
\vspace{0pt}
\centering
\includegraphics[width=\linewidth]{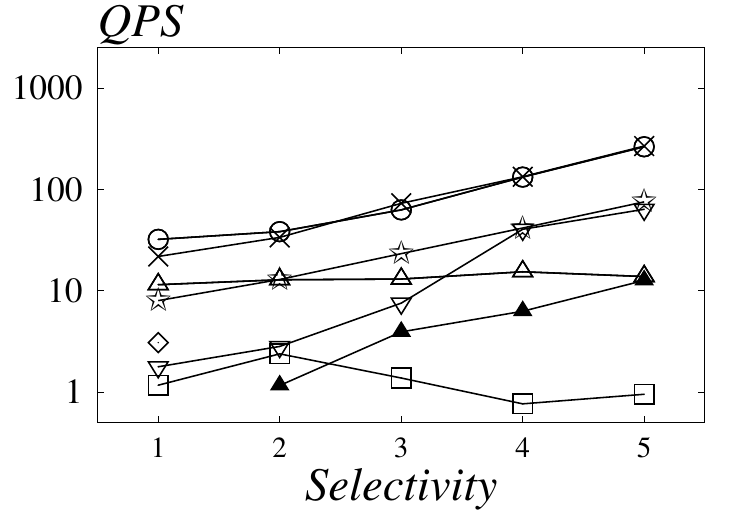}\par
{\small (b) Overlap\par}
\end{minipage}\par
\caption{Impact of selectivity on the YTB-Video.}
\label{exp:YTB-Video_sel}
\end{minipage}\hfill\begin{minipage}[t]{0.50\linewidth}
\vspace{0pt}
\centering
\begin{minipage}[t]{.50\linewidth}
\vspace{0pt}
\centering
\includegraphics[width=\linewidth]{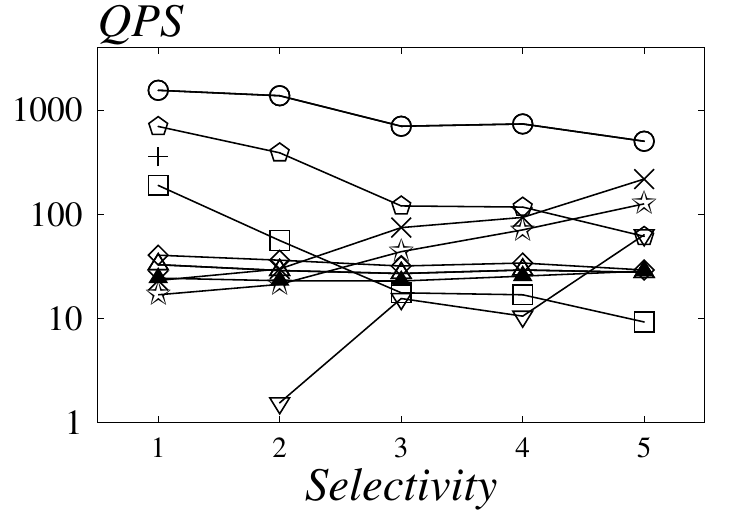}\par
{\small (a) Containment\par}
\end{minipage}\hfill\begin{minipage}[t]{.50\linewidth}
\vspace{0pt}
\centering
\includegraphics[width=\linewidth]{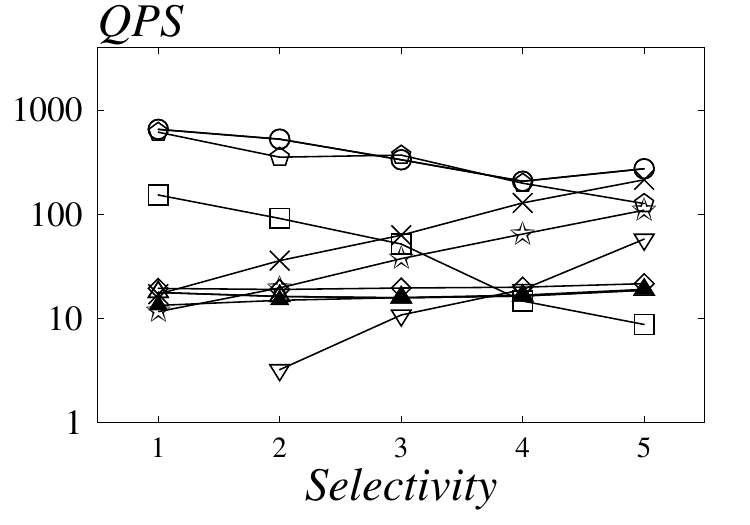}\par
{\small (b) Overlap\par}
\end{minipage}\par
\caption{Impact of selectivity on the Yelp1M.}
\label{exp:yelp_sel}
\end{minipage}
\end{figure}

\begin{figure}[!tbp]
\centering
\includegraphics[width=.85\linewidth]{figures/legend/recall_qps_pole_legend_2.pdf}\par
\begin{minipage}{.50\linewidth}\centering
\begin{minipage}[t]{.50\linewidth}
\vspace{0pt}
\centering
\includegraphics[width=\linewidth]{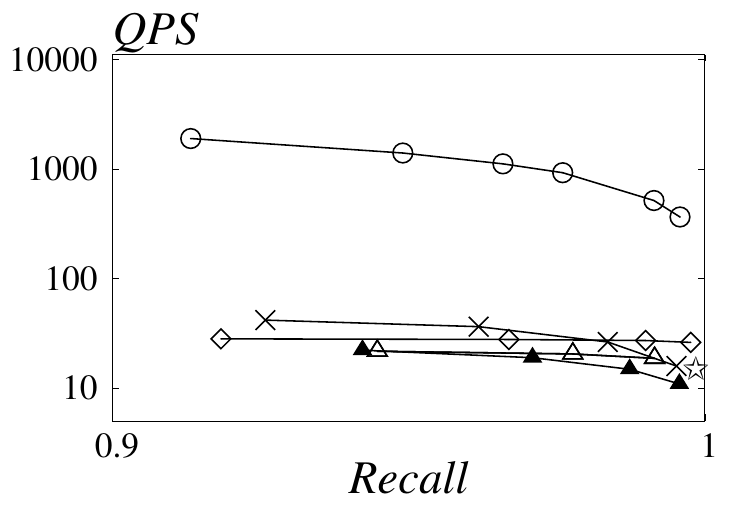}\par
{\small (a) TripClick\par}
\end{minipage}\hfill\begin{minipage}[t]{.50\linewidth}
\vspace{0pt}
\centering
\includegraphics[width=\linewidth]{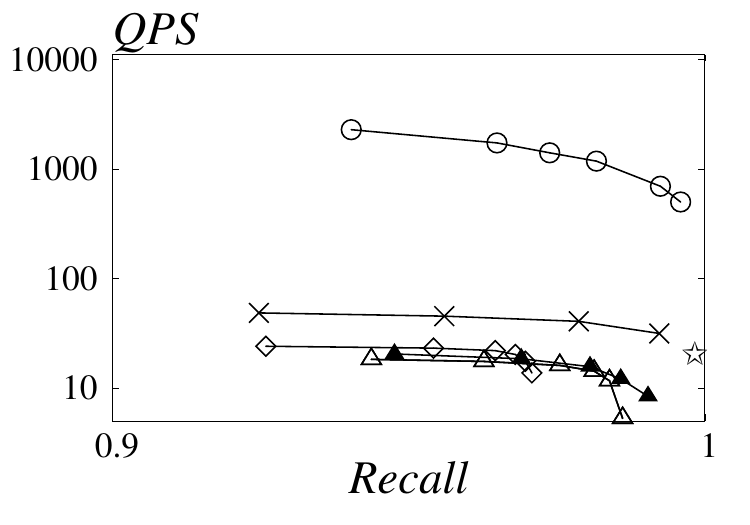}\par
{\small (b) LAION1M\par}
\end{minipage}\par
\end{minipage}\par
\caption{Complex predicate.}
\label{exp:complex_1}
\end{figure}

\begin{figure}[!tbp]
\centering
\includegraphics[width=1\linewidth]{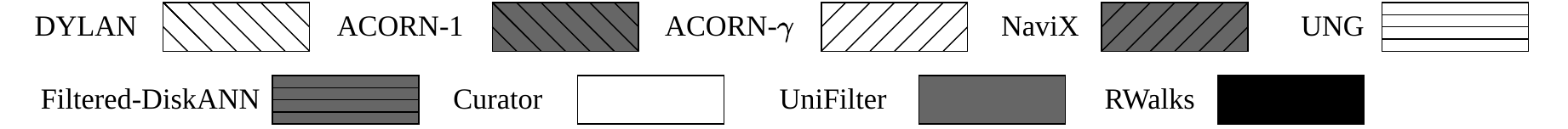}\par
\includegraphics[width=1\linewidth]{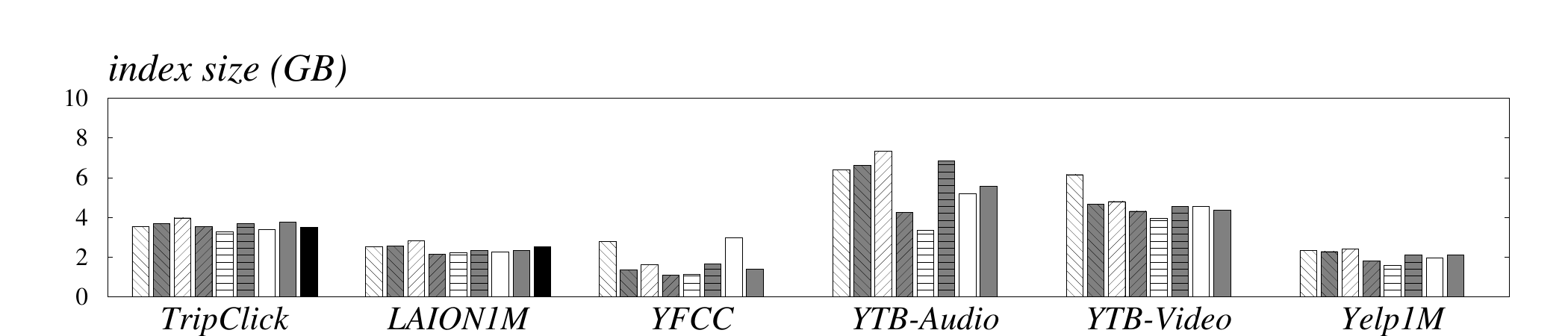}\par
\caption{Memory cost.}
\label{fig:exp_index_memory}
\includegraphics[width=1\linewidth]{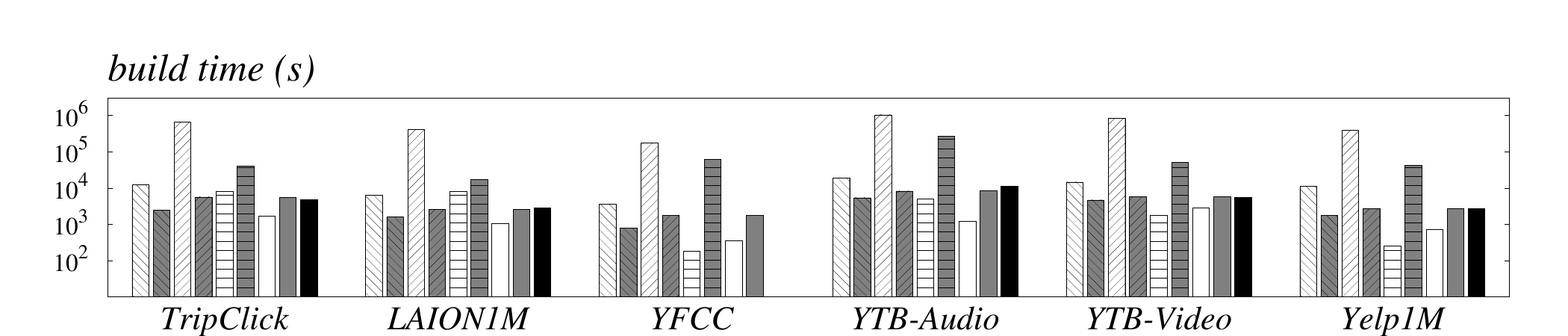}\par
\caption{Indexing time.}
\label{fig:exp_index_time}
\includegraphics[width=1\linewidth]{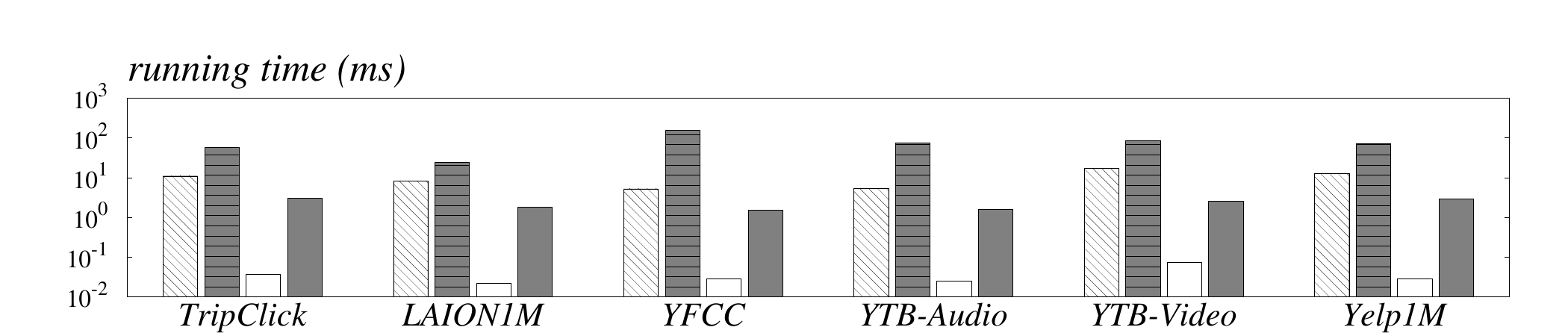}\par
\caption{Insertion time.}
\label{fig:exp_upd_time}
\end{figure}

\header
{\bf Impact of query selectivity. \label{expsub:sel}}
Next, we evaluate the performance of overlap and containment queries under varying selectivity levels across all datasets. Note that the label distribution is discrete, so we cannot generate a large number of queries with a specific selectivity. For datasets except YTB‑Audio and YTB‑Video, we generate queries whose selectivity falls into five intervals: \([2^{-2},2^{-1})\), \([2^{-3},2^{-2})\), \([2^{-4},2^{-3})\), \([2^{-5},2^{-4})\), and \([2^{-6},2^{-5})\). For YTB‑Audio and YTB‑Video, where label frequencies are generally lower, we create queries in the intervals \([2^{-3},2^{-2})\) to \([2^{-7},2^{-6})\) for containment queries. For each interval, we report the QPS achieved at 0.99 recall on TripClick and LAION1M. Most baselines struggle to reach high recall on challenging datasets YFCC, YTB-Audio, YTB-Video, and Yelp1M. Hence, we report QPS at 0.9 recall for these. Due to the label distributions in YFCC and TripClick, it is infeasible to generate containment queries spanning a continuous selectivity range. We therefore report only overlap queries on these two datasets, deferring the results to Appendix~\ref{app:exp-impact-selectivity} for space.

Figs.~\ref{exp:LAION1M_sel}-\ref{exp:yelp_sel} present the evaluation results. On the x-axis, each tick value $i$ represents the interval $[2^{-(i+1)},2^{-i})$. Observe that {\oursolution} maintains consistently high performance across all selectivity levels and datasets. On the LAION1M dataset, UNG and {\oursolution} achieve comparable performance when the selectivity is very low. This is because LAION1M has relatively few labels, and low-selectivity queries involve only a small number of highly similar label sets. UNG's static LNG can capture these relationships in advance by pre‑building cross edges, whereas {\oursolution} incurs overhead to dynamically construct edges across label sets. On the other hand, UNG performs poorly on datasets with more labels.  {\oursolution} shows a clear advantage on datasets with a larger number of labels. On  YTB-Audio, {\oursolution} is the only approach that maintains both high recall and high QPS across all selectivity levels and query types. The cost of Prefilter is mainly determined by $|O_\Phi|$, so its throughput improves as selectivity decreases, yet it remains an order of magnitude slower than {\oursolution} in most settings.

\begin{figure}[!tbp]
\centering
\includegraphics[width=.56\linewidth]{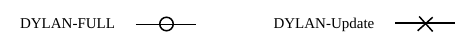}\par
\begin{minipage}[t]{0.50\linewidth}
\vspace{0pt}
\centering
\begin{minipage}[t]{.50\linewidth}
\vspace{0pt}
\centering
\includegraphics[width=\linewidth]{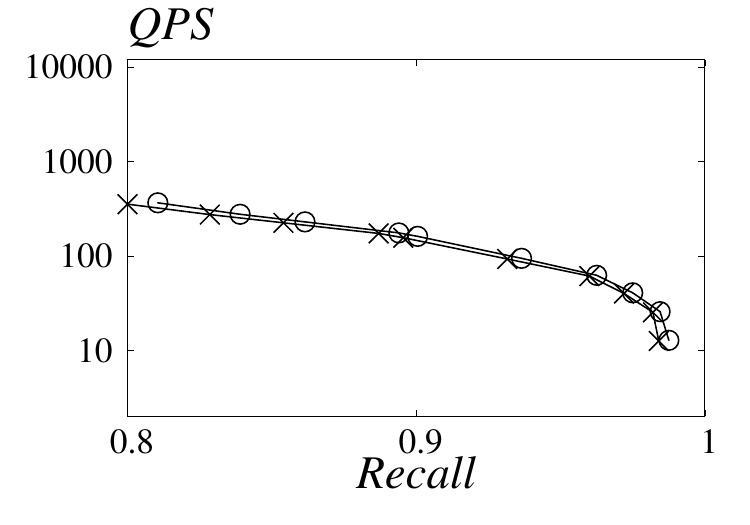}\par
{\small (a) Containment\par}
\end{minipage}\hfill\begin{minipage}[t]{.50\linewidth}
\vspace{0pt}
\centering
\includegraphics[width=\linewidth]{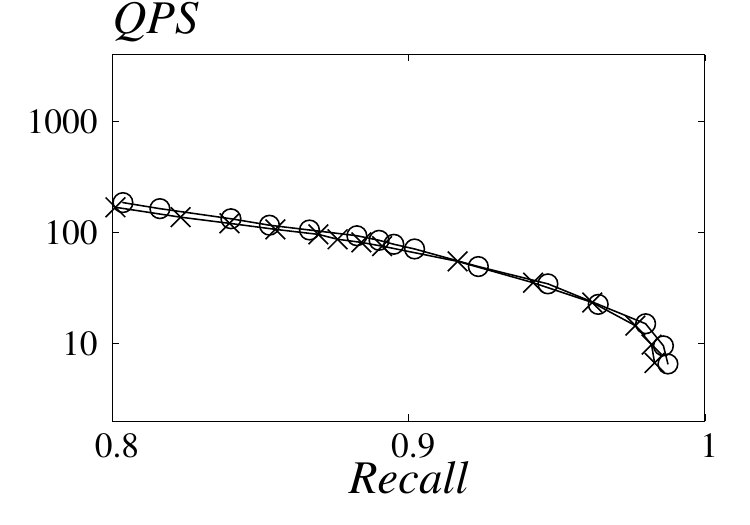}\par
{\small (b) Overlap\par}
\end{minipage}\par
\caption{Impact of insertion-only update on YTB-Video.}
\label{exp:YTB-video_insertion-only}
\end{minipage}\hfill\begin{minipage}[t]{0.50\linewidth}
\vspace{0pt}
\centering
\begin{minipage}[t]{.50\linewidth}
\vspace{0pt}
\centering
\includegraphics[width=\linewidth]{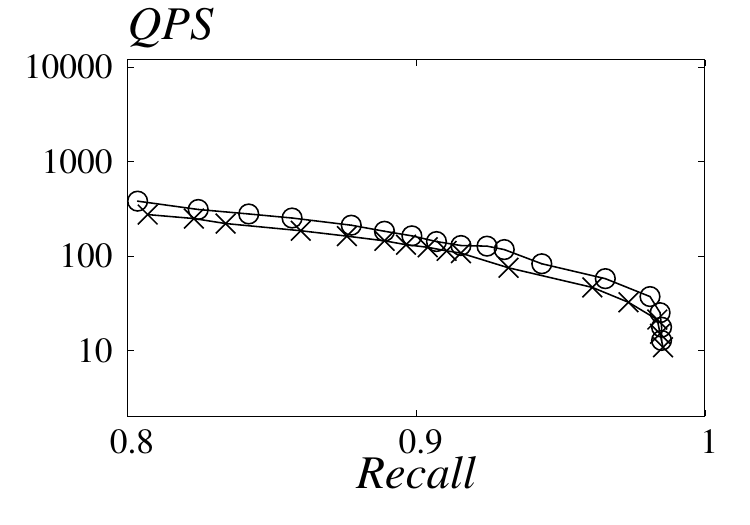}\par
{\small (a) Containment\par}
\end{minipage}\hfill\begin{minipage}[t]{.50\linewidth}
\vspace{0pt}
\centering
\includegraphics[width=\linewidth]{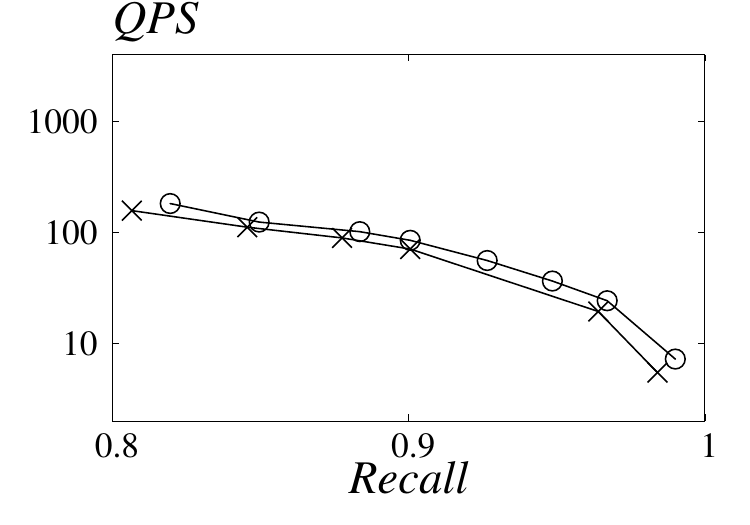}\par
{\small (b) Overlap\par}
\end{minipage}\par
\caption{Impact of mixed update on YTB-Video.}
\label{exp:YTB-video_mixed-upd}
\end{minipage}
\end{figure}

\begin{figure}[!tbp]
\centering
\includegraphics[width=1\linewidth]{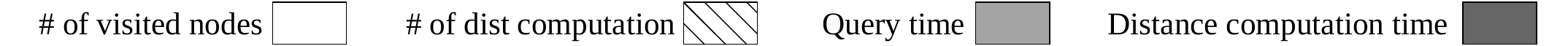}\par
\begin{minipage}{\linewidth}\centering
\begin{minipage}[t]{.50\linewidth}
\vspace{0pt}
\centering
\includegraphics[width=\linewidth]{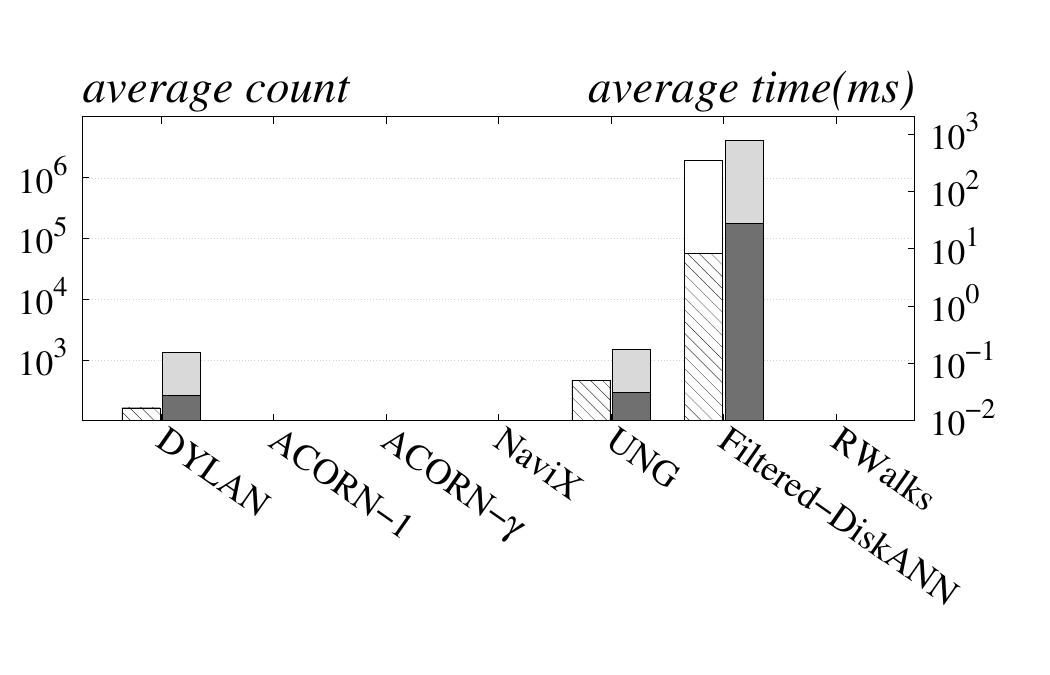}\par
{\small (a) Containment\par}
\end{minipage}\hfill\begin{minipage}[t]{.50\linewidth}
\vspace{0pt}
\centering
\includegraphics[width=\linewidth]{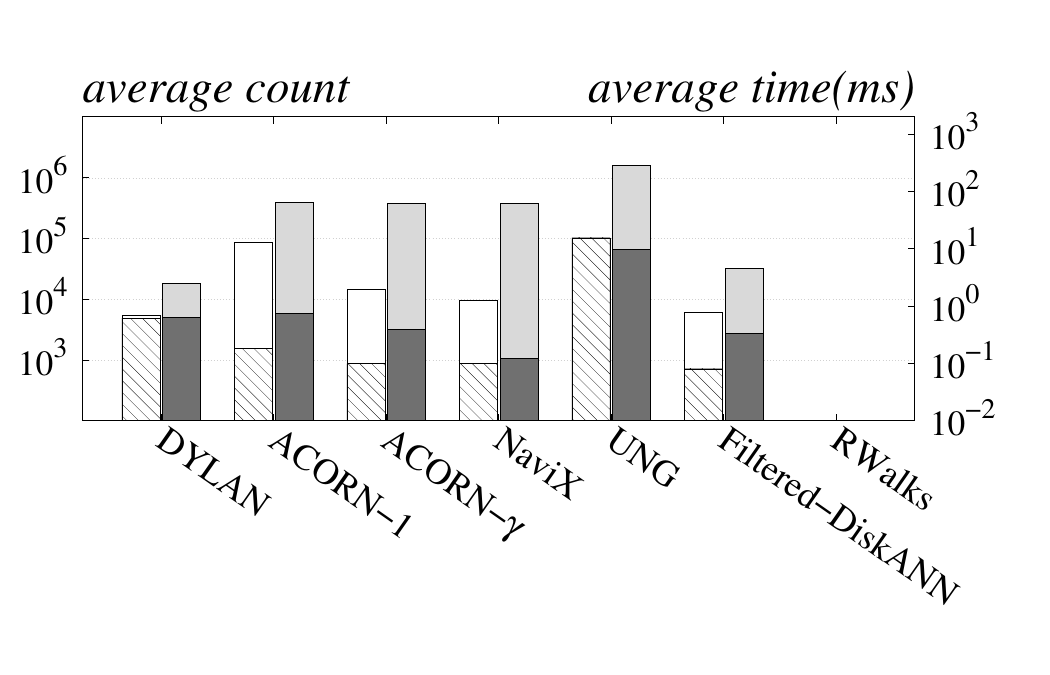}\par
{\small (b) Overlap\par}
\end{minipage}\par
\end{minipage}\par
\caption{Query-cost breakdown on YFCC.}
\label{fig:cost_yfcc}
\end{figure}

\begin{figure}[!tbp]
\centering
\includegraphics[width=1\linewidth]{figures/NDC/distance_legand.pdf}\par
\begin{minipage}{1\linewidth}\centering
\begin{minipage}[t]{.50\linewidth}
\vspace{0pt}
\centering
\includegraphics[width=\linewidth]{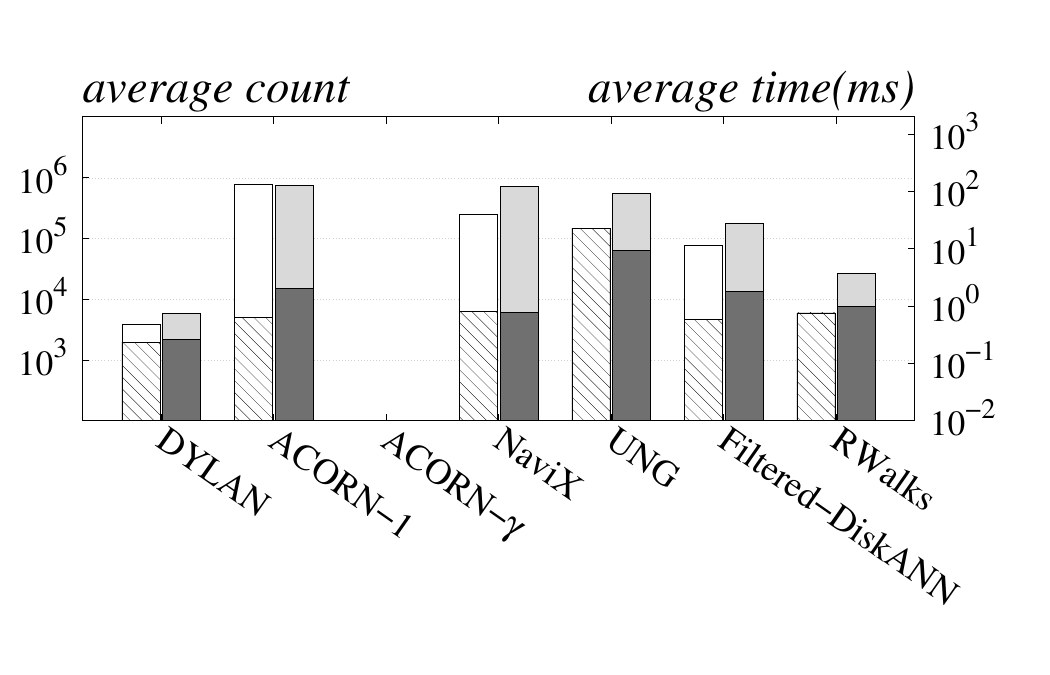}\par
{\small (a) Containment\par}
\end{minipage}\hfill\begin{minipage}[t]{.50\linewidth}
\vspace{0pt}
\centering
\includegraphics[width=\linewidth]{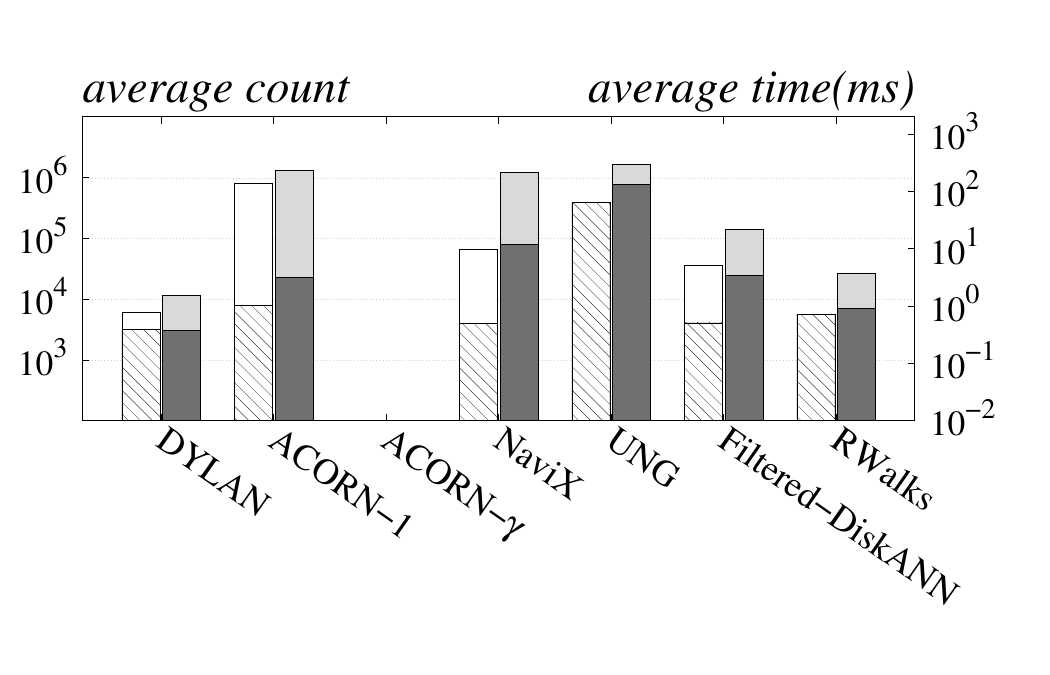}\par
{\small (b) Overlap\par}
\end{minipage}\par
\end{minipage}\par
\caption{Query-cost breakdown on YTB-Audio.}
\label{fig:cost_audio}
\end{figure}

\noindent
{\bf Query performance of complex predicate.} In this experiment, for each query, we randomly select two containment predicates and combine them via disjunction to construct a complex predicate. Fig.~\ref{exp:complex_1} presents the results on the TripClick and LAION1M datasets. Owing to space constraints, the results on additional datasets are provided in Appendix~\ref{app:exp-complex-predicate}. It is worth noting that only Curator and {\oursolution} are label-aware methods capable of handling complex predicates. In comparison with other methods, our approach achieves a speedup of one to nearly two orders of magnitude on LAION1M and TripClick. This result demonstrates that our method retains high performance even in the presence of complex predicates.

\noindent
{\bf Index cost. \label{subsubsec:index_cost}}
Figs.~\ref{fig:exp_index_memory}-\ref{fig:exp_index_time} show the memory consumption and indexing time of all solutions. 
Note that the RAM usage of RWalks on YTB-Audio and YTB-Video is not included in the figure because it consumes 32GB on YTB-Audio and 147GB on YTB-Video, which would dominate the scale. Among methods specifically designed for label‑filtered ANNS, index construction of {\oursolution} is faster than that of Filtered‑DiskANN, but slower than UNG and RWalks, while still remaining within the same order of magnitude. This overhead arises since {\oursolution} spends additional time constructing the hierarchical trie-graph structure to capture label relationships. The higher index building cost of {\oursolution} is compensated by its superior query performance, as demonstrated in the results above.

For TripClick and LAION1M, the memory usage of {\oursolution} is comparable to that of other methods. The higher memory on other datasets is due to the larger number of label sets and more complex label relationships in these datasets, which require additional space to achieve high recall. Recall from Fig.~\ref{fig:YFCC} that our method is the only one that achieves high performance on both query types on YFCC, outperforming other approaches by a factor of 3-5$\times$; on YTB‑Audio (Fig.~\ref{fig:YTB-Audio}), the performance gap reaches two orders of magnitude. 

\begin{figure}[!tbp]
\centering
\includegraphics[width=.515\linewidth]{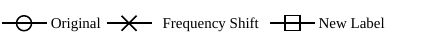}\par
\begin{minipage}[t]{0.50\linewidth}
\vspace{0pt}
\centering
\begin{minipage}[t]{.50\linewidth}
\vspace{0pt}
\centering
\includegraphics[width=\linewidth]{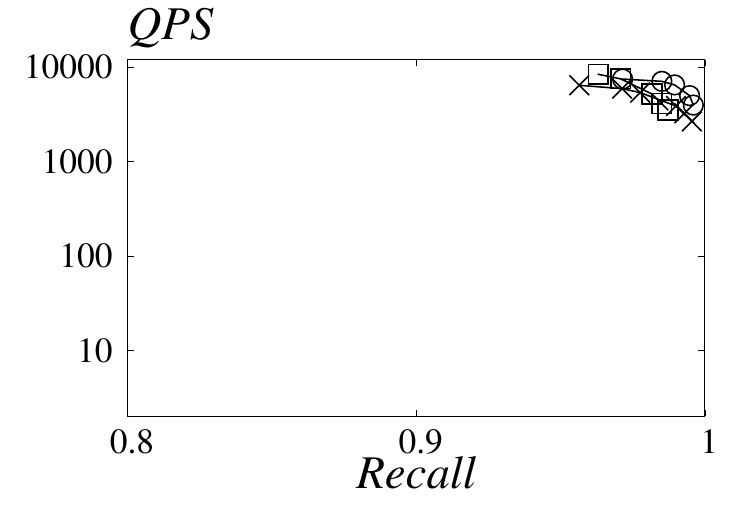}\par
{\small (a) Containment\par}
\end{minipage}\hfill\begin{minipage}[t]{.50\linewidth}
\vspace{0pt}
\centering
\includegraphics[width=\linewidth]{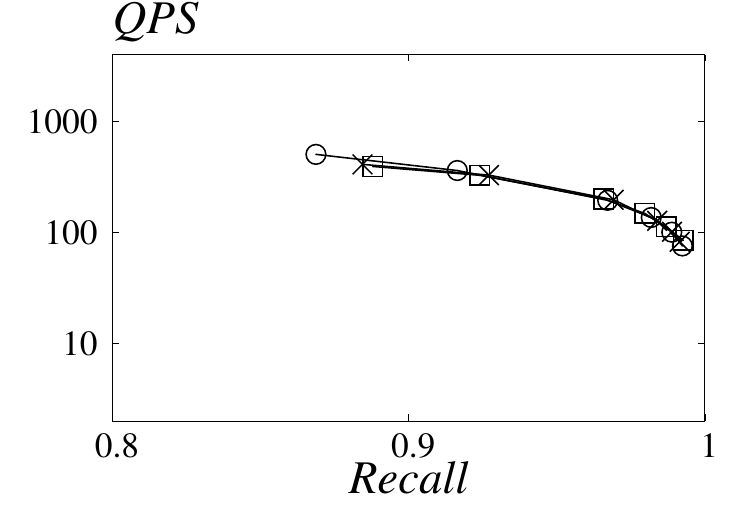}\par
{\small (b) Overlap\par}
\end{minipage}\par
\caption{Robustness: YFCC}
\label{fig:shift_yfcc}
\end{minipage}\hfill\begin{minipage}[t]{0.50\linewidth}
\vspace{0pt}
\centering
\begin{minipage}[t]{.50\linewidth}
\vspace{0pt}
\centering
\includegraphics[width=\linewidth]{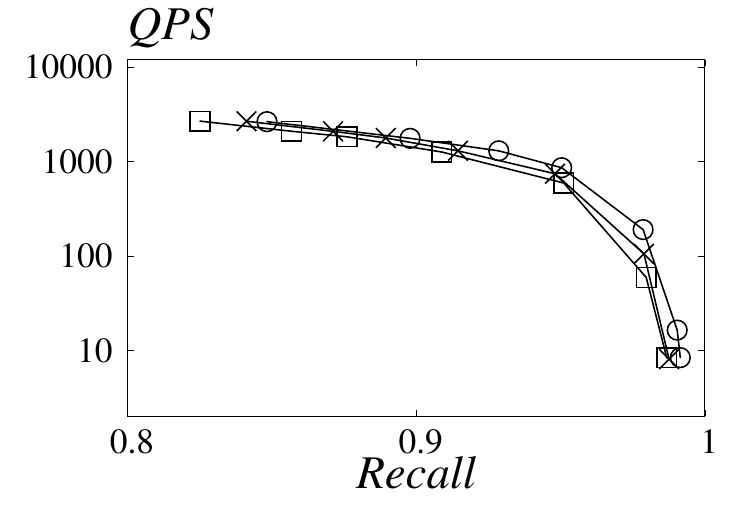}\par
{\small (a) Containment\par}
\end{minipage}\hfill\begin{minipage}[t]{.50\linewidth}
\vspace{0pt}
\centering
\includegraphics[width=\linewidth]{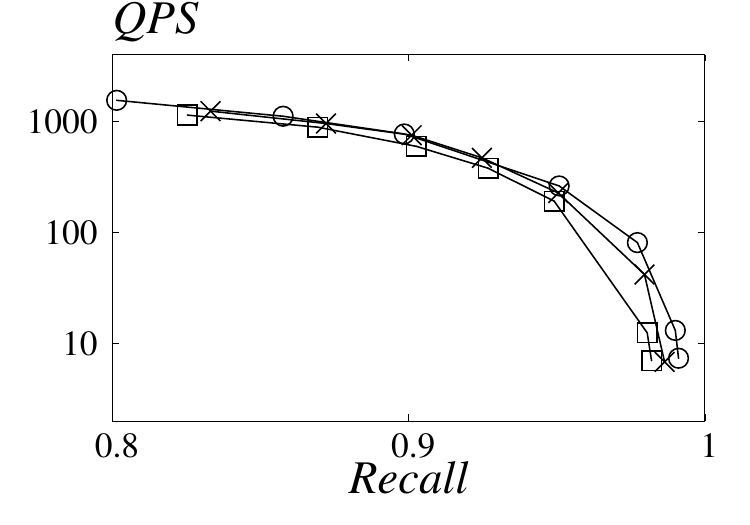}\par
{\small (b) Overlap\par}
\end{minipage}\par
\caption{Robustness: YTB-Audio}
\label{fig:shift_audio}
\end{minipage}
\end{figure}

We measure the memory cost of {\oursolution} by component. Since each vector is
stored only once in the global vector array, materializing an object in multiple
graph indices adds only graph metadata, and the label trie and auxiliary
structures do not dominate the memory cost.
Appendix~\ref{app:memory-breakdown} reports the full breakdown.

\noindent
{\bf Update cost.}
As discussed in Sec.~\ref{sec:intro}, among existing label-aware indices, only Filtered-DiskANN~\cite{GollapudiKSKBRL23} and Curator~\cite{Curator} natively support both label and vector updates. UniFilter caps the size of its auxiliary label graph, so once that capacity is reached, an update performs only the HNSW (i.e., vector) part. Fig.~\ref{fig:exp_upd_time} compares the update time of {\oursolution} with that of other methods: for each dataset, we pre-build the index on 80\% of the objects and then insert the remaining 20\%, reporting the average insertion time. The results show that {\oursolution} is 3--5$\times$ faster than Filtered-DiskANN, which must perform a label-filtered ANNS during each insertion. UniFilter achieves faster update times, but this is because our datasets contain a large number of labels, and UniFilter only updates the HNSW graph. Since Curator is based on an IVF structure, it enables rapid updates, as no graph modifications are necessary---only the cluster assignments of points need to be updated. Both, however, pay for this in search performance, as our earlier experiments show.
We also measure a mixed workload that interleaves the same 20\% insertions with deletions of a randomly selected, disjoint 20\% of the initial build set. Its update cost follows the same trends, and we defer the details to Appendix~\ref{app:impact-of-update}.

\noindent
{\bf Impact of update.} We evaluate two update workloads,
comparing in each case an index built directly on the final object set against
one that reaches the same set through updates. The insertion-only workload
pre-builds on 80\% of the objects and inserts the remaining 20\%; the mixed
workload additionally deletes a disjoint, randomly selected 20\% of the
initial build set. Figs.~\ref{exp:YTB-video_insertion-only}
and~\ref{exp:YTB-video_mixed-upd} report the two workloads on YTB-Video; the
other datasets lead to the same conclusions and are deferred to
Appendix~\ref{app:impact-of-update}. Under both, the recall-QPS curves of the
updated index stay close to those of the directly built one. This confirms the
trade-off targeted by our threshold-based maintenance: keeping a static index
consistent with the data would require a full rebuild per update, at about
$2\times10^4$~s each (Fig.~\ref{fig:exp_index_time}), whereas
{\oursolution} absorbs an insertion in milliseconds
(Fig.~\ref{fig:exp_upd_time}) with essentially no loss in query accuracy.
Insertions go through the maintenance scheme of Sec.~\ref{subsec:update},
which keeps each node's index a 4-approximate one for its region
(Thm.~\ref{thm:insert}), while deletions are even cheaper, since they only mark vertices as invalid and
leave the graph topology untouched. As in standard soft-deletion practice,
invalid vertices accumulate under deletion-heavy workloads; a periodic
compaction can be applied when needed.

\noindent
{\bf Query cost breakdown.} To attribute {\oursolution}'s speedup to its index design, we instrument graph-based search at recall 0.9 and record the numbers of visited points and distance computations, together with a query-time breakdown. The two counts differ because points rejected by the label filter incur no distance computation. On YTB-Audio, {\oursolution} visits 38.6$\times$ fewer points and performs 75.5$\times$ fewer distance computations than UNG, reducing distance-computation time by 37.3$\times$. The remaining throughput gap mainly comes from UNG's preprocessing for finding minimum supersets of the query label set, which is negligible in {\oursolution}. Figs.~\ref{fig:cost_yfcc}-\ref{fig:cost_audio} report the corresponding breakdowns on YFCC and YTB-Audio for both containment and overlap queries, excluding methods that cannot reach 0.9 recall. Across both datasets, the speedup is explained by the number of points the search needs to examine and the number of  distance computations rather than by per-operation efficiency, indicating that it stems from the index design rather than from implementation-level optimizations.

\begin{figure}[!tbp]
\centering
\includegraphics[width=.58\linewidth]{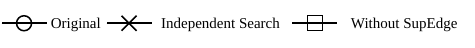}\par
\begin{minipage}{.50\linewidth}\centering
\begin{minipage}[t]{.50\linewidth}
\vspace{0pt}
\centering
\includegraphics[width=\linewidth]{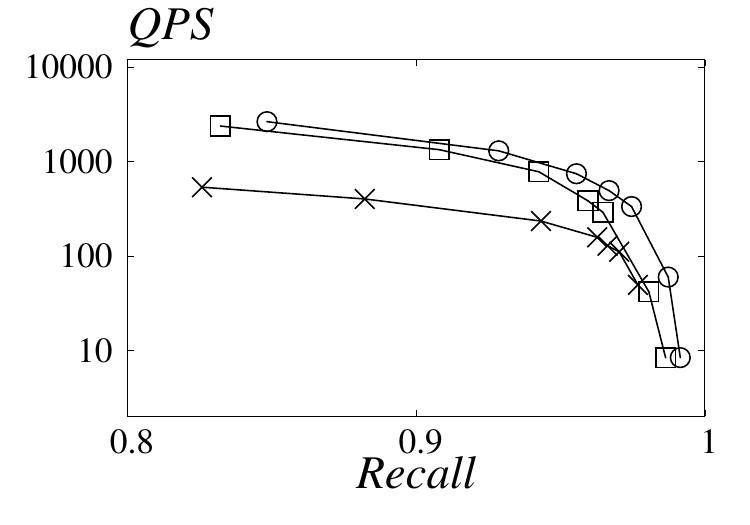}\par
{\small (a) Containment\par}
\end{minipage}\hfill\begin{minipage}[t]{.50\linewidth}
\vspace{0pt}
\centering
\includegraphics[width=\linewidth]{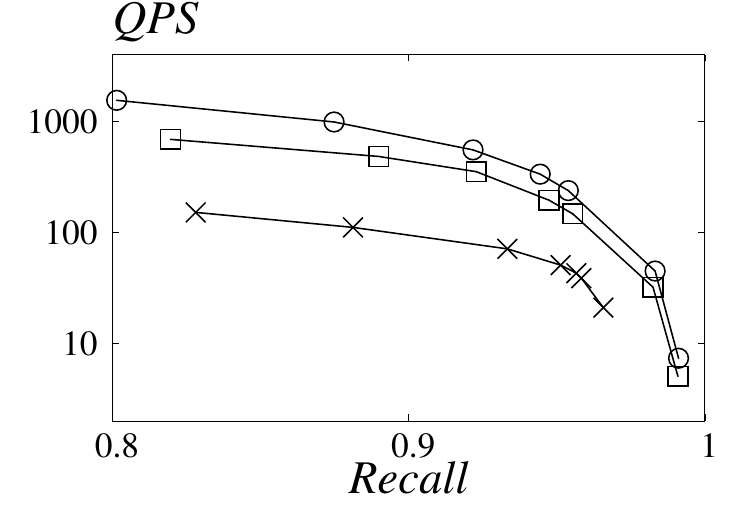}\par
{\small (b) Overlap\par}
\end{minipage}\par
\end{minipage}\par
\caption{Ablation study of optimizations: YTB-Audio.}
\label{fig:ablation_audio}
\end{figure}

\begin{figure}[!tbp]
\centering
\includegraphics[width=.85\linewidth]{figures/legend/recall_qps_pole_legend_2.pdf}\par
\begin{minipage}{.50\linewidth}\centering
\begin{minipage}[t]{.50\linewidth}
\vspace{0pt}
\centering
\includegraphics[width=\linewidth]{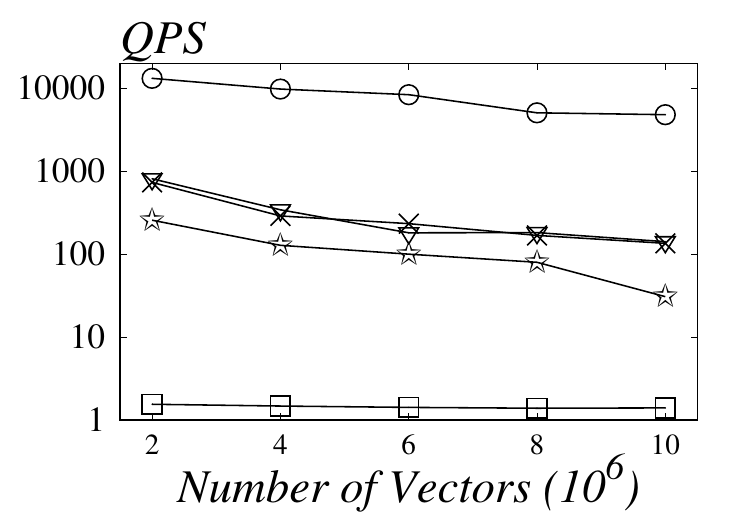}\par
{\small (a) Containment\par}
\end{minipage}\hfill\begin{minipage}[t]{.50\linewidth}
\vspace{0pt}
\centering
\includegraphics[width=\linewidth]{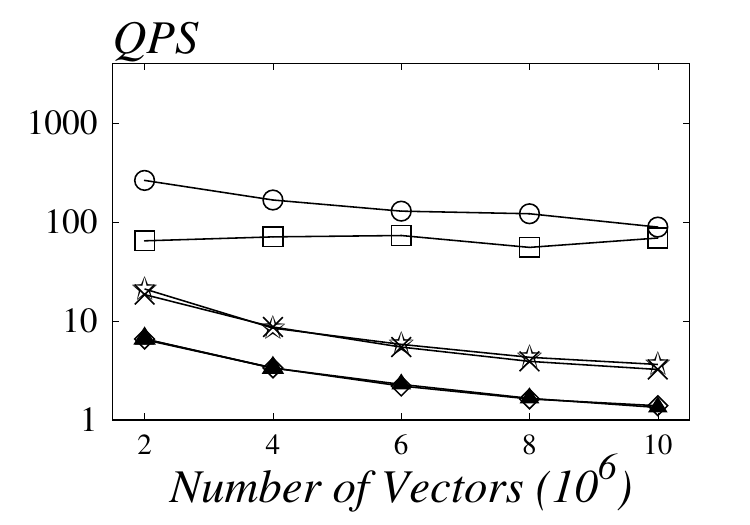}\par
{\small (b) Overlap\par}
\end{minipage}\par
\end{minipage}\par
\caption{Scalability study.}
\label{fig:scalability}
\end{figure}

\noindent
{\bf Robustness study: frequency drift and new labels.}  
We test robustness under frequency shift and new label insertion with 50/50 dynamic workloads on YFCC and YTB-Audio (a 50/50 split rather than the 80/20 used elsewhere, for a
more aggressive shift): half of the data builds the initial index and the other half is inserted, with label ordering determined only from the initial set. For frequency shift, each object is scored by the average of $\log(1+\mathrm{freq}(l))$ over its labels $l$, and the 50\% insertion suffix is sampled without replacement with probability biased toward high-score objects. This makes high-frequency-label objects appear disproportionately during insertion. For new label insertion, we maximize label diversity in the inserted half, introducing 169,496 new labels on YFCC and 3,612 on YTB-Audio (93\% and 94\% of their label universes). Figs.~\ref{fig:shift_yfcc}-\ref{fig:shift_audio} show that the performance remains close to the original, demonstrating the robustness of {\oursolution} to evolving label distributions.

\noindent
\textbf{Ablation study of optimizations.}  Fig.~\ref{fig:ablation_audio} evaluates two optimizations of {\oursolution} on YTB-Audio. The corresponding results on YFCC are deferred to Appendix~\ref{app:ablation-optimization} for space. First, we replace multi-graph search with independent search, which searches
each graph separately and merges the results. The resulting drop in the QPS-recall tradeoff confirms that the key-ancestor graphs are important for bridging otherwise disconnected regions and avoiding local optima. Second, we disable \textsc{supplementEdges}. Its impact is dataset-dependent: it yields a clear improvement on YTB-Audio but little on YFCC, whose materialized graphs are already well connected.

\noindent
{\bf Scalability study.} We evaluate scalability in two dimensions: the number of vectors
and the size of the label universe. 

For the former, we use YFCC10M from the Big ANN
Benchmarks~\cite{bigannBench}, which shares the same data source as YFCC. We vary the
dataset size from 2M to 10M vectors, using the same queries as on YFCC, and report QPS at a
fixed recall of 0.90. Fig.~\ref{fig:scalability} shows that the QPS of {\oursolution}
degrades only mildly as the dataset grows, and that {\oursolution} is the fastest method on
both containment and overlap queries across the entire range, typically by an order of
magnitude over most baselines. The only competitive baseline is Filtered-DiskANN on overlap
queries, which performs well on overlap filters but has no straightforward support for
containment; on containment queries it becomes the slowest method. Hence, no baseline is
competitive on both predicate types, whereas {\oursolution} handles them within a single
index. 

For the latter, we measure how the index size of each method grows with the size of
the label universe across the five 1M-scale datasets, excluding the raw vector storage so
that only the index structures are compared. The results, reported
in Appendix~\ref{app:label-scalability}, are empirically consistent with the Label Scalability column of
Table~\ref{tab:label_aware}: the index size of {\oursolution} and most baselines stays
within a narrow range, while that of RWalks grows sharply.

\begin{figure}[!tbp]
\centering
\begin{minipage}{.50\linewidth}\centering
\begin{minipage}[t]{.50\linewidth}
\vspace{0pt}
\centering
\includegraphics[width=\linewidth]{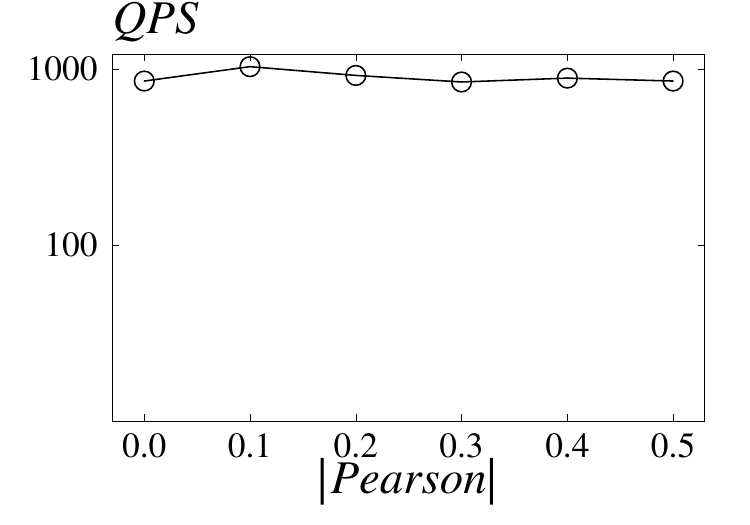}\par
{\small (a) Containment\par}
\end{minipage}\hfill\begin{minipage}[t]{.50\linewidth}
\vspace{0pt}
\centering
\includegraphics[width=\linewidth]{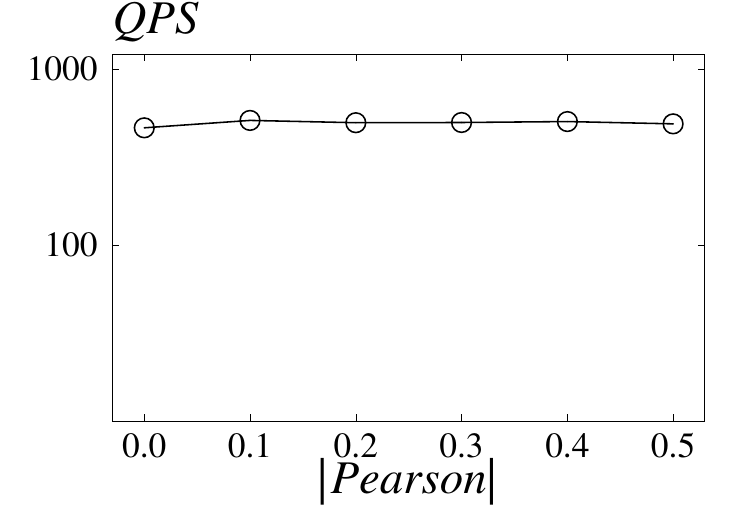}\par
{\small (b) Overlap\par}
\end{minipage}\par
\end{minipage}\par
\caption{Impact of correlation.}
\label{fig:cor_study}
\end{figure}

\noindent{\bf Impact of label-vector correlation.}
{\oursolution} does not assume that objects with similar label sets are close in vector
space. As shown in Table~\ref{tab:Dataset}, the five widely adopted datasets all have
absolute Pearson correlation $|r|\le 0.0428$, whereas the newly added Yelp1M reaches
$0.48$; bucketing object pairs by label-set size does not change this picture, as reported
in Appendix~\ref{app:bucket-correlation}. We therefore use Yelp1M to isolate the effect of
$|r|$: keeping both the vectors and the multiset of label sets unchanged, we vary their
assignment to construct six variants with $|r|$ from approximately $0$ to $0.5$ (higher
values are hard to realize without synthesizing label sets). Fig.~\ref{fig:cor_study}
reports the QPS at recall $0.9$, which is nearly unchanged across all correlation levels
for both query types. This robustness follows from our design. Correlation mainly changes how the qualifying
objects sit in the vector space relative to the trie regions, and either way the
search covers them: when $|r|$ is high, the results concentrate within a region's own
graph; when $|r|$ is low, they are reached across regions via the key-ancestor graphs, so
the effective search graph is the union of the graphs in $\mathcal{G}$ rather than any
single per-region graph.

\noindent
{\bf Impact of parameter $ef_c$. } We also conduct experiments to evaluate the impact of $ef_{c}$ on query performance and indexing cost. Interested readers are referred to Appendix~\ref{app:tune-efc} for detailed results.

%% file: section/4-relatedwork.tex
\section{Related Work}\label{sec:related}
{\bf Approximate nearest neighbor search.}
Existing ANNS solutions can be broadly categorized into three classes: \emph{LSH-based}, \emph{PQ-based}, and \emph{graph-based}. Locality-sensitive hashing (LSH) and its variants, e.g., \cite{IndykM98,AndoniI08,datar2004locality,AndoniILRS15,WeiPLP24}, provide rigorous theoretical error guarantees, but often incur considerable time and space overhead in practice.
Product quantization (PQ)~\cite{JegouDS11} partitions the high-dimensional vector space into multiple low-dimensional subspaces, and quantizes each subspace to compress the vectors. PQ-based methods, e.g. \cite{AndreKS15,JohnsonDJ21,BlalockG17,BabenkoL15,GaoL24} trade result accuracy for significantly lower memory usage, achieving high scalability. 
Graph-based methods \cite{MalkovY20,MalkovPLK14,DongCL11,FuXWC19,SubramanyaDSKK19} build a proximity graph for the dataset, where vertices represent vectors and edges connect similar vectors. %

\header
{\bf ANNS with attribute filtering.} There exist a plethora of research works \cite{engels2024approximate,3639324,GollapudiKSKBRL23,WangYGJXLWGLXYY21,MatsuiHS18,ZhangXCSXCCH00Y23} on designing
generic solutions for attribute-filtered ANNS. Such solutions are often less effective than specialized methods for particular filters~\cite{engels2024approximate,3639324}. Beyond label filters, range-filtered ANNS has also been studied~\cite{jiang2025digra,zhang2025efficient,xu2024irangegraph,3639324}, which is orthogonal to our work.

%% file: section/6-conclusion.tex
\section{Conclusion}

This paper presented {\oursolution}, a dynamic composite index for label-filtered ANNS that
combines a frequency-ordered label trie, selective HNSW-style graph materialization,
cross-region multi-graph search, and threshold-based index maintenance. Experiments on real
datasets show that {\oursolution} achieves superior query efficiency at comparable recall,
especially on label-rich datasets, while maintaining favorable indexing and update-cost
trade-offs. \update{Future work includes an efficient multi-threaded implementation,
designing efficient index for external-memory settings, and considering more complex settings such as joint label and multi-attribute range filters.}

%% file: section/appendix.tex
\appendix
\setlength{\textfloatsep}{6pt plus 1pt minus 1pt}
\captionsetup[figure]{skip=1pt}

\section{Additional experimental results}
\update{
\subsection{Parameter Tuning}\label{app:tune-parameters}
\noindent{\bf Query parameters.}
All queries retrieve $k=10$ results and are executed single-threaded. For
each approximate method, we vary its query-time search parameter to trace
the recall--QPS tradeoff: $L_{\mathrm{search}}$ for Filtered-DiskANN and
UNG, \texttt{efSearch} for NaviX and ACORN, $e\!f$ for RWalks and
{\oursolution}, $L$ for UniFilter, and \texttt{search\_ef} for Curator.
Prefilter evaluates the label predicate exactly and performs an exact $L_2$ scan over the survivors, and therefore
appears as a single point in the recall--QPS plots.

\header
\noindent{\bf Construction parameters.}
We tune the build-time parameters of every method, {\oursolution} included,
on every dataset under one common protocol. Each configuration is scored by
its QPS at recall $0.9$ --- the operating point used throughout
Sec.~\ref{sec:exp} for most experiments --- on containment and on overlap queries, and we select
the configuration maximizing the geometric mean of the two; if only one
query type reaches this recall, the configuration is scored on that type
alone, and if neither does, configurations are ranked by the geometric mean
of their highest attained recalls. The swept grids, with the remaining
parameters fixed to the values recommended by the corresponding papers,
are: $(L_{\mathrm{build}},R)\in\{(96,64),(144,96),(192,128),(240,160)\}$
with $\alpha=1.2$ for Filtered-DiskANN;
$(R,L)\in\{(16,100),(32,100),(32,200),(48,200)\}$ with $\alpha=1.2$, $6$
cross-group edges and $16$ query entry vectors for UNG;
$(\texttt{nlist},B_{\max})\in\{(16,64),(32,128),(64,256),(64,512)\}$ with
beam size $4$ for Curator; and $M\in\{16,32,48,64\}$ with $e\!f_c=400$ for
NaviX, RWalks, ACORN-1 ($\gamma=1$, $M_\beta=2M$), ACORN-$\gamma$
($\gamma=100$, $M_\beta=64$), UniFilter ($M_p=n$), and {\oursolution},
ACORN-1 and ACORN-$\gamma$ being swept independently. We further extend
Filtered-DiskANN's \texttt{detect\_common\_filters} to support both
containment and overlap predicates. Table~\ref{tab:selected_parameters}
reports the selected values. Each baseline is thus evaluated under its own
best configuration rather than a default one, and {\oursolution} receives
no larger construction budget: except on YTB-Video, its selected $M$ is
$16$, no larger than that of any baseline. RWalks is omitted on YFCC, as
its memory consumption exceeds the available 1~TB.
}

\subsection{Impact of Query Selectivity}\label{app:exp-impact-selectivity}

Fig.~\ref{fig:sel_qps} shows the additional results of the impact of filter selectivity on query performance on TripClick and YFCC omitted in Sec.~\ref{sec:exp}. Recall that it is infeasible to generate containment queries spanning a continuous range of target selectivity due to the label distributions in YFCC and TripClick.  We only report overlap queries on these two datasets. These results are consistent with the trends presented in Sec.~\ref{sec:exp}. It can be observed that our method generally outperforms the other approaches, except on YFCC under very low selectivity. 

\subsection{Query Performance of Complex Predicate}\label{app:exp-complex-predicate}
\begin{figure}[!tbp]
\centering
\includegraphics[width=.85\linewidth]{figures/legend/recall_qps_pole_legend_2.pdf}\par
\begin{minipage}[t]{.25\linewidth}
\vspace{0pt}
\centering
\includegraphics[width=\linewidth]{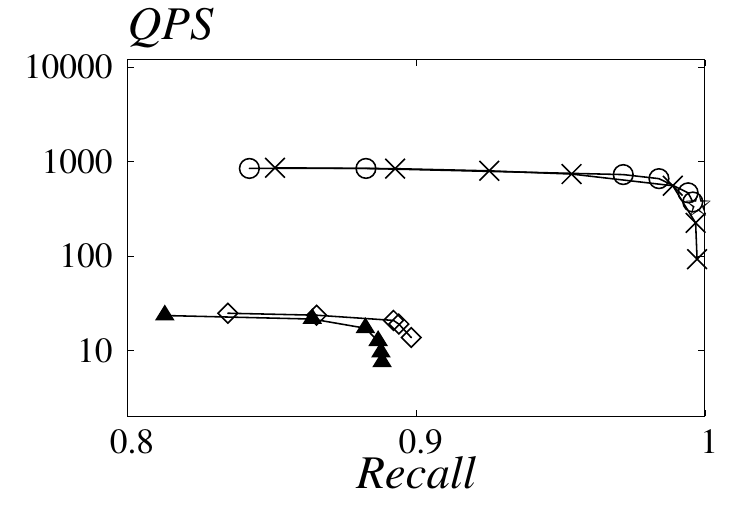}\par
{\small (a) YFCC\par}
\end{minipage}\hfill\begin{minipage}[t]{.25\linewidth}
\vspace{0pt}
\centering
\includegraphics[width=\linewidth]{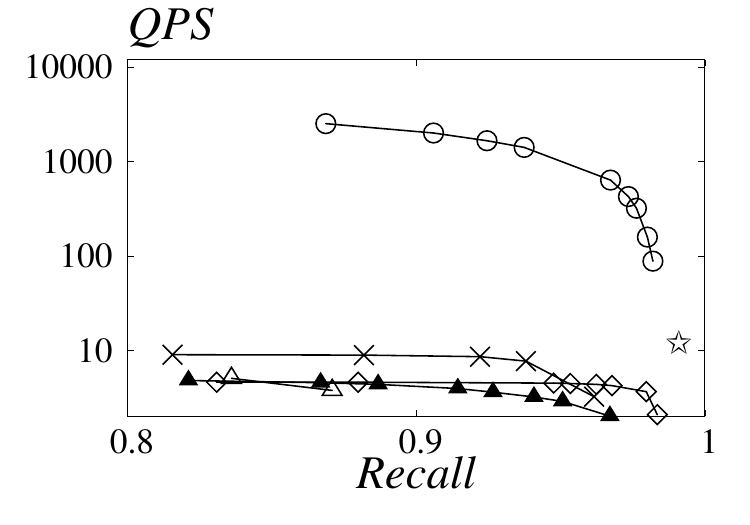}\par
{\small (b) YTB-Audio\par}
\end{minipage}\hfill\begin{minipage}[t]{.25\linewidth}
\vspace{0pt}
\centering
\includegraphics[width=\linewidth]{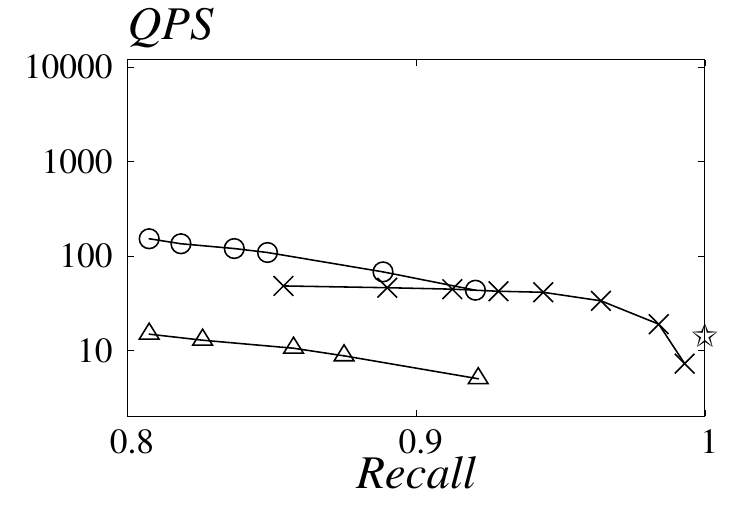}\par
{\small (c) YTB-Video\par}
\end{minipage}\hfill\begin{minipage}[t]{.25\linewidth}
\vspace{0pt}
\centering
\includegraphics[width=\linewidth]{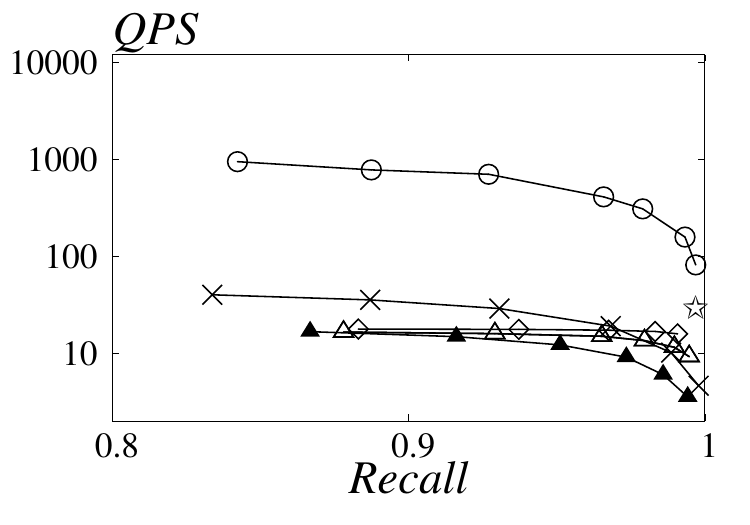}\par
{\small (d) Yelp1M\par}
\end{minipage}\par
\caption{Complex predicate.}
\label{exp:app_complex}
\end{figure}

Fig.~\ref{exp:app_complex} shows the results on the remaining 4 datasets omitted in Sec.~\ref{sec:exp}. Our method achieves performance comparable to Curator on YFCC and YTB-Video, significantly outperforming other approaches. On the YTB-Audio dataset, our method outperforms all other methods, including Curator, by two orders of magnitude. This demonstrates that our method maintains superior performance across all datasets and under any query conditions.

\update{
\subsection{Memory Breakdown of {\oursolution} Index}\label{app:memory-breakdown}
Tab.~\ref{tab:memory_breakdown} reports the memory consumption of {\oursolution} by component. We separately account for the global vector array, object labels, the label trie, materialized graph indices, and other auxiliary structures. The ``Other'' category includes the \textit{belong} arrays and the label inverted lists. Since each vector is stored only once in the global vector array, materializing an object in multiple graph indices only introduces graph metadata rather than additional vector copies. The breakdown also reflects the different characteristics of the datasets. For example, YFCC incurs a relatively large trie cost due to its large label universe and complex label sets, whereas the graph indices constitute the main additional cost on YTB-Audio. Overall, the trie and auxiliary structures remain moderate compared with the vector and graph storage, confirming that the additional structures introduced by {\oursolution} do not dominate its memory consumption.

\input{table/Memory}

\input{table/correlation}

\begin{figure}[!tbp]
\centering
\includegraphics[width=.85\linewidth]{figures/legend/recall_qps_pole_legend_2.pdf}\par
\begin{minipage}[t]{.25\linewidth}
\vspace{0pt}
\centering
\includegraphics[width=\linewidth]{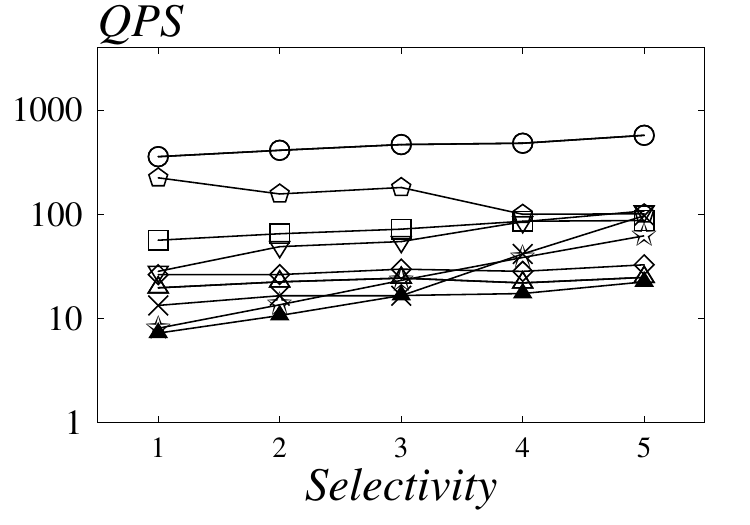}\par
{\small (a) TripClick\par}
\end{minipage}\hspace{.01\linewidth}\begin{minipage}[t]{.25\linewidth}
\vspace{0pt}
\centering
\includegraphics[width=\linewidth]{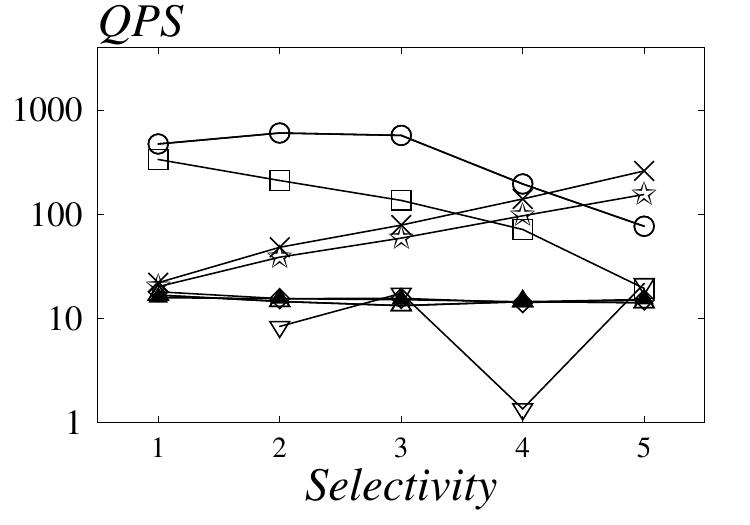}\par
{\small (b) YFCC\par}
\end{minipage}\par
\caption{Impact of selectivity on different datasets.}
\label{fig:sel_qps}
\end{figure}

\begin{figure}[!tbp]
\centering
\includegraphics[width=.85\linewidth]{figures/legend/recall_qps_pole_legend_2.pdf}\par
\includegraphics[width=.25\linewidth]{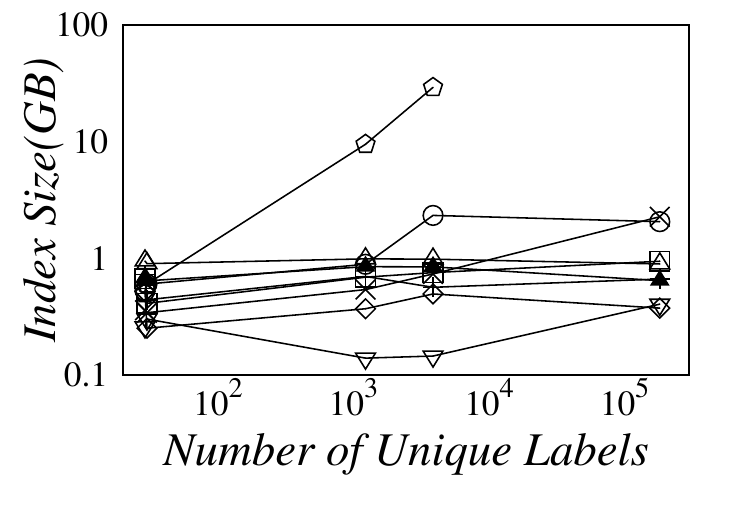}\par
\caption{Index size vs.\ $|\mathcal{L}|$.}
\label{fig:label_scalability}
\end{figure}

\begin{figure}[!tbp]
\centering
\includegraphics[width=.56\linewidth]{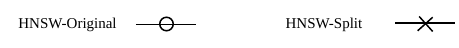}\par
\includegraphics[width=.25\linewidth]{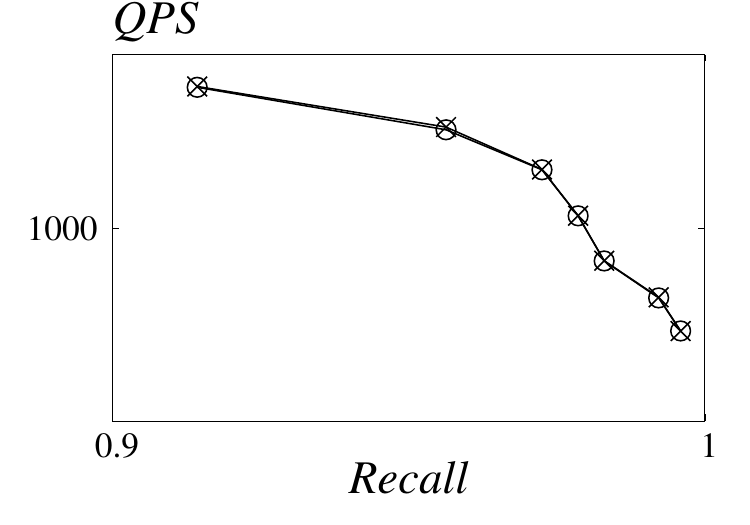}\par
\caption{Memory layout.}
\label{fig:memory_layout}
\end{figure}

\begin{figure}[!tbp]
\centering
\includegraphics[width=.58\linewidth]{figures/ablation/ablationLG.pdf}\par
\begin{minipage}[t]{.25\linewidth}
\vspace{0pt}
\centering
\includegraphics[width=\linewidth]{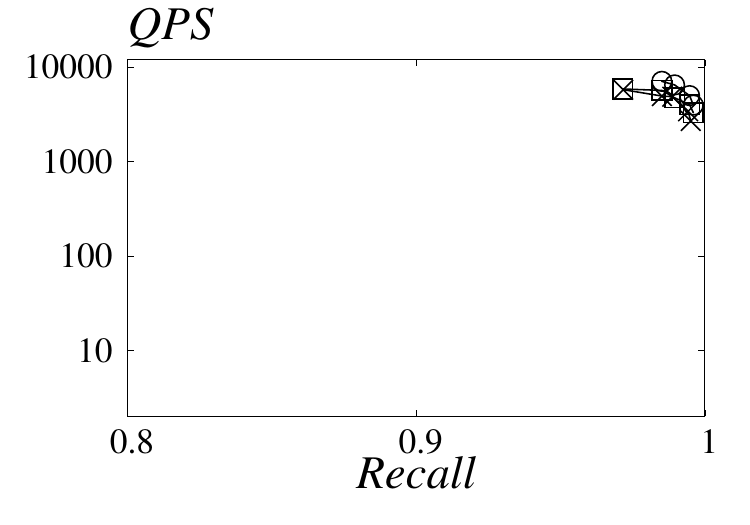}\par
{\small (a) Containment\par}
\end{minipage}\hspace{.01\linewidth}\begin{minipage}[t]{.25\linewidth}
\vspace{0pt}
\centering
\includegraphics[width=\linewidth]{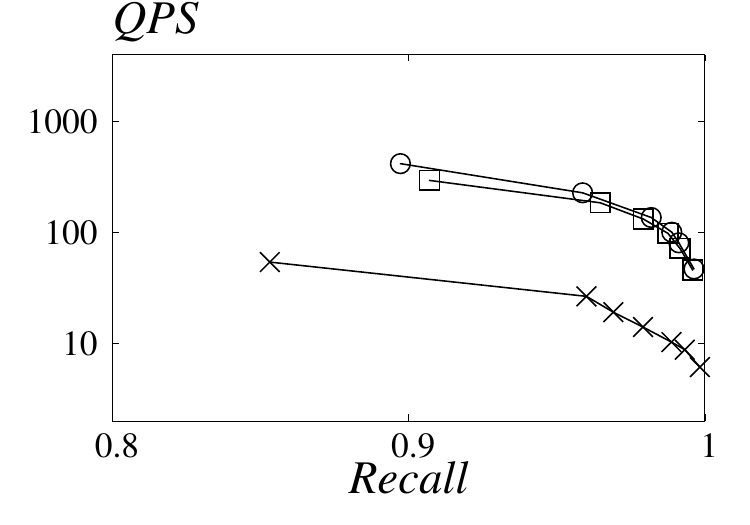}\par
{\small (b) Overlap\par}
\end{minipage}\par
\caption{Ablation study of optimizations: YFCC.}
\label{fig:ablation_YFCC}
\end{figure}

\subsection{Label Scalability}\label{app:label-scalability}
Sec.~\ref{sec:exp} evaluates scalability along two dimensions, the number of
vectors and the size of the label universe, and reports only the former; we
report the latter here. Fig.~\ref{fig:label_scalability} plots the index
size of each method against the number of distinct labels over the five 1M-scale datasets (all datasets except YTB-Audio), whose label
universes span from $29$ labels on TripClick to $181{,}931$ on YFCC. Since
all five datasets hold roughly the same number of vectors, the raw vector
payload is essentially constant across them and would mask the differences
of interest; we therefore exclude it for every method and report only the
index structures built on top of it. 
The index size of {\oursolution} grows by only $4.4\times$ across the entire span and
follows a trend similar to most existing methods, indicating that neither the label trie
nor the auxiliary structures grows substantially with the label universe.
RWalks is
the clear exception: it encodes labels into vector-like representations, so
its index size increases sharply as labels accumulate, which is also why it
exceeds the available 1~TB on YFCC. These results are consistent with the
Label Scalability column of Tab.~\ref{tab:label_aware} and confirm that {\oursolution} scales effectively to large label universes.

\subsection{Impact of Memory Layout}\label{app:memory-layout}
{\oursolution} stores vectors and adjacency lists separately, so reaching a
neighbor's vector requires an extra indirection through its object ID. To
isolate the cost of this indirection, we compare two layouts of a standard
HNSW graph on LAION1M in Figure~\ref{fig:memory_layout}. The first follows
the original implementation of HNSWlib~\cite{Hnswlib}, where each vertex
stores its vector contiguously with its adjacency list, 
avoiding an additional pointer dereference to locate the vector.
The second is the separated layout used by {\oursolution}, where adjacency lists hold only object IDs and all
vectors reside in a global array, so expanding a neighbor requires a second
random access into that array. The two layouts achieve nearly identical
query performance, differing by less than $1\%$: the indirection is
amortized over the distance computation that follows it, and the vector
array is accessed in the same order regardless of layout. In
{\oursolution} the effect is even smaller, since many visited neighbors
fail the label filter and are skipped without any vector access at all.
Together with Fig.~\ref{fig:memory_layout}, this confirms
that the separated layout, which lets an object appear in multiple
materialized graphs without duplicating its vector, does not trade memory
for query latency.

}

\input{table/selected_parameters}

\begin{figure}[!tbp]
\centering
\includegraphics[width=.56\linewidth]{figures/legend/mix_update_legend.pdf}\par
\begin{minipage}[t]{.25\linewidth}
\vspace{0pt}
\centering
\includegraphics[width=\linewidth]{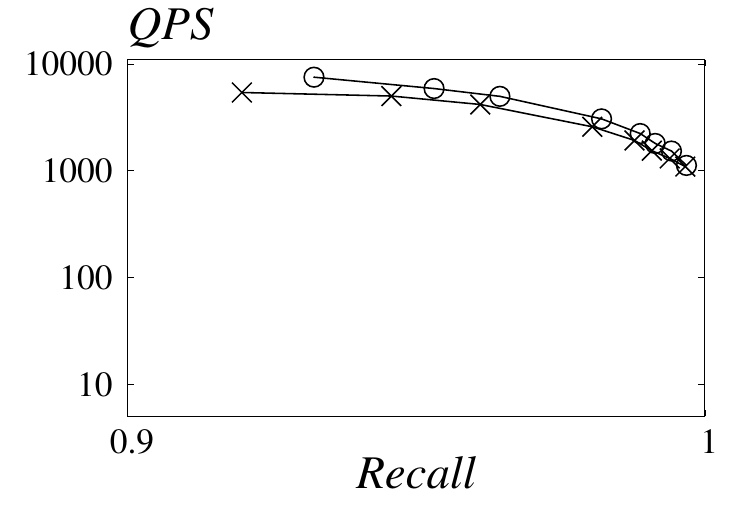}\par
{\small (a) LAION1M\par}
\end{minipage}\hspace{.01\linewidth}\begin{minipage}[t]{.25\linewidth}
\vspace{0pt}
\centering
\includegraphics[width=\linewidth]{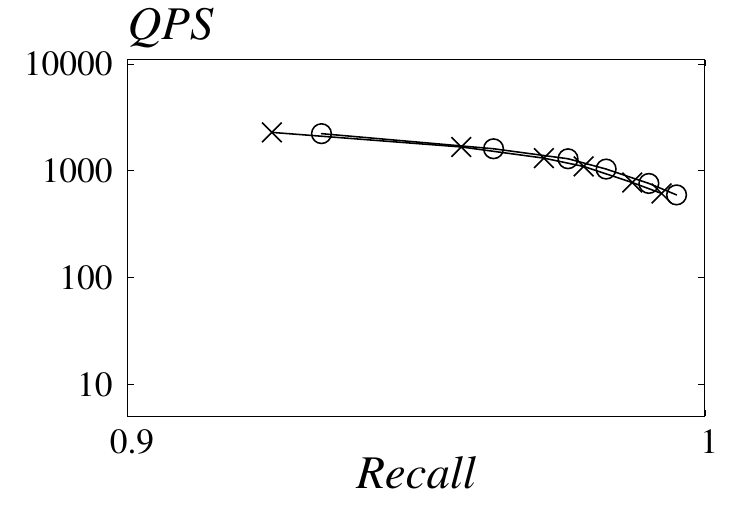}\par
{\small (b) TripClick\par}
\end{minipage}\hspace{.01\linewidth}\begin{minipage}[t]{.25\linewidth}
\vspace{0pt}
\centering
\includegraphics[width=\linewidth]{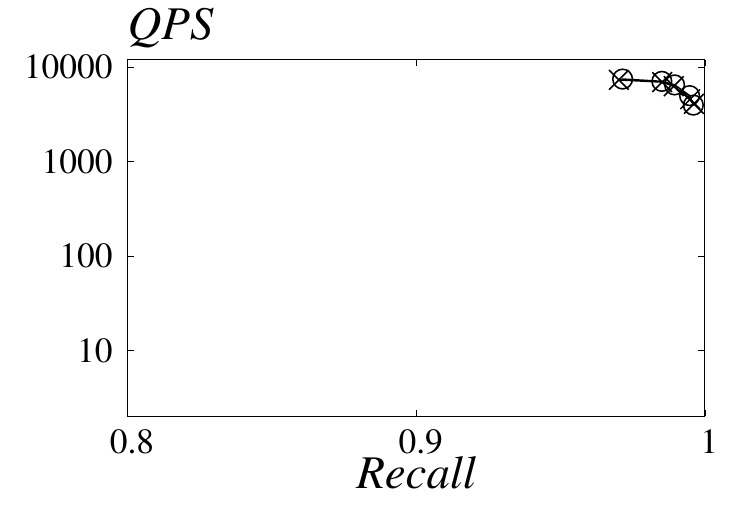}\par
{\small (c) YFCC\par}
\end{minipage}\par
\vspace{3pt}
\begin{minipage}[t]{.25\linewidth}
\vspace{0pt}
\centering
\includegraphics[width=\linewidth]{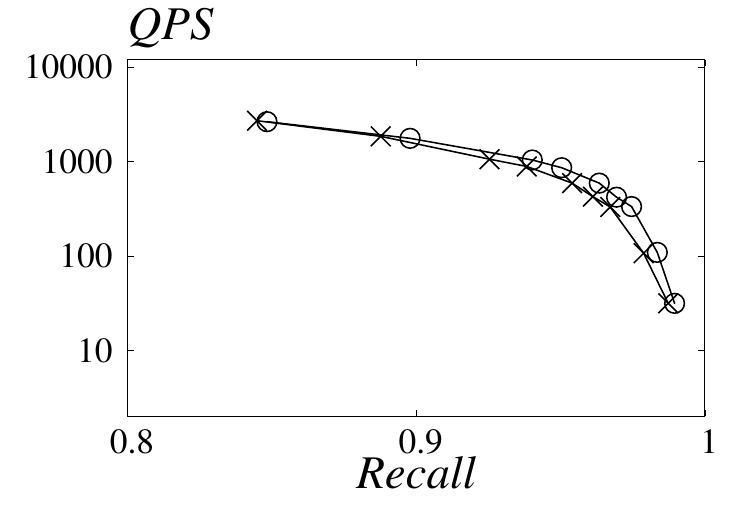}\par
{\small (d) YTB-Audio\par}
\end{minipage}\hspace{.01\linewidth}\begin{minipage}[t]{.25\linewidth}
\vspace{0pt}
\centering
\includegraphics[width=\linewidth]{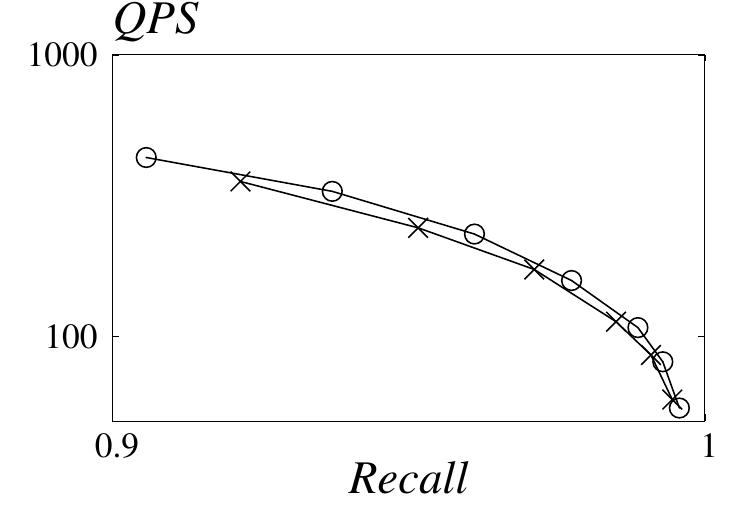}\par
{\small (e) Yelp1M\par}
\end{minipage}\par
\caption{Impact of insertion-only update: containment queries.}
\label{exp:app_upd_cont}
\end{figure}

\begin{figure}[!tbp]
\centering
\includegraphics[width=.56\linewidth]{figures/legend/mix_update_legend.pdf}\par
\begin{minipage}[t]{.25\linewidth}
\vspace{0pt}
\centering
\includegraphics[width=\linewidth]{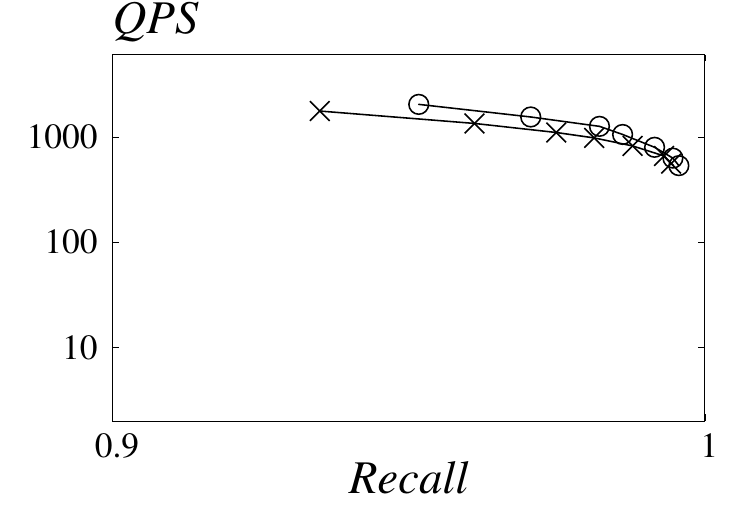}\par
{\small (a) LAION1M\par}
\end{minipage}\hspace{.01\linewidth}\begin{minipage}[t]{.25\linewidth}
\vspace{0pt}
\centering
\includegraphics[width=\linewidth]{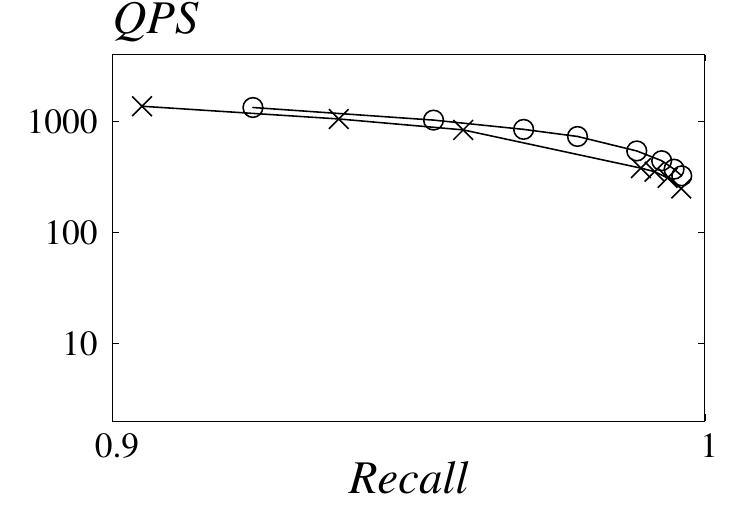}\par
{\small (b) TripClick\par}
\end{minipage}\hspace{.01\linewidth}\begin{minipage}[t]{.25\linewidth}
\vspace{0pt}
\centering
\includegraphics[width=\linewidth]{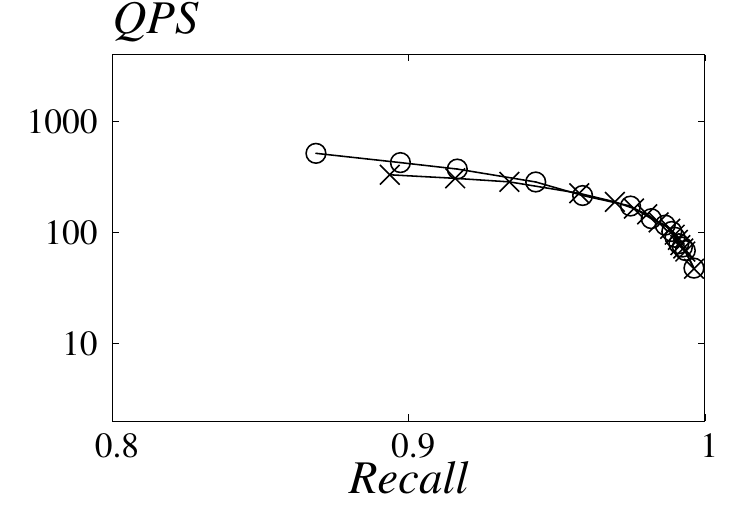}\par
{\small (c) YFCC\par}
\end{minipage}\par
\vspace{3pt}
\begin{minipage}[t]{.25\linewidth}
\vspace{0pt}
\centering
\includegraphics[width=\linewidth]{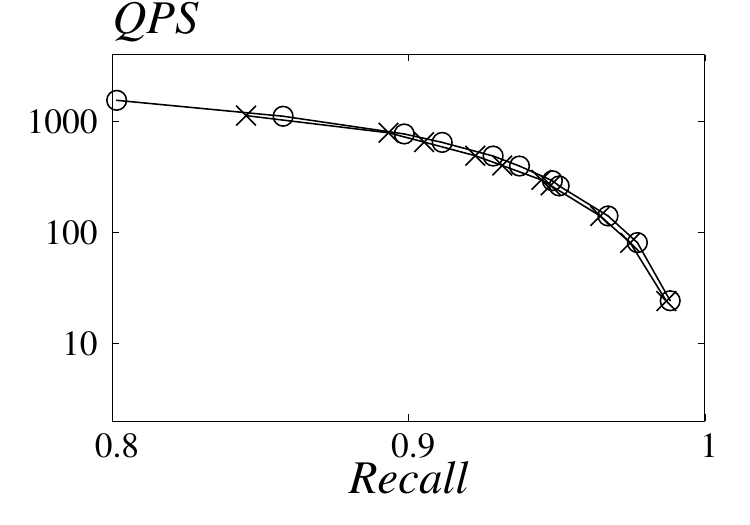}\par
{\small (d) YTB-Audio\par}
\end{minipage}\hspace{.01\linewidth}\begin{minipage}[t]{.25\linewidth}
\vspace{0pt}
\centering
\includegraphics[width=\linewidth]{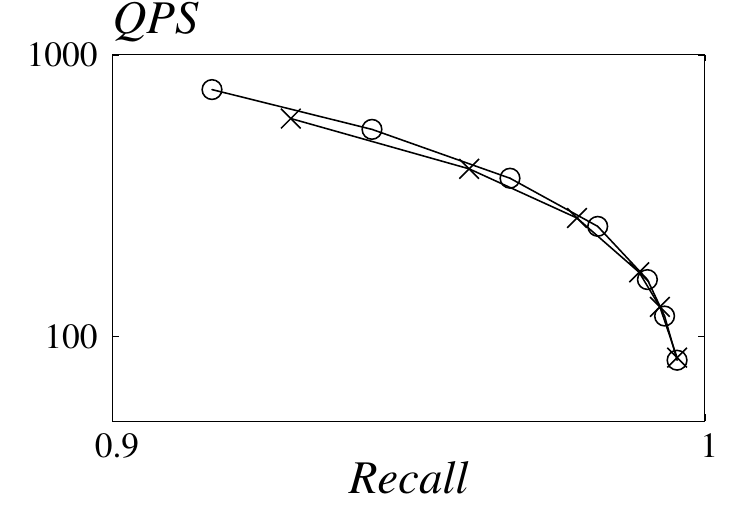}\par
{\small (e) Yelp1M\par}
\end{minipage}\par
\caption{Impact of insertion-only update: overlap queries.}
\label{exp:app_upd_over}
\end{figure}

\begin{figure}[!tbp]
\centering
\includegraphics[width=.56\linewidth]{figures/legend/mix_update_legend.pdf}\par
\begin{minipage}[t]{.25\linewidth}
\vspace{0pt}
\centering
\includegraphics[width=\linewidth]{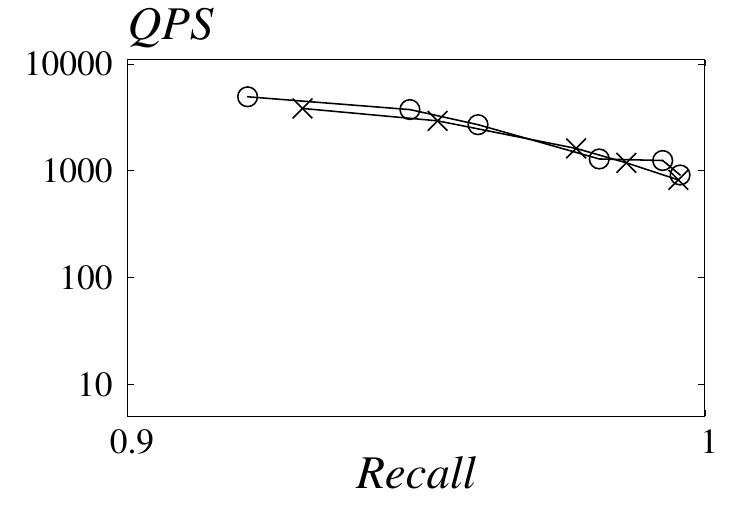}\par
{\small (a) LAION1M\par}
\end{minipage}\hspace{.01\linewidth}\begin{minipage}[t]{.25\linewidth}
\vspace{0pt}
\centering
\includegraphics[width=\linewidth]{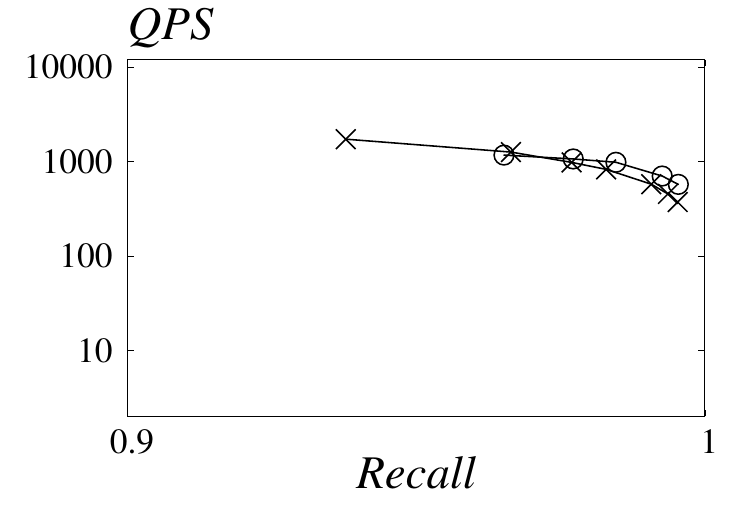}\par
{\small (b) TripClick\par}
\end{minipage}\hspace{.01\linewidth}\begin{minipage}[t]{.25\linewidth}
\vspace{0pt}
\centering
\includegraphics[width=\linewidth]{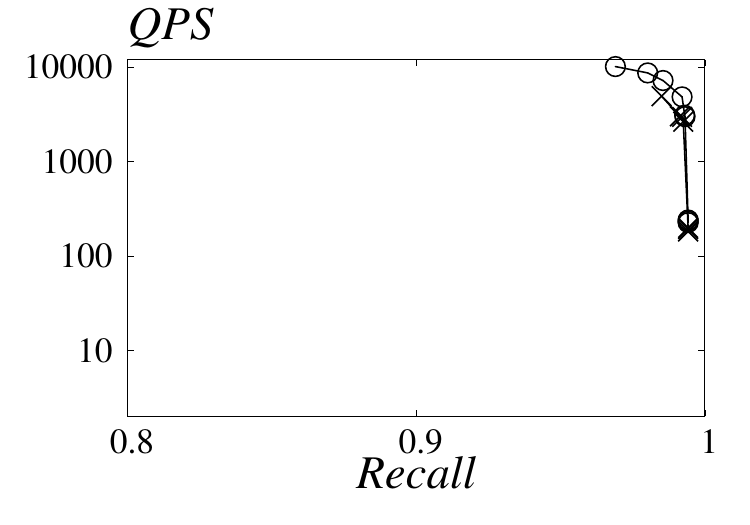}\par
{\small (c) YFCC\par}
\end{minipage}\par
\vspace{3pt}
\begin{minipage}[t]{.25\linewidth}
\vspace{0pt}
\centering
\includegraphics[width=\linewidth]{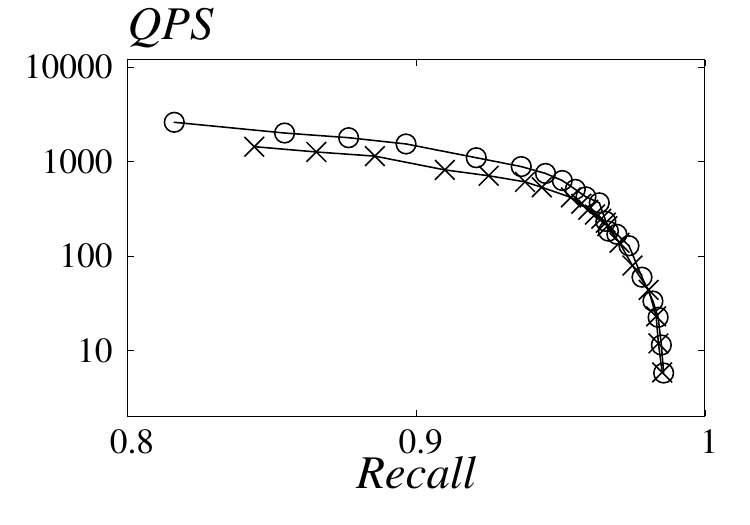}\par
{\small (d) YTB-Audio\par}
\end{minipage}\hspace{.01\linewidth}\begin{minipage}[t]{.25\linewidth}
\vspace{0pt}
\centering
\includegraphics[width=\linewidth]{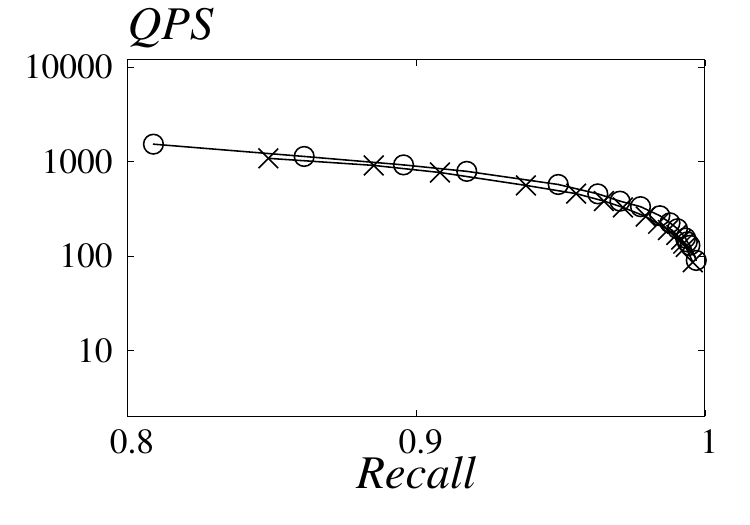}\par
{\small (e) Yelp1M\par}
\end{minipage}\par
\caption{Impact of mixed updates: containment queries.}
\label{exp:app_mix_cont}
\end{figure}

\begin{figure}[!tbp]
\centering
\includegraphics[width=.56\linewidth]{figures/legend/mix_update_legend.pdf}\par
\begin{minipage}[t]{.25\linewidth}
\vspace{0pt}
\centering
\includegraphics[width=\linewidth]{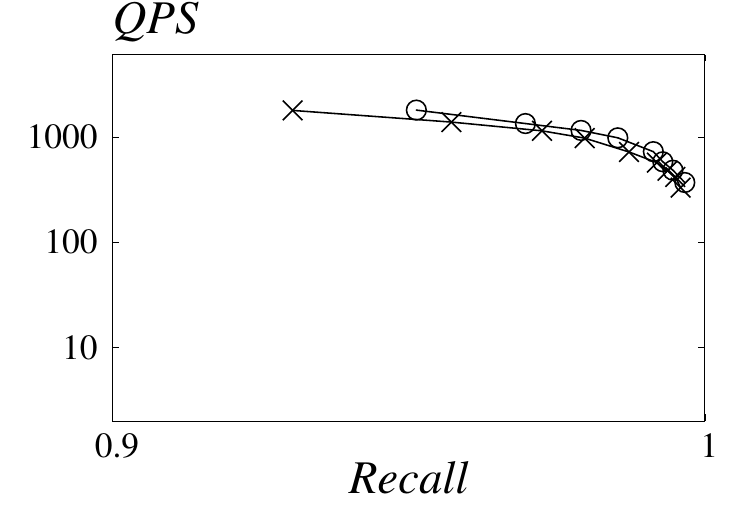}\par
{\small (a) LAION1M\par}
\end{minipage}\hspace{.01\linewidth}\begin{minipage}[t]{.25\linewidth}
\vspace{0pt}
\centering
\includegraphics[width=\linewidth]{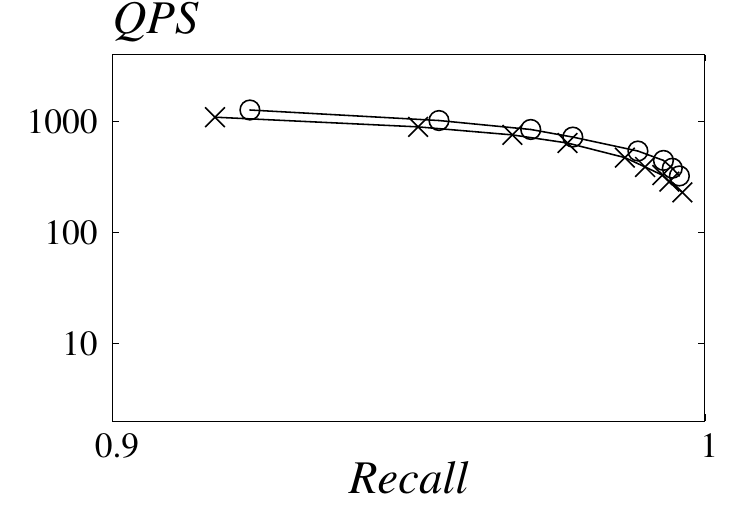}\par
{\small (b) TripClick\par}
\end{minipage}\hspace{.01\linewidth}\begin{minipage}[t]{.25\linewidth}
\vspace{0pt}
\centering
\includegraphics[width=\linewidth]{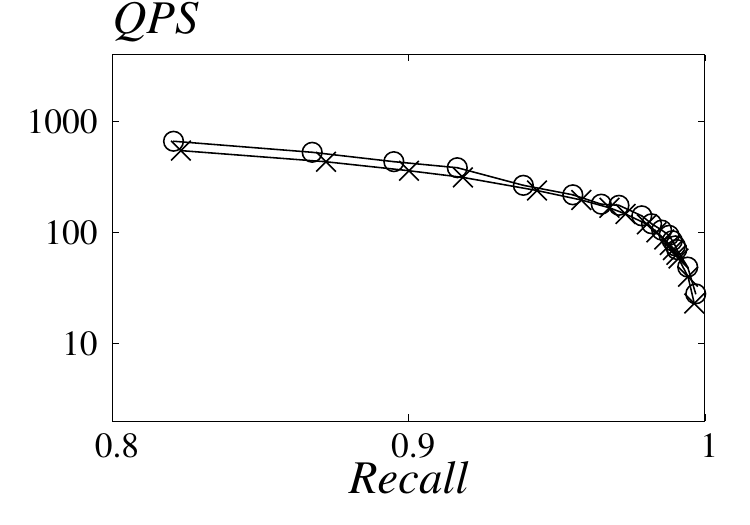}\par
{\small (c) YFCC\par}
\end{minipage}\par
\vspace{3pt}
\begin{minipage}[t]{.25\linewidth}
\vspace{0pt}
\centering
\includegraphics[width=\linewidth]{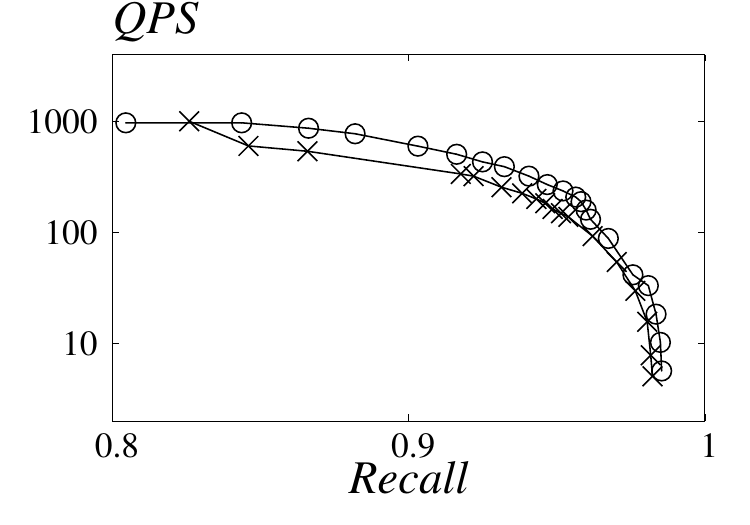}\par
{\small (d) YTB-Audio\par}
\end{minipage}\hspace{.01\linewidth}\begin{minipage}[t]{.25\linewidth}
\vspace{0pt}
\centering
\includegraphics[width=\linewidth]{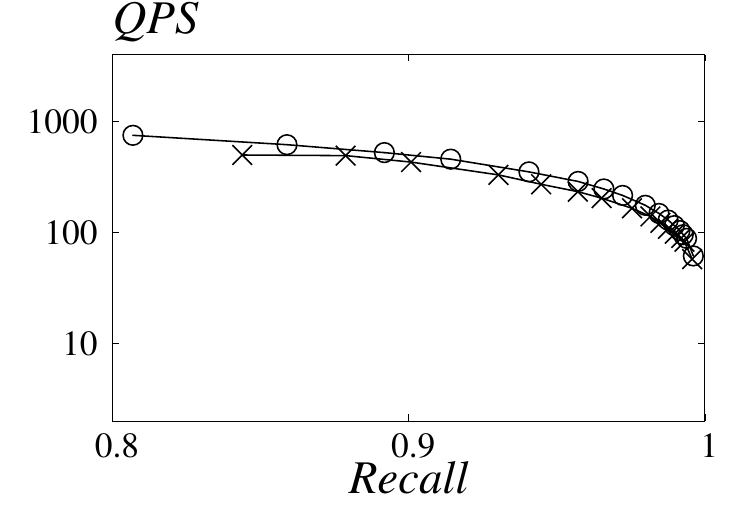}\par
{\small (e) Yelp1M\par}
\end{minipage}\par
\caption{Impact of mixed updates: overlap queries.}
\label{exp:app_mix_over}
\end{figure}

\update{
\subsection{Label-Vector Correlation by Label-Set Size}\label{app:bucket-correlation}
We analyze whether label-set similarity is correlated with vector proximity,
and whether such correlation depends on how many labels the objects carry.
For each dataset, we uniformly sample 1M distinct object pairs and compute
the Pearson correlation $r$ between their label-set Jaccard similarity and
their $L_2$ distance, reporting $|r|$ since negating the distance only flips
the sign. A single overall correlation can be misleading here, as pairs with
few labels and pairs with many labels differ in both their Jaccard values and
their distance distributions: the overall coefficient absorbs these
cross-bucket shifts under a global normalization. We therefore also group the
sampled pairs by $\max(|f_i|,|f_j|)$ into four ranges---$[1,2]$, $[3,5]$,
$[6,10]$, and $>10$---and report the within-bucket correlation for each.

Tab.~\ref{tab:correlation} reports the results. On the five benchmark
datasets, the overall correlations are consistently very small, the largest
being $|r|=0.0428$ on LAION1M, and the bucket-level correlations stay below
$0.0617$ in almost all cases. The only exception is the $[1,2]$ bucket of
YFCC with $|r|=0.1891$, which draws on just $3{,}774$ sampled pairs and
therefore has limited support. Grouping by label-set size thus does not
change the picture: on these datasets, objects with similar label sets are
not systematically closer in the vector space, at any label-set size.
Yelp1M is the exception by construction: its labels are business categories
attached to review text, so an overall $|r|$ of $0.4791$ together with
consistently high bucket-level correlations makes it a complementary
workload on which label similarity does carry information about vector
proximity.
}

\update{
\subsection{Impact of Update}\label{app:impact-of-update}

Sec.~\ref{subsec:comparison-results} reports both update workloads on
YTB-Video only. Here we give the results on the remaining datasets:
Figs.~\ref{exp:app_upd_cont} and~\ref{exp:app_upd_over} for the
insertion-only workload, Figs.~\ref{exp:app_mix_cont}
and~\ref{exp:app_mix_over} for the mixed workload, in each case for
containment and overlap queries respectively, and
Fig.~\ref{fig:exp_mix_cost} for the update time under the mixed workload.
The protocol follows Sec.~\ref{subsec:comparison-results}: an index built
directly on the final object set is compared against one that reaches the
same set through updates. Across all datasets and both workloads, the two
recall-QPS curves stay close, confirming that the threshold-based
maintenance scheme preserves index quality under insertions, and that
lazy deletions leave the graph topology untouched.

\subsection{Ablation study of optimizations}\label{app:ablation-optimization}
Sec.~\ref{subsec:comparison-results} evaluates the two optimizations of
{\oursolution} on YTB-Audio only; Fig.~\ref{fig:ablation_YFCC} reports the
same ablation on YFCC. Replacing multi-graph search with independent search
again degrades the QPS-recall tradeoff, most visibly on overlap queries, where
a filter spans more trie regions and the key-ancestor graphs therefore matter
most for cross-region connectivity. Disabling \textsc{supplementEdges}, in
contrast, has almost no effect on YFCC: its materialized graphs are already
well connected, so the supplemented edges are rarely traversed. This confirms
the dataset-dependence noted in Sec.~\ref{subsec:comparison-results}---the
multi-graph search is the essential component, while
\textsc{supplementEdges} helps only where per-region connectivity is insufficient.

\begin{figure}[!tbp]
\centering
\includegraphics[width=1\linewidth]{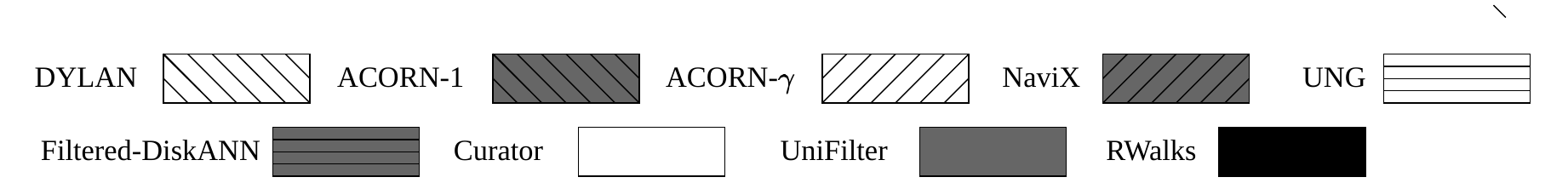}\par
\includegraphics[width=1\linewidth]{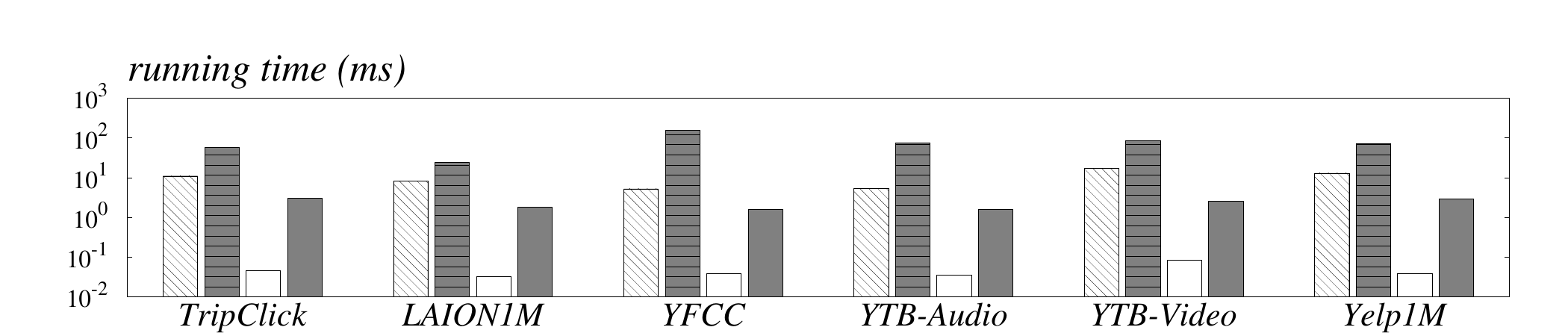}\par
\caption{Mixed update time.}
\label{fig:exp_mix_cost}
\end{figure}

\begin{figure}[!tbp]
\centering
\includegraphics[width=.60\linewidth]{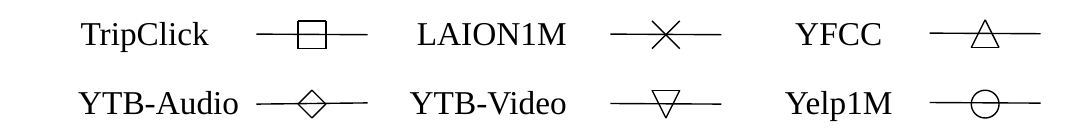}\par
\begin{minipage}[t]{.25\linewidth}
\vspace{0pt}
\centering
\includegraphics[width=\linewidth]{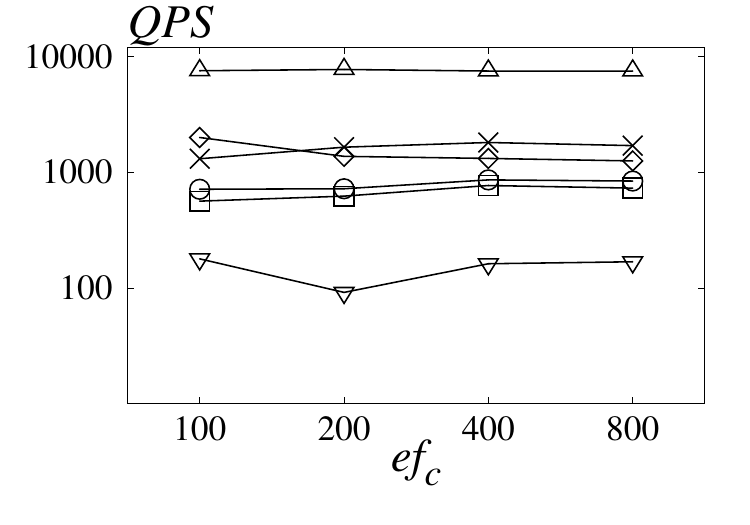}\par
{\small (a) containment\par}
\end{minipage}\hspace{.01\linewidth}\begin{minipage}[t]{.25\linewidth}
\vspace{0pt}
\centering
\includegraphics[width=\linewidth]{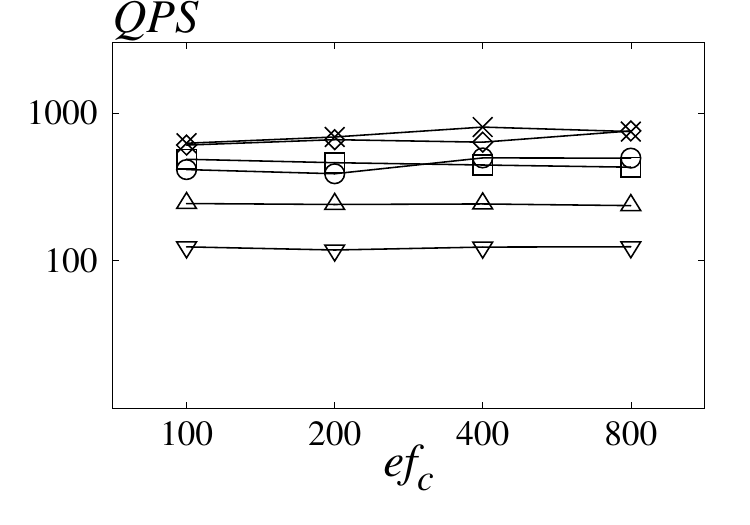}\par
{\small (b) overlap\par}
\end{minipage}\hspace{.01\linewidth}\begin{minipage}[t]{.25\linewidth}
\vspace{0pt}
\centering
\includegraphics[width=\linewidth]{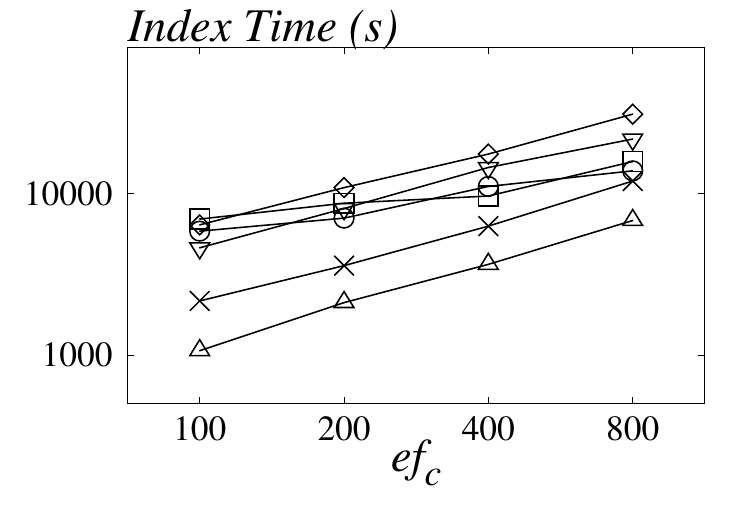}\par
{\small (c) indexing cost\par}
\end{minipage}\par
\caption{Impact of $\boldsymbol{ef_c}$ to query and indexing cost.}
\label{exp:ef_Impact}
\end{figure}

}

\subsection{ Impact of Parameter $ef_c$ on {\oursolution}} \label{app:tune-efc}

Fig.~\ref{exp:ef_Impact} shows the impact of \( ef_c \). As we can observe, increasing \( ef_c \) leads to higher indexing time but strengthens the quality of the graph, resulting in better performance. This improvement in connectivity enhances query performance. Based on the experimental results, we set \( ef_{c} = 400 \) as the default parameter as it gains a nice trade-off between the query performance and indexing cost.

\section{Proofs}\label{app:proofs}

{\bf Proof of Lemma \ref{lemma:index_quality}. } Let $v=u.\mathsf{index}$. By construction, $v$ is an ancestor of $u$, so $O_u\subseteq O_v$, and $u$ shares $v.G$ only when $\mathsf{log\_size}(u)=\mathsf{log\_size}(v)$. Hence, for some $k$,
\[
2^k\le |O_u|\le |O_v|<2^{k+1},
\]
which implies $|O_v|<2|O_u|$. Therefore, $v.G$ is a $2$-approximate index for $O_u$ by Definition~\ref{def:alpha-approximate-index}.

\header
{\bf Proof of Lemma \ref{lemma:unique-heavy}. }We prove by contradiction. Assume that node \(u\) has two distinct children \(c_1\) and \(c_2\) such that \(\log_2(|O_{c_1}|) = \log_2(|O_u|)\) and \(\log_2(|O_{c_2}|) = \log_2(|O_u|)\). Without loss of generality, let \(|O_{c_1}| \le |O_{c_2}|\). Since \(O_u\) is the set of all objects in the subtree of \(u\) and \(c_1, c_2\) are disjoint subtrees, we have \(|O_u| \ge |O_{c_1}| + |O_{c_2}| \ge 2|O_{c_1}|\). Consequently,
$\log_2(|O_u|) \ge \log_2(2|O_{c_1}|) = \log_2(|O_{c_1}|) + 1 = \log _2(|O_u|) + 1$,
which is a contradiction. Therefore, the assumption is false, and the lemma holds.

\header
{\bf Proof of Lemma \ref{lemma:appear}. }For the object $o$, it only belongs to the graph indexes of its ancestors in the Trie; hence there are at most $\O(|f_o|)$ such nodes. Moreover, because a node $u$ maintains its own index ($u.index=u$) only when $\mathsf{log\_size}(u) \neq \mathsf{log\_size}(u.p)$, the number of distinct $\mathsf{log\_size}$ values along the root-to-leaf path is $\O(\log n)$. Therefore, $o$ can belong to at most $\O(\log n)$ such indexes. Combining the two bounds yields the result.

\header
{\bf Proof of Theorem \ref{thm:build}. }The space complexity of our index consists of the following components: the Trie structure, the inverted lists \(I\), the \(\mathsf{belong}\) array, and all the HNSW graph indexes. Each object \(o\) is inserted into the Trie by traversing (or creating) at most \(|f_o|\) nodes. Hence the Trie and the associated inverted lists require \(\O\bigl(\sum_{o \in O} |f_o|\bigr)\) space. The \(belong_{\mathbf{o},i}\) indicates whether object \(o\) belongs to a graph of logarithmic size \(i\). Since \(i\) can take at most \(\O(\log n)\) distinct values, the total size of the \(\mathsf{belong}\) array is \(\O(n \log n)\). Through Lemma \ref{lemma:appear}, the overall space occupied by all graph indexes is \( \O\Bigl(\sum_{o \in O} \min\{|f_o|,\log n\}\, M\Bigr) \)
Summing the three parts, the total space complexity is \[ \O\Bigl(n\log n + \sum_{o \in O} |f_o| + \sum_{o \in O} \min\{|f_o|,\log n\}\, M\Bigr). \]

For construction time, we first build the Trie and the inverted lists, which takes \(\O\bigl(\sum_{o \in O} |f_o|\bigr)\) time. Initializing and updating the \(\mathsf{belong}\) array costs \(\O(n\log n)\). The dominant cost is the HNSW insertion: each object \(o\) is inserted into at most \(\min\{|f_o|,\log n\}\) graphs, and each insertion has complexity \(c_{\text{upd}}\) (the time to insert one point into a plain HNSW graph). 
For lines 5–8 of Algo. \ref{algo:index_build}, each element \(o\) is subject to the \texttt{supplementEdges} operation at most \(\min\{l_{\text{th}}, |f_o|\}\) times, where the cost of each such operation is \(c_q\), which denotes the cost of performing one query on DYLAN.
Hence the total construction time is
\[ \O\Bigl(n\log n + \sum_{o \in O} |f_o| + \sum_{o \in O} \min\{|f_o|,\log n\}\, (c_{\text{upd}}+c_q)\Bigr).\]

\header
{\bf Proof of Theorem \ref{thm:find_nodes}. } For containment queries, we access ancestors of nodes in $I_{\max}$ and examine the prefix of each node. For overlap queries, we access ancestors of nodes from the inverted list of every label in $f_q$ and also examine the prefix of each node. The number of ancestors of a node $u$ is $|\text{prefix}(u)|$, and each ancestor can be accessed in $\O(1)$ time. 
Now consider a complex query. For any node \(u \in I_l\) where \(l \in \mathcal{L}_{\Phi}\), if \(u\) is inspected during a merge step in ComplexSearch, the associated cost is \(\text{prefix}(u)\). In the worst-case scenario, every merge operation in ComplexSearch includes \(u\). Since a complex predicate comprising \(|\Phi|\) containment or overlap queries entails \(|\Phi|\) merge operations, the overall complexity, when combined with the costs of ContainmentSearch and OverlapSearch, is given by
$\O\left( \sum_{l \in \mathcal{L}_{\Phi}} \sum_{u \in I_l} |\text{prefix}(u)| \cdot |\Phi| \right)$.
Therefore, the theorem holds.

\header
{\bf Proof of Theorem \ref{thm:insert}. }Operations on the Trie and updates to the inverted lists take \(\O(|f_o|)\) time. Traversing the path from the root to the leaf also costs \(\O(|f_o|)\). For HNSW insertion and supplementEdge, an object belongs to at most \(\O(\min\{|f_o|, \log n\})\) graphs, so the total insertion cost is \(\O(\min\{|f_o|, \log n\} \cdot (c_{\text{upd}}+c_q))\). Adjusting the index assigned to a node may involve copying or deleting an index; one such operation on node \(u\) costs \(\O(|O_u|M)\) and is triggered only when a change in \(|O_u|\) alters \(\mathsf{log\_size}(u)\). After one such change, the next copy/delete on this index occurs after \(\Omega(|O_u|)\) insertions into \(O_u\). Hence the amortized cost per insertion contributed by copying/deleting is \(\O(M)\). Since an insertion affects at most \(|f_o|\) subtrees, the total amortized overhead from copying/deleting is \(\O(|f_o|M)\).

Rebuilding the index for node \(u\) costs \(\O(|O_u|c_{\text{upd}})\). After a rebuild, \(u.G\) contains at most \(2^{\mathsf{log\_size}(u)+1}-1\) nodes, and the next rebuild is triggered after \(\Omega(|O_u|)\) further insertions. Therefore, the amortized cost per insertion from rebuilding is \(\O(c_{\text{upd}})\). Because an insertion touches at most \(\min\{|f_o|,\log n\}\) graphs, the total amortized rebuilding overhead is \(\O(\min\{|f_o|,\log n\}\cdot c_{\text{upd}})\). Summing all parts yields an amortized time complexity per update of 
\(
\O\bigl(|f_o| M + \min\{|f_o|, \log n\} \cdot (c_{\text{upd}} +c_q)\bigr).
\)
Moreover, the graph \(u.\text{index}.G\) is allowed to contain at most \(2^{\mathsf{log\_size}(u)+2}-1\) vertices, so
\[
\frac{|O_{u.\text{index}.G}|}{|O_u|} \le \frac{2^{\mathsf{log\_size}(u)+2}-1}{2^{\mathsf{log\_size}(u)}} < 4,
\] 
assuring that the index remains a \(4\)-approximate index for \(O_u\).

%% file: table/Memory.tex
\begin{table}[t]
\centering
\caption{DYLAN memory cost breakdown by component (GB)}
\vspace{-4mm}
\label{tab:memory_breakdown}
\scalebox{0.88}{
\begin{CRfit}\begin{tabular}{lrrrrrr}
\toprule
Dataset & Vector & Labels & Trie & Graph & Other & Total \\
\midrule
TripClick & 3.02 & 0.06 & 0.01 & 0.41 & 0.06 & 3.56 \\
LAION1M & 1.91 & 0.05 & 0.01 & 0.48 & 0.06 & 2.52 \\
YFCC & 0.72 & 0.07 & 1.09 & 0.77 & 0.15 & 2.80 \\
YTB-Audio & 2.39 & 0.27 & 0.28 & 3.08 & 0.38 & 6.38 \\
YTB-Video & 3.82 & 0.05 & 0.09 & 2.12 & 0.08 & 6.15 \\
Yelp1M & 1.43 & 0.06 & 0.03 & 0.71 & 0.08 & 2.32 \\ 
\bottomrule
\end{tabular}\end{CRfit}
}
\end{table}

%% file: table/correlation.tex
\begin{table}[t] 
\vspace{-1mm}
\centering \small \caption{Absolute Pearson correlation $|r|$ between label-set Jaccard similarity and $L_2$ distance. Object pairs are grouped by $\max(|f_i|,|f_j|)$.} \label{tab:correlation} \vspace{-4mm}
\begin{CRfit}\begin{tabular}{lccccc} \toprule Dataset & $[1,2]$ & $[3,5]$ & $[6,10]$ & $>10$ & Overall \\ \midrule TripClick & 0.0055 & 0.0078 & 0.0528 & -- & 0.0055 \\ LAION1M & 0.0408 & 0.0617 & -- & -- & 0.0428 \\ YFCC & 0.1891 & 0.0028 & 0.0073 & 0.0012 & 0.0217 \\ YTB-Audio & 0.0018 & 0.0011 & 0.0028 & 0.0031 & 0.0009 \\ YTB-Video & 0.0216 & 0.0209 & 0.0217 & 0.0047 & 0.0206 \\ Yelp1M & 0.5742 & 0.5089 & 0.4711 & 0.4458 & 0.4791 \\ \bottomrule \end{tabular}\end{CRfit} \end{table} 

%% file: table/selected_parameters.tex
\begin{table*}[t]
    \centering
    \caption{Selected construction parameters for each method and dataset.
Here $R$ and $M$ denote the graph out-degree, $L_{\mathrm{build}}$ and $L$
denote the construction candidate size, and $\mathtt{nlist}$ and $B_{\max}$
denote Curator's branching factor and maximum leaf size. ACORN-1 and
ACORN-$\gamma$ are tuned independently, and ``--'' indicates that the method is omitted due to out-of-memory.}\vspace{-4mm}
    \label{tab:selected_parameters}
    \scalebox{0.92}{%
    \begin{CRfit}\begin{tabular}{lccccccccc}
        \toprule
        \multirow{2}{*}{Dataset}
        & Filtered-DiskANN & UNG & NaviX & ACORN-1 & ACORN-$\gamma$
        & Curator & RWalks & UniFilter & {\oursolution} \\
        \cmidrule(lr){2-2}\cmidrule(lr){3-3}\cmidrule(lr){4-4}\cmidrule(lr){5-5}\cmidrule(lr){6-6}\cmidrule(lr){7-7}\cmidrule(lr){8-8}\cmidrule(lr){9-9}\cmidrule(lr){10-10}
        & $(L_{\mathrm{build}},R)$ & $(R,L)$ & $M$ & $M$ & $M$
        & $(\mathtt{nlist},B_{\max})$ & $M$ & $M$ & $M$ \\
        \midrule
        LAION1M   & $(144,96)$  & $(48,200)$ & $48$ & $48$ & $48$ & $(64,512)$ & $48$ & $16$ & $16$ \\
        TripClick & $(192,128)$ & $(32,200)$ & $64$ & $48$ & $48$ & $(64,512)$ & $32$ & $16$ & $16$ \\
        YFCC      & $(240,160)$ & $(16,100)$ & $48$ & $48$ & $48$ & $(64,512)$ & --   & $32$ & $16$ \\
        YTB-Audio & $(240,160)$ & $(48,200)$ & $48$ & $64$ & $64$ & $(64,512)$ & $48$ & $48$ & $16$ \\
        YTB-Video & $(240,160)$ & $(48,200)$ & $64$ & $64$ & $64$ & $(64,512)$ & $64$ & $16$ & $64$ \\
        Yelp1M    & $(240,160)$ & $(16,100)$ & $48$ & $64$ & $64$ & $(64,512)$ & $48$ & $64$ & $16$ \\
        \bottomrule
    \end{tabular}\end{CRfit}}
\end{table*}